\documentclass[11pt,a4paper]{article}
\pdfoutput=1
\usepackage{amssymb,amsmath,amsfonts, mathtools, mathrsfs}
\usepackage[utf8]{inputenc} 
\usepackage[dvipsnames]{xcolor}
\usepackage{graphicx}
\usepackage{caption}
\usepackage{subcaption}
\usepackage{enumerate}
\usepackage{dsfont}
\usepackage[sort,compress]{cite}
\usepackage{verbatim}
\usepackage{amsfonts,euscript,amssymb,stmaryrd,braket}
\usepackage{tikz}
\usetikzlibrary{arrows,decorations.markings,patterns}
\usepackage{slashed}
\definecolor{hyperref}{RGB}{026,028,185}
\usepackage[bookmarks=true,colorlinks=true,linkcolor=hyperref,citecolor=hyperref,urlcolor=hyperref,bookmarksnumbered]{hyperref}
\usepackage{cite}
\usepackage{gensymb}

\def\clock{{\count0=\time
           \divide\count0 60
           \ifnum\count0<10 0\fi\the\count0
           \multiply\count0 -60 \advance\count0 \time
           :\ifnum\count0<10 0\fi \the\count0
         }}
\newcommand{\timestamp}{{\small\vbox{\hbox{\tt\jobname.tex}
\hbox{\the\day/\the\month/\the\year, \clock}}}}

\newcommand{\bea}{\begin{eqnarray}}
\newcommand{\eea}{\end{eqnarray}}

\newcommand{\be}{\begin{equation}}
\newcommand{\ee}{\end{equation}}

\makeatletter
\let\old@startsection=\@startsection
\let\oldl@section=\l@section
\renewcommand{\@startsection}[6]{\old@startsection{#1}{#2}{#3}{#4}{#5}{#6\mathversion{bold}}}
\renewcommand{\l@section}[2]{\oldl@section{\mathversion{bold}#1}{#2}}
\makeatother

\numberwithin{equation}{section}

\usepackage{color}

\begin{document}
\renewcommand{\thefootnote}{\arabic{footnote}}

\overfullrule=0pt
\parskip=2pt
\parindent=12pt
\headheight=0in \headsep=0in \topmargin=0in \oddsidemargin=0in

\vspace{ -3cm} \thispagestyle{empty} \vspace{-1cm}
\begin{flushright} 
\footnotesize
\textcolor{red}{\phantom{print-report}}
\end{flushright}

\begin{center}
\vspace{1.2cm}
{\Large\bf \mathversion{bold}

Holographic entanglement entropy in AdS$_4$/BCFT$_3$

 \vspace{0.015cm}

and the Willmore functional

}

 \vspace{0.8cm} 
{
Domenico Seminara$^{a,}$\footnote[1]{seminara@fi.infn.it},
Jacopo Sisti$^{b,}$\footnote[2]{jsisti@sissa.it}
and Erik Tonni$^{b,}$\footnote[3]{erik.tonni@sissa.it}
}
 \vskip  0.8cm

\small
{\em
$^{a}$
Dipartimento di Fisica, Universit\`a di Firenze and INFN Sezione di Firenze, Via G. Sansone 1, 50019 Sesto Fiorentino,
Italy
  \vskip 0.05cm
$^{b}$ 
SISSA and INFN, via Bonomea 265, 34136, Trieste, Italy 
}
\normalsize

\end{center}

\vspace{0.3cm}
\begin{abstract}

We study the holographic entanglement entropy of spatial regions 
having arbitrary shapes in the AdS$_4$/BCFT$_3$ correspondence with static gravitational backgrounds, 
focusing on the subleading term with respect to the area law term in its expansion as the UV cutoff vanishes. 
An analytic expression depending on the unit vector normal to the minimal area surface anchored to the entangling curve is obtained.
When the bulk spacetime is a part of AdS$_4$, this formula becomes 
the Willmore functional with a proper boundary term evaluated on the minimal surface viewed 
as a submanifold of a three dimensional flat Euclidean space with boundary. 
For some smooth domains, the analytic expressions of the finite term are reproduced,
including the case of a disk disjoint from a boundary which is either flat or circular. 
When the spatial region contains corners adjacent to the boundary, the subleading term 
is a logarithmic divergence whose coefficient is determined by a corner function that is known analytically,
and this result is also recovered. 
A numerical approach is employed to construct extremal surfaces anchored to entangling curves with arbitrary shapes.
This analysis is used both to check some analytic results
and to find numerically the finite term of the holographic entanglement entropy 
for some ellipses at finite distance from a flat boundary.

\end{abstract}

\newpage

\tableofcontents

 \newpage
\section{Introduction}
\label{sec intro}

Entanglement has attracted an intense research activity during the last two decades
in quantum field theory, quantum gravity, quantum many-body systems and quantum information 
(see \cite{ent-reviews} for reviews).
Among the entanglement indicators, the entanglement entropy plays a dominant role
because it quantifies the entanglement of a bipartition when the entire quantum system is in a pure state.

Given the Hilbert space $\mathcal{H} $ associated to a quantum system in the state characterised by the density matrix $\rho$, 
and assuming that it is bipartite as $\mathcal{H} = \mathcal{H}_A  \otimes \mathcal{H}_B$,
the $A$'s reduced density matrix is $\rho_A = \textrm{Tr}_{\mathcal{H}_B} \rho$ and
the entanglement entropy between $A$ and $B$ is defined as the Von Neumann entropy of $\rho_A$, namely
$ S_A =  -\,\textrm{Tr} (\rho_A \log \rho_A)$.
Similarly, the entanglement entropy $S_B$ is the Von Neumann entropy of $B$'s reduced density matrix $\rho_B = \textrm{Tr}_{\mathcal{H}_A} \rho\,$.
When $\rho$ is a pure state, we have $S_A = S_B$.
Hereafter we only consider spatial bipartitions where $A$ is a spatial region and $B$ its complement. 

In the approach to quantum gravity based on the gauge/gravity correspondence, 
a crucial result was found by Ryu and Takayanagi \cite{RT}, 
who proposed the holographic formula to compute the entanglement entropy of a $d+1$ dimensional CFT at strong coupling
with a gravitational dual description characterised by an asymptotically AdS$_{d+2}$ spacetime. 
This prescription has been extended to time dependent backgrounds in \cite{Hubeny:2007xt}.
Recently, an interesting reformulation of the holographic entanglement entropy through some peculiar flows
has been proposed in \cite{Freedman:2016zud} and explored further in \cite{Headrick:2017ucz}.

In this manuscript, for simplicity, only static spacetimes are considered. 
By introducing the coordinate $z > 0$ along the holographic direction in the gravitational spacetime,
the dual CFT$_{d+1}$ is defined on the conformal boundary at $z=0$.
Given a region $A$ in a spatial slice  of the CFT$_{d+1}$,
its holographic entanglement entropy at strong coupling is obtained from the area of 
the $d$ dimensional minimal area hypersurface $\hat{\gamma}_A$ anchored to the boundary of $A$
(i.e. such that $\partial \hat{\gamma}_A = \partial A$) and  homologous to $A$ \cite{Headrick:2007km}.
Since the asymptotically AdS$_{d+2}$ gravitational spacetime is noncompact along the holographic direction
and $\hat{\gamma}_A $ reaches its boundary, the area of $\hat{\gamma}_A$ diverges.
This divergence is usually  regularised by introducing a cutoff $\varepsilon > 0$ in the holographic direction $z$ (i.e. $z\geqslant \varepsilon$)
such that $\varepsilon \ll \textrm{Area}(\partial A)$, which corresponds to the gravitational dual of the UV cutoff in the CFT$_{d+1}$.
Denoting by $\hat{\gamma}_\varepsilon \equiv \hat{\gamma}_A \cap \{  z\geqslant \varepsilon  \}$ the restriction of $\hat{\gamma}_A$ to $z\geqslant \varepsilon$, 
the holographic entanglement entropy is given by 
\be
\label{RT formula intro dim generic}
S_A = \frac{\textrm{Area}(\hat{\gamma}_\varepsilon )}{4G_{\textrm{\tiny N}}}
\ee
being $G_{\textrm{\tiny N}}$ the $d+2$ dimensional gravitational Newton constant.

By expanding the r.h.s. of \eqref{RT formula intro dim generic} as $\varepsilon \to 0^+$,
the leading divergence is $O(1/\varepsilon ^{d-1})$ and its coefficient is proportional to the area of
the hypersurface $\partial A  \cap \partial B$ which separates $A$ and $B$ (entangling surface).
The terms subleading with respect to the area law provide important information.
For instance, in $d=3$ and for smooth $\partial A$, a logarithmic divergence occurs and its
coefficient contains the anomaly coefficients of the CFT$_4$ \cite{Solodukhin:1994yz}.

In this manuscript, we focus on $d=2$, where (\ref{RT formula intro dim generic}) becomes
\be
\label{RT formula bdy intro}
S_A 
= \frac{ L^2_{\textrm{\tiny AdS}} }{4 G_{\textrm{\tiny N}}}\;  \mathcal{A}[\hat{\gamma}_\varepsilon ]
\ee
where the dependence on the AdS radius $L_{\textrm{\tiny AdS}} $ has been factored out
and the area $\mathcal{A}[\hat{\gamma}_\varepsilon ]$ of the two dimensional surface $\hat{\gamma}_\varepsilon $ 
must be evaluated by setting $L^2_{\textrm{\tiny AdS}} =1$.
In AdS$_4$/CFT$_3$, the minimal area surface $\hat{\gamma}_A$ is anchored to the entangling curve $\partial A  = \partial B$
and the expansion of $\mathcal{A}[\hat{\gamma}_\varepsilon ]$ as $\varepsilon  \to 0$ reads
\be
\label{area expansion intro}
 \mathcal{A}[\hat{\gamma}_\varepsilon ]
\,=\,
\frac{P_{A}}{\varepsilon} 
- F_A
+ o(1)
\ee
where $P_A =\textrm{length}(\partial A ) = \textrm{length}(\partial B )$ is the perimeter of $A$.

In three dimensional quantum field theories, the subleading term with respect to the area law in $S_A$ is finite for smooth entangling curves and it contains relevant information. 
For instance, when $A$ is a disk,  it has been shown that this term decreases along a renormalization group flow going from an ultraviolet to an infrared fixed point
\cite{F-theorem}.
In a CFT$_3$, when $A$ contains corners, the subleading term with respect to the area law is a logarithmic divergence
whose coefficient is determined by a model dependent corner function \cite{Drukker:1999zq, Casini:2006hu, Hirata:2006jx}.
The limit of this function as the corner disappears provides the coefficient characterising the two point function of the stress tensor
\cite{Bueno:2015rda, Faulkner:2015csl}.

These important results tell that it is useful to study the shape dependence of the subleading term with respect to the area law in $S_A$.
Nonetheless, it is very difficult to get analytic expressions valid for generic shapes, even for simple quantum field theories.  
This problem has been tackled for the holographic entanglement entropy in AdS$_4$/CFT$_3$.
Interesting results have been obtained for regions given by small perturbations of the disk and for star shaped domains \cite{Hubeny:2012ry}.
When $A$ has a generic shape, analytic expressions for $F_A$ can be written where the Willmore functional \cite{Thomsen, willmorebound, willmorebook} plays an important role. 
The first result has been found in \cite{Babich:1992mc} for the static case where the gravitational background is AdS$_4$.
This analysis has been further developed in \cite{AM} and then extended to a generic asymptotically AdS$_4$ spacetime in \cite{Fonda:2015nma}.
In \cite{Fonda:2015nma} the analytic results have been also checked against numerical data obtained with
{\it Surface Evolver} \cite{evolverpaper, evolverlink}, which has been first employed to study the holographic entanglement entropy in \cite{Fonda:2014cca}.
The analytic expressions for $F_A$ found in \cite{Babich:1992mc, AM, Fonda:2015nma} hold also when 
$A$ is made by disjoint regions. 
We remark that, in CFT$_3$, it is very difficult to find analytic results about the entanglement entropy of disjoint regions \cite{Cardy:2013nua}. 
Also in CFT$_2$, where the conformal symmetry is more powerful, 
few analytic results are available when  the subsystem   is made by disjoint intervals  \cite{2dCFT-disjoint-interval}.

Conformal field theories in the presence of boundaries (BCFTs) have been largely studied in the literature \cite{bcft2, McAvity:1993ue,Solodukhin:2015eca,Deutsch:1978sc}
and also their gravitational duals through the gauge/gravity correspondence
(which is called AdS/BCFT in these cases) have been constructed \cite{pre-ads/bcft, Takayanagi:2011zk, Fujita:2011fp, Nozaki:2012qd, Miao:2017gyt, Astaneh:2017ghi, FarajiAstaneh:2017hqv, Azeyanagi:2007qj, holog-kondo}.
These gravitational backgrounds are part of asymptotically AdS spacetimes delimited by a hypersurface $\mathcal{Q}$ extended in the bulk 
whose boundary coincides with the boundary of the dual BCFT \cite{Takayanagi:2011zk, Fujita:2011fp, Nozaki:2012qd, Miao:2017gyt, Astaneh:2017ghi, FarajiAstaneh:2017hqv}.

We are interested in the shape dependence of the holographic entanglement entropy in AdS/BCFT
through the prescription (\ref{RT formula intro dim generic}).
Given a spatial region $A$ in a spatial slice of the BCFT, 
the holographic entanglement entropy is determined by the minimal area hypersurface $\hat{\gamma}_A$ anchored to the entangling surface $\partial A \cap \partial B$.
Whenever $\partial A$ intersects the boundary of the BCFT, 
we have $\partial A \cap \partial B \subsetneq \partial A$ 
and the area of $\partial A \cap \partial B$ occurs in the leading divergence (area law term).
Another peculiar feature of extremal hypersurfaces in the context of AdS/BCFT is that $\hat{\gamma}_A$ may intersect $\mathcal{Q}$ 
(with a slight abuse of notation, in the following we will denote in the same way $\mathcal{Q}$  and its spatial section).
It is important to remark that, since $\hat{\gamma}_A \cap \mathcal{Q}$ is not fixed, the extremization of the area functional leads
to the condition that $\hat{\gamma}_A$ intersects $\mathcal{Q}$ orthogonally. 
Furthermore, as discussed above, in order to evaluate the holographic entanglement entropy we have to introduce the UV cutoff $\varepsilon$
and consider the area of the restricted hypersurface $\hat{\gamma}_\varepsilon \equiv \hat{\gamma}_A \cap \{  z\geqslant \varepsilon  \}$
because $\hat{\gamma}_A$ reaches the conformal boundary of an asymptotically AdS space.

In this manuscript, we consider the holographic entanglement entropy in  AdS$_4$/BCFT$_3$ of spatial regions $A$ having an arbitrary shape.
For the sake of simplicity, we will consider static backgrounds in AdS$_4$/BCFT$_3$,
which provide the simplest arena where the shape dependence plays an important role. 
The holographic entanglement entropy is computed through (\ref{RT formula bdy intro}), 
where the minimal area surface $\hat{\gamma}_A$ is anchored to the entangling curve and it can intersect orthogonally $\mathcal{Q}$
(see also footnote 11 of \cite{Headrick:2017ucz}).

The expansion of the area $\mathcal{A}[\hat{\gamma}_\varepsilon ]$ 
of the two dimensional surface $\hat{\gamma}_\varepsilon $ in (\ref{RT formula bdy intro}) as $\varepsilon  \to 0$ reads
\be
\label{hee bdy intro}
\mathcal{A}[\hat{\gamma}_\varepsilon]
= 
\frac{P_{A,B}}{\varepsilon} 
- F_A
+ o(1)
\ee
being  $P_{A,B} = \textrm{length}(\partial A \cap \partial B) \leqslant P_A$ the length of the entangling curve.
When $\partial A$ is smooth, $F_A$ is finite.
It is worth considering the configurations whose subleading term $F_A$ can be computed analytically.
For instance, the case of an infinite strip parallel to a flat boundary has been considered in 
\cite{Nagasaki:2011ue, Miao:2017gyt, Seminara:2017hhh}.

When $A$ contains corners, the subleading term $F_A$ diverges logarithmically and the coefficient of this
divergence is determined by different kinds of corner functions, depending on the position of the tips of the corners.
For  the corners whose tip is not on the boundary of the BCFT$_3$, the well known corner function of \cite{Drukker:1999zq}
must be employed. 
If the tip of the corner is located on the boundary of the BCFT$_3$, the corner functions encode also the boundary conditions characterising the BCFT$_3$. 
In the context of AdS$_4$/BCFT$_3$, these corner functions have been studied analytically in \cite{Seminara:2017hhh}
by computing the holographic entanglement entropy of an infinite wedge.
For instance, when $A$ is an infinite wedge adjacent to a flat boundary, the holographic entanglement entropy is given by (\ref{RT formula bdy intro}) with
\be
\label{area wedge intro exp}
\mathcal{A}[\hat{\gamma}_\varepsilon]
= 
\,
\frac{L}{\varepsilon} 
- F_{\alpha}(\omega) \, \log(L / \varepsilon)
+ O(1)
\ee
where $L \gg \varepsilon$ is an infrared cutoff, $\omega$ is the opening angle of the wedge and the subindex $\alpha$ 
denotes the fact that the corner function  $F_{\alpha}(\omega)$  depends on the boundary conditions in a highly non trivial way.
The analytic expression of $F_{\alpha}(\omega)$ has been checked numerically by employing  Surface Evolver \cite{Seminara:2017hhh}.

\subsection{Summary of the results}

In this manuscript, we study the subleading term $F_A$ of the
holographic entanglement entropy in AdS$_4$/BCFT$_3$ (see (\ref{RT formula bdy intro}) and (\ref{hee bdy intro})) 
for entangling curves  having arbitrary shapes.

After a brief description of the AdS/BCFT setup \cite{Takayanagi:2011zk, Fujita:2011fp, Nozaki:2012qd, Miao:2017gyt, Astaneh:2017ghi, FarajiAstaneh:2017hqv},
in Sec.\;\ref{sec HEE bdy} we adapt the method employed in  \cite{Babich:1992mc, AM, Fonda:2015nma} for the holographic entanglement entropy in AdS$_4$/CFT$_3$ to this case.
This analysis leads to writing $F_A$ as a functional evaluated on the surface $\hat{\gamma}_\varepsilon$ embedded in a three dimensional Euclidean  space with boundary which is asymptotically flat close to the boundary.
This result holds for any static gravitational background and for any region, even when it is made by disjoint domains.
Focusing on the simplest AdS$_4$/BCFT$_3$ setup, 
where the gravitational background is a part of $\mathbb{H}_3$
and the asymptotically flat space  is a part of $\mathbb{R}^3$,
in Sec.\;\ref{sec ads4} we observe that the functional obtained for $F_A$ becomes the Willmore functional
\cite{Thomsen, willmorebound, willmorebook} with a proper boundary term
evaluated on the surface $\hat{\gamma}_\varepsilon$ embedded in $\mathbb{R}^3$.
In the remaining part of the manuscript, further simplifications are introduced 
by restricting to BCFT$_3$'s whose spatial slice is either a half plane (see Sec.\;\ref{sec flat bdy}) or a disk (see Sec.\;\ref{sec circular bdy}).

The analytic expression found for $F_A$ is checked by considering some particular regions such that 
the corresponding $F_A$ can be found analytically. 
In Sec.\;\ref{sec strip adjacent} we recover the result for an infinite strip parallel to a flat boundary  \cite{Nagasaki:2011ue, Miao:2017gyt, Seminara:2017hhh}.
When $A$ is a finite region with smooth $\partial A$  that does not intersect the boundary, $F_A$ is finite. 
The simplest configuration to consider is a disk disjoint  from a boundary which is either flat or circular. 
In Sec.\;\ref{sec disk} we compute $F_A$ analytically for these configurations 
and check the results against numerical data obtained through Surface Evolver.
We remark that Surface Evolver is a very powerful tool in this analysis because 
it allows to study numerically any kind of region $A$, even when it is made by disjoint connected domains
or when it contains corners (see \cite{Fonda:2014cca, Seminara:2017hhh} for some examples in AdS$_4$).
In Sec.\;\ref{sec domain disjoint} Surface Evolver is employed to find numerically $F_A$ corresponding to some 
ellipses disjoint from a flat boundary.

In Sec.\;\ref{sec corners} we check that the result derived in Sec.\;\ref{sec ads4} for $F_A$ can be applied also when $A$ contains corners
by considering the explicit cases of a half disk (see Sec.\;\ref{sec app half disk}) and an infinite wedge adjacent to the flat boundary (see Sec.\;\ref{sec wedge}).
The analytic expression for the  corner function $F_{\alpha}(\omega)$ found in \cite{Seminara:2017hhh} is recovered
from the general expression of $F_A$ obtained in Sec.\;\ref{sec ads4}.

In Appendix\;\ref{app:mapping} we report the mappings that are employed to study the disk disjoint from a flat boundary.
The Appendix\;\ref{app:disk} contains the technical details for the derivation of the analytic results presented in Sec.\;\ref{sec disk}
about a disk concentric to a circular boundary.
In Appendix\;\ref{app: wedge adj} we discuss the details underlying the derivation of the corner function of \cite{Seminara:2017hhh}
through the general formula for $F_A$ of Sec.\;\ref{sec flat bdy}.
In Appendix\;\ref{app aux_domains} we further discuss the auxiliary surfaces corresponding to some extremal surfaces
occurring in the manuscript.

\newpage

\section{Holographic entanglement entropy in AdS$_4$/BCFT$_3$}
\label{sec HEE bdy}

In this section we provide an analytic formula for the subleading term $F_A$
of the holographic entanglement entropy in AdS$_4$/BCFT$_3$ which is valid for any region $A$ and any static background. 
In Sec.\;\ref{sec general bg} we derive the general formula and in Sec.\;\ref{sec ads4} we describe how it simplifies when the gravitational background is a part of AdS$_4$,
focusing on the simplest cases where the boundary of a spatial slice of the BCFT$_3$ is either an infinite line or
a circle.

Following \cite{Takayanagi:2011zk}, we consider the gravitational background dual to a BCFT$_{d+1}$ 
given by an asymptotically AdS$_{d+2}$ spacetime $\mathcal{M}$ restricted by the occurrence of a $d+1$ dimensional hypersurface $\mathcal{Q}$ 
in the bulk whose boundary coincides with the boundary of the BCFT$_{d+1}$.
Hence the boundary of $\mathcal{M}$ is the union of $\mathcal{Q}$ and the conformal boundary where the BCFT$_{d+1}$ is defined. 
The gravitational action for the $d+2$ dimensional metric $G_{ab}$ in the bulk reads  \cite{Takayanagi:2011zk, Fujita:2011fp}
\be
\label{gravi-action-general}
\mathcal{I} = 
\frac{1}{16 \pi  G_{\textrm{\tiny N}}} 
\int_{\mathcal{M}} \sqrt{-\,G}\, \big( R- 2\Lambda \big) 
+
\frac{1}{8 \pi  G_{\textrm{\tiny N}}} 
\int_{\mathcal{Q}} \sqrt{-H} \,K
+ \mathcal{I}_\mathcal{Q}
\ee
being $\Lambda = - \tfrac{d(d+1)}{2 L^2_{\textrm{\tiny AdS}}}$ the negative cosmological constant,
$H_{ab}$ the induced metric on $\mathcal{Q}$ and $K = H^{ab} K_{ab}$ the trace of the extrinsic curvature $K_{ab}$ of $\mathcal{Q}$.
The boundary term $\mathcal{I}_\mathcal{Q}$ describes some matter fields localised on $\mathcal{Q}$.
The boundary term due to the fact that $\partial M$ is non smooth \cite{Hayward:1993my} along the boundary of the BCFT$_{d+1}$
and the ones introduced by the holographic renormalisation procedure \cite{holog-ren} 
have been omitted because they are not relevant in our analysis. 
We will focus only on static backgrounds. 

While in Sec.\;\ref{sec general bg}  a generic $\mathcal{Q}$ is considered, for the remaining part of the manuscript 
we focus on the simplest case where $\mathcal{I}_\mathcal{Q}$ in (\ref{gravi-action-general}) is given by 
\be
\label{gravi-action-Tconstant}
\mathcal{I}_\mathcal{Q} 
\,=\, 
-  \,\frac{T}{8 \pi  G_{\textrm{\tiny N}}} 
\int_{\mathcal{Q}} \sqrt{-H} 
\ee
being $T$ a constant real parameter characterising the hypersurface $\mathcal{Q}$.
Different proposals have been made to construct $\mathcal{Q}$ \cite{Takayanagi:2011zk, Miao:2017gyt, Astaneh:2017ghi},
but they will not be discussed here because our results can be employed independently 
of the way underlying the construction of $\mathcal{Q}$.

In this manuscript, we consider the holographic entanglement entropy in AdS$_4$/BCFT$_3$ with static gravitational backgrounds.

Given a two dimensional region $A$ in the spatial slice of the BCFT$_3$, the corresponding holographic entanglement entropy 
is given by (\ref{RT formula bdy intro}) and (\ref{hee bdy intro}), as discussed in Sec.\;\ref{sec intro}.
The minimal area surface $\hat{\gamma}_A$ is anchored to the entangling curve $\partial A \cap \partial B$
and, whenever $\hat{\gamma}_A \cap \mathcal{Q} \neq \emptyset$, 
these two surfaces are orthogonal along their intersection. 
We remind that the  expansion (\ref{hee bdy intro}) is defined by first introducing the UV cutoff $\varepsilon $ 
and then computing the area of the part of $\hat{\gamma}_A$ restricted to $z\geqslant \varepsilon$,
namely  $\hat{\gamma}_\varepsilon \equiv \hat{\gamma}_A \cap \{  z\geqslant \varepsilon  \}$.
By employing the method of \cite{Babich:1992mc, AM, Fonda:2015nma},
in Sec.\;\ref{sec general bg} we find  an analytic expression for the subleading term
$F_A$ in (\ref{hee bdy intro}) that is valid for any region $A$ and for any static gravitational background.

\subsection{Static backgrounds}
\label{sec general bg}

In the AdS$_4$/BCFT$_3$ setup described above,
let us denote by $\mathcal{M}_3$ the three dimensional Euclidean space with metric $g_{\mu\nu}$
obtained by taking a constant time slice of the static asymptotically AdS$_4$ gravitational background.
The boundary of $\mathcal{M}_3$
is the union of the conformal boundary, which is the constant time slice of the spacetime where the BCFT$_3$ is defined, 
and the surface $\mathcal{Q}$ delimiting the gravitational bulk.

Let us consider a two dimensional surface $\gamma$ embedded into $\mathcal{M}_3$ 
whose boundary $\partial \gamma$ is made by either one or many disjoint closed curves. 
Denoting by $n_\mu$ the spacelike unit vector normal to $\gamma$,
the metric induced on $\gamma$ (first fundamental form) and the extrinsic curvature of $\gamma$ (second fundamental form) 
are given respectively by
\be
\label{fund forms def}
h_{\mu\nu} = g_{\mu\nu} - n_\mu  n_\nu 
\;\;\qquad\;\;
K_{\mu\nu} =  h_\mu^{\;\;\alpha}h_\nu^{\;\;\beta}\,  \nabla_\alpha n_\beta
\ee
where $\nabla_\alpha$ is the torsionless covariant derivative compatible with $g_{\mu\nu}$.

In our analysis $g_{\mu\nu} $ is conformally equivalent to the metric $\tilde{g}_{\mu\nu} $
corresponding to a Euclidean  space $\widetilde{\mathcal{M}}_3$ which is asymptotically flat near the conformal boundary, 
namely 
\be
\label{conformal metric sec2}
g_{\mu\nu}  \,= \,e^{2\varphi} \, \tilde{g}_{\mu\nu} 
\ee
where $\varphi$ is a function of the coordinates. 
The two dimensional surface $\gamma$ is also a submanifold of $\widetilde{\mathcal{M}}_3$.
Denoting by $\tilde{n}_{\mu}$ the spacelike unit vector normal to $\gamma \subset \widetilde{\mathcal{M}}_3$,
we have that $n_{\mu} = e^{\varphi} \, \tilde{n}_{\mu}$.
The fundamental forms in (\ref{fund forms def}) can be written in terms of 
the fundamental forms  $\tilde{h}_{\mu\nu}$ and $\widetilde{K}_{\mu\nu}$ characterising 
the embedding $\gamma \subset \widetilde{\mathcal{M}}_3$ as follows
\be
\label{forms weyl laws}
h_{\mu\nu} \,= \,e^{2\varphi} \,\tilde{h}_{\mu\nu}
\;\; \qquad \;\;
K_{\mu\nu} = e^{\varphi}  \big(  \widetilde{K}_{\mu\nu} + \tilde{h}_{\mu\nu} \,\tilde{n}^\lambda \partial_\lambda \varphi  \big)
\ee
The area $\mathcal{A}[\gamma]$ of the surface $\gamma$ can be written as \cite{Fonda:2015nma}
\begin{eqnarray}
\label{area willmore generic}
\mathcal{A}[\gamma]
&=&
\oint_{\partial \gamma} 
\tilde{b}^\mu  \partial_\mu \varphi\, d\tilde{s} \,
 +
 \frac{1}{4}  \int_{\gamma} 
  \big(\textrm{Tr} K\big)^2 \,d\mathcal{A}
\\
\rule{0pt}{.7cm}
& &
- \, \int_{\gamma} 
\left(
\, \frac{1}{4} \big(\textrm{Tr} \widetilde{K}\big)^2
+\widetilde{\nabla}^2\varphi 
- e^{2\varphi}
- \tilde{n}^\mu \tilde{n}^\nu \, \widetilde{\nabla}_\mu \widetilde{\nabla}_\nu \varphi
+ \big( \tilde{n}^\lambda \partial_\lambda \varphi \big)^2
\right)
 d\tilde{\mathcal{A}}
 \nonumber
\end{eqnarray}
where $\widetilde{\nabla}_\alpha$ is the torsionless covariant derivative compatible with $\tilde{g}_{\mu\nu}$ 
and $\tilde{b}^\mu $ is the unit vector on $\partial \gamma$ that is tangent to $\gamma$, 
orthogonal to $\partial \gamma$ and outward pointing with respect to $\gamma$.
The area element $d\mathcal{A}  = \sqrt{h} \, d\Sigma$  of $\gamma \subset \mathcal{M}_3$ and
the area element $d\tilde{\mathcal{A}}  = \sqrt{\tilde{h}} \, d\Sigma$ of 
$\gamma \subset \widetilde{\mathcal{M}}_3$ are related as $d\mathcal{A} = e^{2\varphi} d\tilde{\mathcal{A}}$, 
being $d\Sigma = d\sigma_1 d\sigma_2$, where $\sigma_i$ are some local coordinates on $\gamma$.

If part of $\gamma$ belongs to the conformal boundary at $z=0$,
the area \eqref{area willmore generic} in infinite because of the behaviour of the metric $h_{\mu\nu}$ near the conformal boundary.
In order to regularise the area, one introduces the UV cutoff $\varepsilon$ and considers the part of $\gamma$ given by
$\gamma_\varepsilon \equiv \gamma \cap \{z \geqslant \varepsilon \}$.
The curve $\partial \gamma_\varepsilon$ can be decomposed as 
$\partial \gamma_\varepsilon = \partial \gamma_{\mathcal{Q}} \cup \partial \gamma_\parallel$, 
where $\partial \gamma_{\mathcal{Q}}  \equiv \gamma_\varepsilon \cap \mathcal{Q}$ 
and $\partial \gamma_\parallel \equiv \gamma_\varepsilon \cap \{z=\varepsilon\}$
are not necessarily closed lines. 
Consequently, for the surfaces $\gamma_\varepsilon $ the boundary term in (\ref{area willmore generic}) can be written as 
\be
\label{boundary_contr}
\oint_{\partial \gamma_\varepsilon} 
\tilde{b}^\mu  \partial_\mu \varphi\, d\tilde{s} 
\;= 
\int_{\partial \gamma_\parallel } 
\tilde{b}^\mu  \partial_\mu \varphi\, d\tilde{s} \;
+
\int_{\partial \gamma_{\mathcal{Q}} } 
\tilde{b}^\mu  \partial_\mu \varphi\, d\tilde{s} 
\ee

Let us consider the integral over $\partial \gamma_\parallel $ in the r.h.s. of this expression.
Since in our analysis $\varphi = -\log(z) + O(z^a)$ with $a>1$ as $z \to 0$,
we need to know the behaviour of the component $\tilde{b}^z$ at $z=\varepsilon$ 
as $\varepsilon \to 0$.
If $\tilde{b}^z = -\,1+ o(\varepsilon)$, for the integral over $\partial \gamma_\parallel $ in (\ref{boundary_contr}) we obtain the following expansion 
\begin{equation}
\label{integ line eps}
\int_{\partial \hat{\gamma}_\parallel} 
\tilde{b}^\mu  \partial_\mu \varphi\, d\tilde{s}  
\,=\,
 \frac{P_{A,B}}{ \varepsilon} + o(1)
\end{equation}
as $\varepsilon \to 0$, being  $P_{A,B} = \textrm{length}(\partial A \cap \partial B)$ the length of the entangling curve. 
The above expansion  for $\tilde{b}^z$ holds for any surface, not necessarily minimal,  which  intersects the conformal boundary orthogonally  \cite{AM}.
Hereafter we will consider only this class of surfaces, which includes also 
the extremal surfaces, which are compelled to intersect orthogonally the conformal boundary \cite{Babich:1992mc, AM, Graham:1999pm, Fonda:2015nma}.

By plugging (\ref{integ line eps}) into (\ref{boundary_contr}) first and then substituting the resulting expression into (\ref{area willmore generic}), 
for the area of the surfaces $\gamma_\varepsilon $ we find the following expansion
\begin{eqnarray}
\label{area willmore generic v1}
\mathcal{A}[\gamma_\varepsilon ]
&\,=\,&
 \frac{P_{A,B}}{ \varepsilon} 
 +
 \int_{\partial \gamma_{\mathcal{Q}} } 
\tilde{b}^\mu  \partial_\mu \varphi\, d\tilde{s} \,
 +
 \frac{1}{4}  \int_{\gamma_\varepsilon} 
  \big(\textrm{Tr} K\big)^2 \,d\mathcal{A}
\\
\rule{0pt}{.7cm}
& &
- \, \int_{\gamma_\varepsilon} 
\left(
\, \frac{1}{4} \big(\textrm{Tr} \widetilde{K}\big)^2
+\widetilde{\nabla}^2\varphi 
- e^{2\varphi}
- \tilde{n}^\mu \tilde{n}^\nu \, \widetilde{\nabla}_\mu \widetilde{\nabla}_\nu \varphi
+ \big( \tilde{n}^\lambda \partial_\lambda \varphi \big)^2
\right)
 d\tilde{\mathcal{A}}\,
 + o(1)
 \nonumber
\end{eqnarray}
as $\varepsilon \to 0$.
We remark that (\ref{area willmore generic v1})
also holds for surfaces $\gamma_\varepsilon$ that are not extremal of the area functional.
Furthermore, no restrictions are imposed along the curve $\partial \gamma_\mathcal{Q}$.

In order to compute the holographic entanglement entropy in AdS$_4$/BCFT$_3$ through (\ref{RT formula bdy intro}),
we must consider  the minimal area surface $\hat{\gamma}_A$ which is anchored to the entangling curve $\partial A \cap \partial B$.
This implies that $\hat{\gamma}_A$  intersects the surface $\mathcal{Q}$ orthogonally, whenever $\hat{\gamma}_A \cap \mathcal{Q} \neq \emptyset$.
The expression (\ref{area willmore generic v1}) significantly simplifies for the extremal surfaces $\hat{\gamma}_\varepsilon \equiv \hat{\gamma}_A \cap \{z \geqslant \varepsilon \}$
(with a slight abuse of notation, sometimes we denote by $\hat{\gamma}_A$ also extremal surfaces which are not the global minimum).
The local extrema of the area functional are the solutions of the following equation
\be
\label{minimal area condition}
\textrm{Tr} K  = 0
\hspace{.5cm} \Longleftrightarrow \hspace{.5cm}
\big( \textrm{Tr}\widetilde{K} \big)^2 = 4 (\tilde{n}^\lambda \partial_\lambda \varphi )^2
\ee
which, furthermore, intersect orthogonally $\mathcal{Q}$ whenever $\hat{\gamma}_A \cap \mathcal{Q} \neq \emptyset$.
The second expression in (\ref{minimal area condition}) has been obtained by using the second formula in (\ref{forms weyl laws}).

Plugging the extremality condition (\ref{minimal area condition}) into (\ref{area willmore generic v1}), 
we find the expansion of $\mathcal{A}[\hat{\gamma}_\varepsilon]$ as $\varepsilon \to 0$, 
which provides the holographic entanglement entropy of a region $A$ in AdS$_4$/BCFT$_3$
for static gravitational backgrounds. 
It reads
\be
\label{area willmore bdy minimal generic}
\mathcal{A}[\hat{\gamma}_\varepsilon]
\,=\,
 \frac{P_{A,B}}{ \varepsilon} 
 +
\int_{\partial \hat{\gamma}_{\mathcal{Q}}} 
\tilde{b}^\mu  \partial_\mu \varphi\, d\tilde{s} \,
-
\int_{\hat{\gamma}_\varepsilon} 
\left(
\, \frac{1}{2} \big(\textrm{Tr} \widetilde{K}\big)^2
+\widetilde{\nabla}^2\varphi 
- e^{2\varphi}
- \tilde{n}^\mu \tilde{n}^\nu \, \widetilde{\nabla}_\mu \widetilde{\nabla}_\nu \varphi
\right)
 d\tilde{\mathcal{A}}\,
   + o(1)
\ee
where the leading divergence gives the expected area law term for the holographic entanglement entropy in AdS$_4$/BCFT$_3$.
Comparing (\ref{area willmore bdy minimal generic}) with the expansion (\ref{hee bdy intro}) expected for $\mathcal{A}[\hat{\gamma}_\varepsilon]$,
we find that the subleading term is given by 
\be
\label{F_A generic}
F_A 
=
\int_{\hat{\gamma}_\varepsilon} 
\left(
\, \frac{1}{2} \big(\textrm{Tr} \widetilde{K}\big)^2
+\widetilde{\nabla}^2\varphi 
- e^{2\varphi}
- \tilde{n}^\mu \tilde{n}^\nu \, \widetilde{\nabla}_\mu \widetilde{\nabla}_\nu \varphi
\right)
 d\tilde{\mathcal{A}}
 -
 \int_{\partial \hat{\gamma}_{\mathcal{Q}}} 
\tilde{b}^\mu  \partial_\mu \varphi \, d\tilde{s} 
\ee

This is the main result of this manuscript.
According to (\ref{F_A generic}), the subleading term 
is made by two contributions:
an integral over the whole minimal surface $\hat{\gamma}_\varepsilon$ 
and a line integral over the curve $\partial \hat{\gamma}_{\mathcal{Q}}=\hat{\gamma}_\varepsilon \cap \mathcal{Q}$.
We remark that the definition of $\mathcal{Q}$ has not been employed in the derivation of  (\ref{F_A generic}).

The integrand of the surface term in  \eqref{F_A generic} is the same obtained in \cite{Fonda:2015nma},
where this analysis has been applied for the holographic entanglement entropy in AdS$_4$/CFT$_3$.
The holographic entanglement entropy in AdS$_4$/BCFT$_3$  includes the additional term 
given by the line integral over $\partial \hat{\gamma}_\mathcal{Q}$.
This term can be written in a more geometrical form by considering the transformation rule of the
geodesic curvature $k$ under Weyl transformations  (see e.g. \cite{AM})
\be
k\,=\,e^{-\varphi}\big(\tilde k +\tilde b^\mu\partial_\mu\varphi\big)
\ee 
This formula allows to write the line integral over $\partial \gamma_{\mathcal{Q}} $ in (\ref{F_A generic}) as follows
\be
\label{geodesic curvature rel}
\int_{\partial \gamma_{\mathcal{Q}} } 
\tilde{b}^\mu  \partial_\mu \varphi\, d\tilde{s} \,
\,=\,
 \int_{\partial \gamma_{\mathcal{Q}} }  k\, d s
 -
 \int_{\partial \gamma_{\mathcal{Q}} }  \tilde{k}\, d\tilde{s}  
\ee

In this manuscript, we consider backgrounds such that $\varphi = -\log(z)$ in (\ref{conformal metric sec2}). 
In these cases, the first and the last term of the integrand in the surface integral in (\ref{F_A generic}) become respectively
\be
\label{explicit exp}
\big(\textrm{Tr} \widetilde{K}\big)^2  = \frac{4(\tilde{n}^{z})^2}{z^2} 
\qquad
\tilde{n}^\mu \tilde{n}^\nu \, \widetilde{\nabla}_\mu \widetilde{\nabla}_\nu \varphi
=
\frac{(\tilde{n}^{z})^2}{z^2}+\frac{1}{z} \, \widetilde{\Gamma}^z_{\mu\nu} \, \tilde{n}^\mu \tilde{n}^\nu 
\ee
where the first expression has been obtained from the second expression in (\ref{minimal area condition}) 
and $\widetilde{\Gamma}^z_{\mu\nu}$ are some components of the Christoffel 
connection compatible with  $\tilde{g}_{\mu\nu}$.

\subsection{AdS$_4$}
\label{sec ads4}

In the remaining part of the manuscript, we focus on the simple gravitational background given by a part of AdS$_4$ 
delimited by $\mathcal{Q}$ and the conformal boundary, which provides the gravitational background dual to the ground state of the BCFT$_3$.
The metric of AdS$_4$ in Poincar\'e coordinates reads
\be
\label{ads4 metric}
ds^2 
=
\frac{1}{z^2} \,\Big( -dt^2+ dz^2 + dx^2 + dy^2  \,\Big)
\ee
where $z>0$, while the range of the remaining coordinates is $\mathbb{R}$.
The metric induced on a $t=\textrm{const}$ slice of AdS$_4$ is the one characterising the three dimensional 
Euclidean hyperbolic space $\mathbb{H}_3$
\be
\label{H3 metric}
ds^2 
=
\frac{1}{z^2} \,\Big( dz^2 + dx^2 + dy^2  \,\Big)
\ee

Specialising the results of Sec.\,\ref{sec general bg} to this background, 
we have $\mathcal{M}_3 = \mathbb{H}_3$, i.e. $g_{\mu\nu} = \tfrac{1}{z^2}\, \delta_{\mu\nu}$,
which means that $\tilde{g}_{\mu\nu} = \delta_{\mu\nu} $ and $\varphi = -\log(z)$.
In this case, drastic simplifications occur (\ref{F_A generic}) because
$\widetilde{\nabla}^2\varphi - e^{2\varphi} = 0$ and all the components of the connection $\widetilde{\Gamma}^z_{\mu\nu}$ vanish identically. 
Thus, when the gravitational bulk is a proper subset of $\mathbb{H}_3$ delimited by the surface $\mathcal{Q}$ and the conformal boundary,
the expression (\ref{F_A generic}) for $F_A$ reduces to
\be
\label{F_A willmore ads}
F_A 
\,=\,
\frac{1}{4}
\int_{\hat{\gamma}_\varepsilon} 
\big(\textrm{Tr} \widetilde{K}\big)^2
\,d\tilde{\mathcal{A}}\,
+
\int_{\partial \hat{\gamma}_{\mathcal{Q}}} 
\frac{\tilde{b}^z}{z} \; d\tilde{s} 
\; =
\int_{\hat{\gamma}_\varepsilon} 
\frac{(\tilde{n}^{z})^2}{z^2} 
\; d\tilde{\mathcal{A}}\,
+
\int_{\partial \hat{\gamma}_{\mathcal{Q}}} 
\frac{\tilde{b}^z}{z} \, d\tilde{s} 
\ee
The surface integral over $\hat{\gamma}_\varepsilon$ in the first expression is the Willmore functional of $\hat{\gamma}_\varepsilon \subset \mathbb{R}^3$.
Notice that the curves $\partial \hat{\gamma}_{\mathcal{Q}}$ corresponding to some configurations may intersect the plane given by $z=\varepsilon$.

When $A$ contains corners, the expression (\ref{F_A willmore ads}) diverges logarithmically as $\varepsilon \to 0$.
In AdS$_4$/CFT$_3$, the emergence of the logarithmic divergence from the Willmore functional for domains with corners 
has been studied in  \cite{Fonda:2015nma}, where the corner function found \cite{Drukker:1999zq} has been recovered. 
In this AdS$_4$/BCFT$_3$ setup, the occurrence of a logarithmic divergence in (\ref{F_A willmore ads}) for singular domains will be discussed in Sec.\,\ref{sec corners}
and the corner function found in \cite{Seminara:2017hhh} will be obtained. 

When the entangling curve is a smooth and closed line that does not intersect the spatial boundary of the BCFT$_3$, 
the limit $\varepsilon \to 0$ of (\ref{F_A willmore ads}) provides the following finite expression
\be
\label{F_A willmore ads-finite}
F_A 
\,=\,
\frac{1}{4}
\int_{\hat{\gamma}_A} 
\big(\textrm{Tr} \widetilde{K}\big)^2
\,d\tilde{\mathcal{A}}\,
+
\int_{\partial \hat{\gamma}_{\mathcal{Q}}} 
\frac{\tilde{b}^z}{z} \, d\tilde{s} 
\; =
\int_{\hat{\gamma}_A} 
\frac{(\tilde{n}^{z})^2}{z^2} 
\; d\tilde{\mathcal{A}}\,
+
\int_{\partial \hat{\gamma}_{\mathcal{Q}}} 
\frac{\tilde{b}^z}{z} \; d\tilde{s} 
\ee
which will be largely employed throughout this manuscript.

Hereafter we will focus on BCFT$_3$'s whose spatial slice is either the half plane bounded by a straight line or the disk.
In Sec.\;\ref{sec flat bdy} and Sec.\;\ref{sec circular bdy} some details about these two setups are discussed.


\subsubsection{Flat boundary}
\label{sec flat bdy}

\begin{figure}[t] 
\vspace{-.8cm}
\hspace{.0cm}
\includegraphics[width=1.04\textwidth]{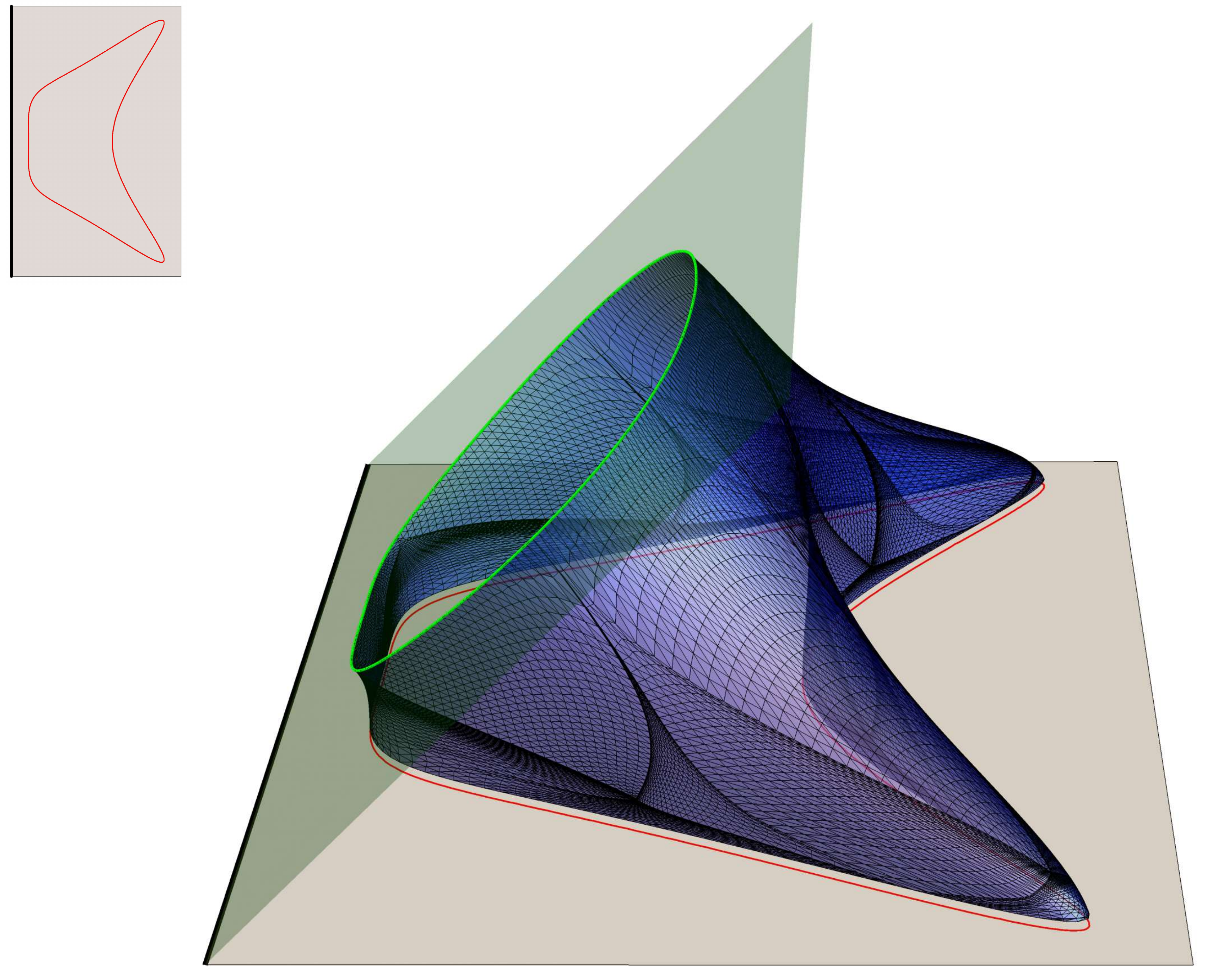}
\vspace{-.2cm}
\caption{\label{fig_advertising}
\small
Extremal surface $\hat{\gamma}_\varepsilon$ constructed with Surface Evolver from a spatial domain $A$ in the right half plane
(the grey half plane) whose $\partial A$ is the red curve, which is also highlighted in the inset. 
The gravitational bulk is the part of $\mathbb{H}_3$ defined by (\ref{AdS4 regions}), 
whose boundary is made by the conformal boundary at $z=0$ (the grey half plane) and $\mathcal{Q}$ 
(the green half plane defined in (\ref{brane-profile})).
Here $\alpha = 3\pi/4$.
The green curve corresponds to $\partial \hat{\gamma}_{\mathcal{Q}} = \hat{\gamma}_\varepsilon \cap \mathcal{Q}$,
and $\hat{\gamma}_\varepsilon $ intersects $\mathcal{Q}$ orthogonally along this curve. 
}
\end{figure}

Let us consider a BCFT$_3$ defined in a spacetime whose generic spatial slice  is the half plane $\{(x,y) \in \mathbb{R}^2 ,x\geqslant 0 \}$
bounded by the straight line $x=0$
(see the grey horizontal half plane and the straight solid black line in Fig.\;\ref{fig_advertising}).
When the term (\ref{gravi-action-Tconstant})  occurs in the gravitational action (\ref{gravi-action-general}),
it has been found in \cite{Takayanagi:2011zk, Fujita:2011fp} that the spatial section of the  gravitational background is given by $\mathbb{H}_3$, whose metric is (\ref{H3 metric}), 
bounded by the following half plane $\mathcal{Q}$ in the bulk 
\be
\label{brane-profile}
\mathcal{Q} \, :  \;\;\; 
x  =  - \,(\cot \alpha ) \,z
\qquad
\alpha \in (0,\pi)
\qquad
z\geqslant 0
\ee
(the green half plane in Fig.\;\ref{fig_advertising})
whose boundary coincides with the straight line $x=0$ bounding the spatial slice of the BCFT$_3$.
The angular parameter $\alpha$ provides the slope of the half plane $\mathcal{Q}$ and it is related to the constant $T$ in (\ref{gravi-action-Tconstant})
as 
$T= (2/ L_{\textrm{\tiny AdS}} ) \cos\alpha$.
In particular, a $t=\textrm{const}$ slice of the gravitational bulk is the part of $\mathbb{H}_3$ defined by 
\be
\label{AdS4 regions}
x\geqslant - \,(\cot \alpha ) \,z
\ee

The term $F_A$ in the holographic entanglement entropy can be easily obtained
by specialising (\ref{F_A willmore ads}) to this AdS$_4$/BCFT$_3$ setup.
We remark that, for this case, the line integral over $\partial \hat{\gamma}_{\mathcal{Q}}$ in (\ref{F_A willmore ads}) simplifies 
because $\tilde{b}^z = - \cos \alpha$ for all the points of $\partial \hat{\gamma}_{\mathcal{Q}}$.
Furthermore, $\tilde k=0$ in (\ref{geodesic curvature rel}) in this setup,
i.e. $\partial \hat{\gamma}_{\mathcal{Q}}$ is a geodesic of $\hat{\gamma}_\varepsilon \in \mathbb{R}^3$.
Thus, for any region $A$ in the half plane $x \geqslant 0$, we find
\be
\label{F_A willmore ads Q-plane}
F_A \,=\,
\int_{\hat{\gamma}_\varepsilon} 
\frac{(\tilde{n}^{z})^2}{z^2} 
\, d\tilde{\mathcal{A}}\,
- 
( \cos \alpha )
\int_{\partial \hat{\gamma}_{\mathcal{Q}}} 
\frac{1}{z} \, d\tilde{s} 
\ee
The two integrals in this expression are always positive, but their relative sign depends on the slope $\alpha$.
In particular, when $\alpha \geqslant \pi/2$ we have $F_A >0$,
while $F_A$ can be negative when $\alpha < \pi/2$
(see e.g. the expression (\ref{area half-disk final main}) for the half disk adjacent to the flat boundary considered in Sec.\;\ref{sec app half disk}).

When $\partial A$ is a closed and smooth curve that does not intersect the boundary $x=0$, 
the limit $\varepsilon \to 0$ of (\ref{F_A willmore ads Q-plane}) is finite and one finds 
\be
\label{F_A willmore ads Q-plane finite}
F_A \,=\,
\int_{\hat{\gamma}_A} 
\frac{(\tilde{n}^{z})^2}{z^2} 
\, d\tilde{\mathcal{A}}\,
- 
( \cos \alpha )
\int_{\partial \hat{\gamma}_{\mathcal{Q}}} 
\frac{1}{z} \, d\tilde{s} 
\ee
which  corresponds to (\ref{F_A willmore ads-finite}) specialised to this setup.

In Fig.\;\ref{fig_advertising} we show an explicit example where (\ref{F_A willmore ads Q-plane finite}) can be applied.
The entangling curve $\partial A$ is the red curve in the $z=0$ half plane also highlighted  in the inset. 
Surface Evolver has been employed to construct $\hat{\gamma}_\varepsilon$, as done in \cite{Seminara:2017hhh} for other regions in 
this AdS$_4$/BCFT$_3$ setup.


\subsubsection{Circular boundary}
\label{sec circular bdy}

The second setup is given by a BCFT$_3$ defined on a spacetime whose $t=\textrm{const}$ slice is a disk of radius $R_{\mathcal{Q}}$,
that  can be conveniently described by introducing the polar coordinates $(\rho,\phi)$ with the origin in the center of the disk, 
namely such that $0\leqslant \rho \leqslant R_{\mathcal{Q}}$ and $0 \leqslant  \phi < 2\pi$.
This disk can be mapped into the half plane $\{(x,y) \in \mathbb{R}^2 ,x\geqslant 0 \}$ considered in Sec.\;\ref{sec flat bdy},
as discussed in Appendix\;\ref{app:mapping}.
In terms of the polar coordinates in the conformal boundary, the metric of $\mathbb{H}_3$ reads
$ds^2 = (dz^2 + d\rho^2 + \rho^2 d\phi^2) /z^2$, being $z>0$ the holographic coordinate.

For a BCFT$_3$ defined in the above disk of radius $R_{\mathcal{Q}}$,
the gravitational background dual to the ground state is a region of $\mathbb{H}_3$ delimited by a surface $\mathcal{Q}$ invariant under rotations 
about the $z$-axis, whose boundary is the circle $\mathcal{C}_\mathcal{Q}$  given by $(\rho,z)=(R_{\mathcal{Q}} ,0)$.
When the term (\ref{gravi-action-Tconstant})  occurs in the gravitational action (\ref{gravi-action-general}),
the profile of $\mathcal{Q}$ can be found as the image of the half plane (\ref{brane-profile})
through the conformal map (\ref{mapping}) described in Appendix\;\ref{app:mapping}.
The result reads \cite{Takayanagi:2011zk, Fujita:2011fp} 
\be
\label{Qbrane disk}
\rho=\sqrt{(R_\mathcal{Q} \, \csc\alpha)^2 - (z-R_\mathcal{Q} \cot\alpha )^2}
\ee
(see also (\ref{brane_trasf})), 
which corresponds to a spherical cap  $\mathcal{Q}$ centered in $(\rho, z) = (0, R_\mathcal{Q} \cot\alpha)$
with radius $R_{\mathcal{Q}} / \sin \alpha$
(see the green surface in the left panel of Fig.\;\ref{fig_ampolla}).
When $\alpha =\pi/2$, this spherical cap  becomes the hemisphere $\rho^2 + z^2 = R_\mathcal{Q}^2$.
By introducing the angular coordinate $\theta$ as $ \tan \theta = z/\rho$, 
from (\ref{Qbrane disk}) we find that the coordinates of a point of $\mathcal{Q}$ are $(\rho, z)= R_\mathcal{Q}  \big( Q_\alpha(\theta), Q_\alpha(\theta)\tan\theta\big)$
with
\be
\label{Qbrane circ bdy}
Q_\alpha(\theta)
\,\equiv\,
\cos\theta 
\left(
\cot \alpha \, \sin \theta +
\sqrt{1+ (\cot \alpha \, \sin \theta)^2}
\;\right)
=
\frac{\sqrt{\zeta^2 +(\sin \alpha)^2}
+ \zeta \cos \alpha }{\left(\zeta^2+1\right) \sin\alpha}
\ee
where in the last step we have introduced $\zeta \equiv \tan \theta$, that will be employed also in Sec.\,\ref{sec disk disjoint circ}.

In this AdS$_4$/BCFT$_3$ setup, $F_A$ is given by (\ref{F_A willmore ads}) (or (\ref{F_A willmore ads-finite}) whenever it can be applied).
We remark that typically $\tilde{b}^z$ is not constant along $\partial \hat{\gamma}_{\mathcal{Q}}$. 
Instead, this simplification occurs when $A$ is a disk sharing the origin with $\mathcal{C}_\mathcal{Q}$ (see the left panel of Fig.\;\ref{fig_ampolla}).


\section{Infinite strip adjacent to the boundary}
\label{sec strip adjacent}

In this section we focus on the holographic entanglement entropy of infinite strips parallel to the flat boundary,
in the AdS$_4$/BCFT$_3$ setup described in Sec.\;\ref{sec flat bdy}.
We show that the formula (\ref{F_A willmore ads Q-plane finite}) 
reproduces the result for $F_A$ computed in 
\cite{Nagasaki:2011ue,Miao:2017gyt,Seminara:2017hhh} by means of a straightforward computation of the area
for the corresponding minimal surfaces.

An infinite strip $A$ of width $\ell$ adjacent to the boundary can be studied by taking the rectangular domain
$(x,y) \in \mathbb{R}^2$ such that $0\leqslant x \leqslant \ell$ and $-L_\parallel / 2\leqslant y \leqslant L_\parallel / 2$
in the regime of $L_\parallel  \gg \ell \gg \varepsilon$.
In this limit, the invariance under translations in the $y$ direction can be assumed.
The corresponding minimal surfaces $\hat{\gamma}_A$ have been studied in \cite{Seminara:2017hhh} in the whole regime of $\alpha \in (0,\pi)$,
by employing the partial results previously obtained in \cite{Nagasaki:2011ue,Miao:2017gyt}.

The minimal surface $\hat{\gamma}_A$ intersects the $z=0$ half plane orthogonally along the line $x=\ell$, which is a component of $\partial \hat{\gamma}_A$. 
In this case $P_{A,B} = L_\parallel$ in (\ref{area willmore bdy minimal generic}),
therefore the leading linear divergence (area law term) in the expansion of $\mathcal{A}[\hat{\gamma}_\varepsilon]$ as $\varepsilon \to 0$ is
$L_\parallel / \varepsilon$.
We are mainly interested in the subleading term $F_A$, which depends on the entire surface.
Because of the invariance under translations in the $y$ direction, $\hat{\gamma}_A$ is characterised by its section at $y = \textrm{const}$.

\begin{figure}[t] 
\vspace{-.8cm}
\hspace{-.4cm}
\includegraphics[width=1.04\textwidth]{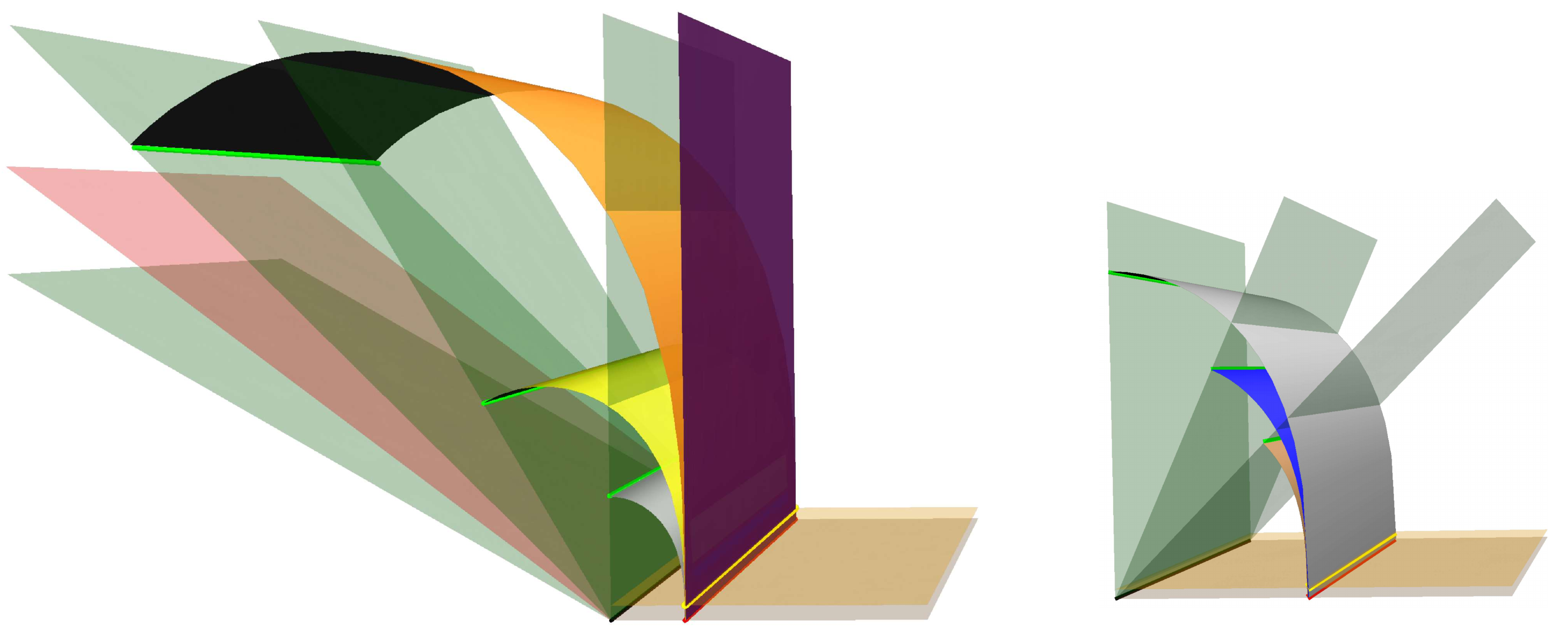}
\vspace{-.4cm}
\caption{\label{fig_strips}
\small
Minimal area surfaces  corresponding to the same infinite strip $A$ in the $z=0$ half plane (the grey half plane)
which is adjacent to the boundary $x=0$ (straight solid black line).
The entangling curve is the straight solid red line.
The yellow horizontal plane is given by $z=\varepsilon$.
The green half planes correspond to $\mathcal{Q}$ in (\ref{brane-profile}) for different values of $\alpha$
and the red half plane is $\mathcal{Q}$  with  $\alpha = \alpha_c$.
For $\alpha > \alpha_c$ we show $\hat{\gamma}_{A}^{\textrm{\tiny \,con}}$ 
and $\partial \hat{\gamma}_{\mathcal{Q}} =\hat{\gamma}_{A}^{\textrm{\tiny \,con}}\cap \mathcal{Q}$
are highlighted (green straight lines).
The vertical purple half plane corresponds to $\hat{\gamma}_{A}^{\textrm{\tiny \,dis}}$.
In the left panel $0< \alpha \leqslant \pi/2$ and in the right panel $\pi/2\leqslant  \alpha <\pi$.
}
\end{figure}

When $\alpha \leqslant \pi/2$, two surfaces $\hat{\gamma}_{A}^{\textrm{\tiny \,dis}}$ and $\hat{\gamma}_{A}^{\textrm{\tiny \,con}}$ 
extremise the area functional (see the left panel in Fig.\;\ref{fig_strips}); 
therefore their areas must be compared  to find the global minimum \cite{Miao:2017gyt, Seminara:2017hhh}.
The surface $\hat{\gamma}_{A}^{\textrm{\tiny \,dis}}$ is the half plane $x=\ell$ (the purple half plane in Fig.\;\ref{fig_strips}), 
which remains orthogonal to the $z=0$ plane and does not intersect $\mathcal{Q}$ at a finite value of $z$.
Instead, the surface $\hat{\gamma}_{A}^{\textrm{\tiny \,con}}$ intersects  the half plane $\mathcal{Q}$ orthogonally at a finite value $z_\ast$ of the coordinate $z$.
When $\alpha > \pi/2$ (see the right panel in Fig.\;\ref{fig_strips}), the solution $\hat{\gamma}_{A}^{\textrm{\tiny \,dis}}$ does not exist;
hence the global minimum is given by $\hat{\gamma}_{A}^{\textrm{\tiny \,con}}$.

The extremal surface $\hat{\gamma}_{A}^{\textrm{\tiny \,con}}$ for a given $\alpha \in (0,\pi)$ is characterised by the following profile \cite{Seminara:2017hhh}
\be
\label{strip profile final}
P_\theta 
=
\big( x(\theta)\,, z(\theta)   \big)  
\,=\,
\frac{\ell}{\mathfrak{g}(\alpha)} 
\left(
   \mathbb{E}\big( \pi/4-\alpha/2 \, |\, 2 \big) - \frac{\cos\alpha}{\sqrt{\sin \alpha}} 
   +
  \mathbb{E}\big( \pi/4-\theta/2 \, |\, 2 \big) 
  \, , \, \sqrt{\sin \theta} \;
  \right)
\ee
where $\theta \in [0, \pi-\alpha]$ is the angular parameter such that $\theta=0$ corresponds to $z=0$
and 
\be
\label{g function def main}
\mathfrak{g}(\alpha) 
\,\equiv\,
\mathbb{E}\big( \pi/4-\alpha/2 \, |\, 2 \big) - \frac{\cos\alpha}{\sqrt{\sin \alpha}} 
+ 
\frac{\Gamma\big(\tfrac{3}{4}\big)^2}{\sqrt{2\pi}}
\ee 
being $\mathbb{E}(x|y)$ the incomplete elliptic integral of the second kind (we adopt the convention of Mathematica for the elliptic function throughout this manuscript).  
From (\ref{strip profile final}) we can easily obtain $z_\ast = z(\pi-\alpha)$ given by
\be
\label{z_ast strip main}
z_\ast = \frac{\sqrt{\sin \alpha}}{\mathfrak{g}(\alpha)}\; \ell
\ee
which characterises the position of the straight green lines corresponding to $\hat{\gamma}_{A}^{\textrm{\tiny \,con}} \cap \mathcal{Q}$ in Fig.\;\ref{fig_strips}.
Since we must have $z_\ast >0$, from (\ref{z_ast strip main}) one observes  that $\hat{\gamma}_{A}^{\textrm{\tiny \,con}}$ is well defined when $\mathfrak{g}(\alpha) > 0$.
It is straightforward to notice that $\mathfrak{g}(\alpha)$ has only one zero for $\alpha \in (0,\pi)$ given by $\alpha_c \simeq \pi /4.85$.
Thus, when $\alpha \leqslant \alpha_c $ the solution $\hat{\gamma}_{A}^{\textrm{\tiny \,con}}$ 
does not exist and the global minimum is $\hat{\gamma}_{A}^{\textrm{\tiny \,dis}}$ (the purple half plane in Fig.\;\ref{fig_strips}), 
as discussed in \cite{Miao:2017gyt, Seminara:2017hhh}.

The $O(1)$ term in the expansion of $\mathcal{A}[\hat{\gamma}_\varepsilon]$ as $\varepsilon \to 0$ for $\alpha \in (0,\pi)$ reads \cite{Seminara:2017hhh}
\be
\label{F_A strip adj}
F_A
\,=\,
L_\parallel \, \frac{a_0(\alpha)}{ \ell } 
  \hspace{.2cm}  \qquad \hspace{.2cm} 
a_0(\alpha) =
\left\{\begin{array}{ll}
-\,\mathfrak{g}(\alpha)^2 \hspace{.7cm} &   \alpha \geqslant \alpha_c
\\
\rule{0pt}{.5cm}
\;\;\;\; 0 &   \alpha  \leqslant \alpha_c
\end{array}\right.
\ee

The main observation of this section is that the non trivial expression for $F_A$ corresponding to the regime $\alpha \geqslant \alpha_c$ in (\ref{F_A strip adj}) 
can be recovered by evaluating \eqref{F_A willmore ads Q-plane finite} for $\hat{\gamma}_{A}^{\textrm{\tiny \,con}}$
as surface embedded in $\mathbb{R}^3$. 
The surface $\hat{\gamma}_{A}^{\textrm{\tiny \,con}}$ is described by the constraint $\mathcal{C} =0$, being $\mathcal{C}\equiv z-z(x)$,
and its unit normal vector $\tilde n_\mu = (\tilde{n}_z, \tilde{n}_x, \tilde{n}_y) $ can be found by 
first computing $\partial_\mu \mathcal{C}$ and then normalising the resulting vector. 
We find $\tilde n_\mu  = (1,-z',0)/\sqrt{1+(z')^2}$.
The area element in the surface integral occurring in (\ref{F_A willmore ads Q-plane finite})
reads $d\tilde{\mathcal{A}} = \sqrt{1+(z')^2}\,dx\,dy$ in this case. 
Combining these observations, we get
\be
\label{strip integral surf}
\int_{\hat{\gamma}_A}  
\frac{(\tilde{n}^{z})^2}{z^2} 
\, d\tilde{\mathcal{A}}
\,=\,
\int_{\hat{\gamma}_A}   
\frac{dx\,dy}{z^2\,\sqrt{1+(z')^2}}
\ee
where we have not used yet the fact that $z(x)$ corresponds to $\hat{\gamma}_A$.
Specifying (\ref{strip integral surf}) to the profile \eqref{strip profile final},
we find $\sqrt{1+(z')^2}=1/\sin\theta$ and $dx=\ell\,\sqrt{\sin\theta} \,d\theta / (2\mathfrak{g}(\alpha))$.
By employing these observations, (\ref{strip integral surf}) becomes
\be
\label{strip surf term}
\int_{\hat{\gamma}_A}  
\frac{(\tilde{n}^{z})^2}{z^2} 
\, d\tilde{\mathcal{A}}
\,=\,
L_\parallel \,\frac{2\mathfrak{g}(\alpha)}{\ell}\int_{0}^{\pi-\alpha} \sqrt{\sin \theta}\,d\theta 
\,=\,
L_\parallel\,
\frac{\mathfrak{g}(\alpha)}{\ell} 
\left(
\mathbb{E}\big( \pi/4-\alpha/2 \, |\, 2 \big)
+
\frac{\Gamma( \frac{3}{4})^2}{\sqrt{2 \pi}} 
\,\right)
\ee

The integral over the line $\partial \hat{\gamma}_{\mathcal{Q}}$ in (\ref{F_A willmore ads Q-plane finite})
significantly simplifies for these domains because $\partial \hat{\gamma}_{\mathcal{Q}}$ is the straight line 
given by $(z,x,y) = (z_\ast , x_\ast, y)$ with $-L_\parallel / 2\leqslant y \leqslant L_\parallel / 2$, 
where $(x_\ast , z_\ast) = P_{\pi-\alpha}$ can be read from (\ref{strip profile final})
and it corresponds to the green straight lines in Fig.\;\ref{fig_strips}.
Thus, the line integral in (\ref{F_A willmore ads Q-plane finite}) gives
\be
\label{strip Qbdy term}
\int_{\partial \hat{\gamma}_{\mathcal{Q}}} 
\frac{1}{z} \, d\tilde{s} 
\,=\,
\frac{L_\parallel}{z_\ast}
\,=\,
\frac{\mathfrak{g}(\alpha)}{\ell\,\sqrt{\sin \alpha}}\, L_\parallel\,
\ee
where (\ref{z_ast strip main}) has been used in the last step.

Plugging (\ref{strip surf term}) and (\ref{strip Qbdy term}) into the general expression (\ref{F_A willmore ads Q-plane finite}),
for an infinite strip of width $\ell$ adjacent to the boundary we find
\be
\label{strip FA sum}
F_A \big|_{\hat{\gamma}_{A}^{\textrm{\tiny \,con}}} 
=\,
L_\parallel\,
\frac{\mathfrak{g}(\alpha)}{\ell}
\left[
\left(
\mathbb{E}\big( \pi/4-\alpha/2 \, |\, 2 \big)
+
\frac{\Gamma( \frac{3}{4})^2}{\sqrt{2 \pi}} 
\,\right)
-\frac{\cos\alpha}{\sqrt{\sin \alpha}}
\,\right]
\,=\,
L_\parallel\,
\frac{\mathfrak{g}(\alpha)^2}{\ell}
\ee
where  the last result has been obtained by employing (\ref{g function def main}).
Notice that both the terms in (\ref{F_A willmore ads Q-plane finite}) 
provide non trivial contributions.

From the results discussed in this section, it is straightforward to find $F_A$ when $A$ is an infinite strip parallel to the flat boundary 
and at a finite distance from it through the formula \eqref{F_A willmore ads Q-plane finite}, 
recovering the result presented in Sec.\;5.3 of \cite{Seminara:2017hhh}.
In the analysis of this configuration, we find it instructive to employ the extremal surfaces anchored to two infinite parallel strips in the plane \cite{Tonni:2010pv}
as discussed in Appendix\;\ref{app aux_domains}.


\section{Disk disjoint from the boundary}
\label{sec disk}

In this section we study the holographic entanglement entropy of a disk $A$ at a finite distance from the boundary. 

\begin{figure}[t] 
\vspace{-.8cm}
\hspace{-.8cm}
\includegraphics[width=1.1\textwidth]{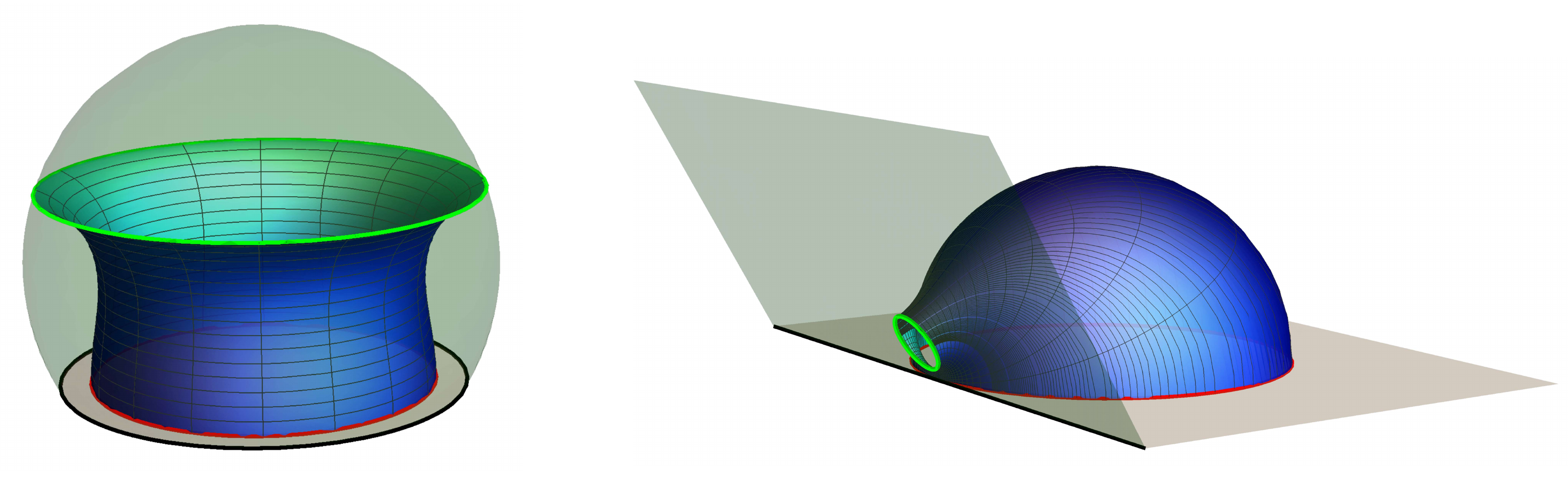}
\vspace{-.8cm}
\caption{\label{fig_ampolla}
\small
Left: 
Extremal area surface $\hat \gamma_A^{\,\text{\tiny con}}$ anchored to a disk $A$ disjoint from a circular concentric boundary
(see Sec.\;\ref{sec circular bdy} and Sec.\;\ref{sec disk disjoint circ})
where $\mathcal{Q}$ (green spherical dome) is described by \eqref{Qbrane disk}. 
Here $\alpha=\pi/3$ and $R_\circ/R_\mathcal{Q}\sim 0.85$, 
which corresponds to $r_{\circ,\textrm{\tiny \,min}}$ (see Sec.\;\ref{sec disk profiles}).
Right: 
Extremal surface $\hat \gamma_A^{\,\text{\tiny con}}$ anchored to a disk disjoint from a flat boundary
(see Sec.\;\ref{sec flat bdy} and Sec.\;\ref{subsec:disk disjoint}). 
Here $\alpha=\pi/3$ and $d/R$ can be obtained from the first expression in (\ref{delta})
with the value of $R_\circ/R_{\mathcal{Q}}$ of the left panel
because the two configurations shown in these panels are related through (\ref{mapping}).
}
\end{figure}

In the setup described in Sec.\,\ref{sec circular bdy}, in Sec.\,\ref{sec disk disjoint circ} we consider
the case of a disk $A$ concentric to the circular boundary 
because the symmetry of this configuration allows us to obtain an analytic expression for  the profile characterising the minimal surface $\hat{\gamma}_A$
(in the left panel of Fig.\;\ref{fig_ampolla} we show an example of $\hat{\gamma}_A$).
The corresponding area $\mathcal{A}[\hat{\gamma}_\varepsilon]$ is computed in two ways:
by the direct evaluation of the integral  and by specifying the general formula (\ref{F_A willmore ads Q-plane finite}) to this case. 
In Sec.\,\ref{subsec:disk disjoint}, by employing the second transformation in (\ref{mapping}) and the analytic 
results presented in Sec.\,\ref{sec disk disjoint circ}, we study the holographic entanglement entropy of
a disk disjoint from the flat boundary in the setup introduced in Sec.\,\ref{sec flat bdy} 
(see the right panel of Fig.\;\ref{fig_ampolla} for an example of $\hat{\gamma}_A$ in this setup).
The two wo configurations in Fig.\;\ref{fig_ampolla} have the same $\alpha$ and are related through the map (\ref{mapping}) 
discussed in Appendix\;\ref{app:mapping}.

\subsection{Disk disjoint from a circular concentric boundary}
\label{sec disk disjoint circ}

In the AdS$_4$/BCFT$_3$ setup introduced in Sec.\,\ref{sec circular bdy},
let us consider a disk $A$ with radius $R_\circ < R_\mathcal{Q}$
which is concentric to the boundary of the spatial slice of the spacetime. 
In Sec.\;\ref{sec disk profiles} we obtain an analytic expression for the profile characterising $\hat{\gamma}_A$
and in Sec.\;\ref{sec area disk concentric} we evaluate the corresponding area $\mathcal{A}[\hat{\gamma}_\varepsilon]$. 
In the following we report only the main results of this analysis.
Their detailed derivation, which is closely related to the evaluation of the holographic entanglement entropy of an annulus in AdS$_4$/CFT$_3$ 
\cite{Fonda:2014cca, Drukker:2005cu} 
has been presented in Appendix\;\ref{app:disk}.

\subsubsection{Profile of the extremal surfaces}
\label{sec disk profiles}

\begin{figure}[t] 
\vspace{-.8cm}
\hspace{-.6cm}
\includegraphics[width=1.05\textwidth]{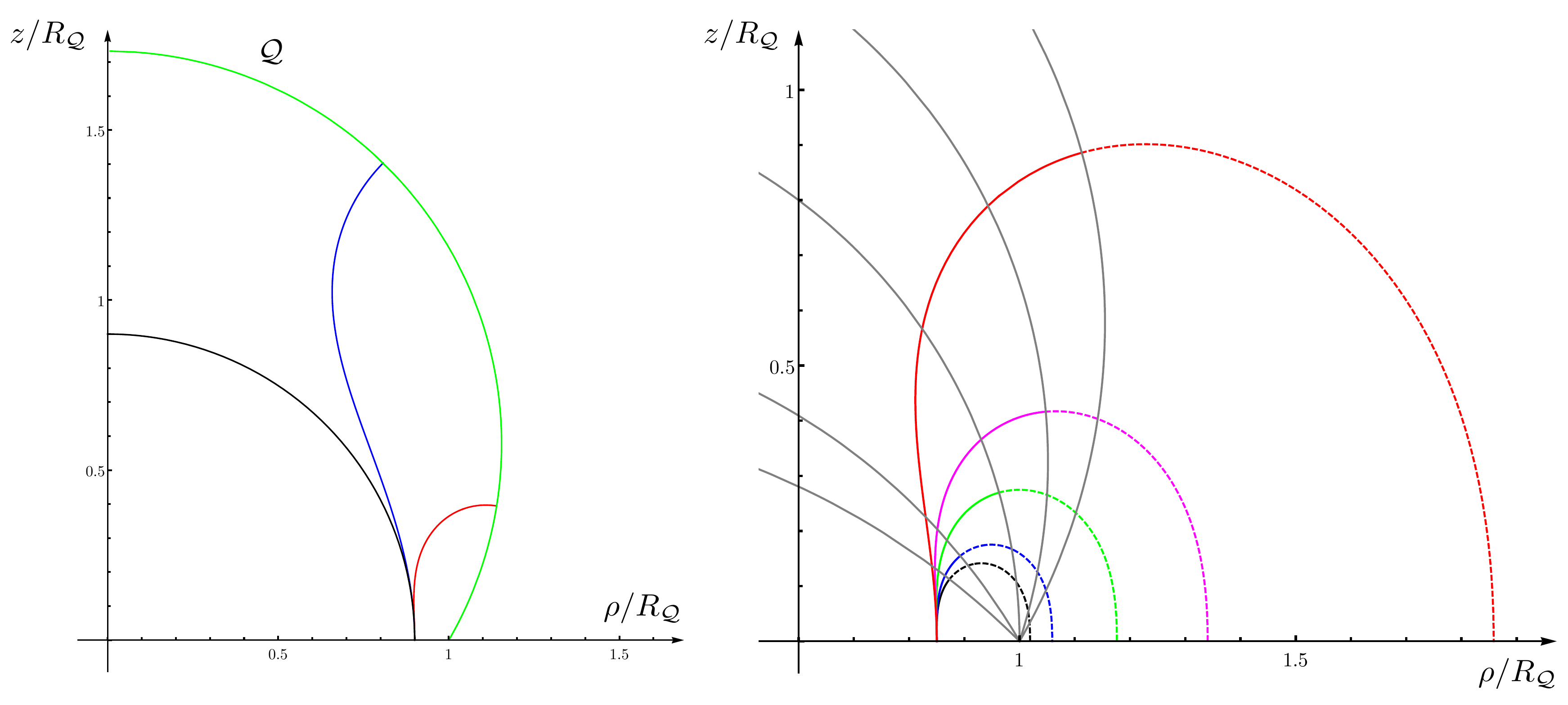}
\vspace{-.8cm}
\caption{\label{fig_3profiles}
\small
Sections of the extremal surfaces anchored to a disk $A$ of radius $R_\circ$ disjoint from a circular concentric boundary with radius $R_{\mathcal{Q}}$ 
(see Sec.\;\ref{sec disk profiles}).
Left: Profiles corresponding to the three extremal surfaces in the case of ${R_\circ/R_{\mathcal{Q}}=0.9}$ and $\alpha=\pi/3$.
The green curve represents $\mathcal{Q}$.
The black curve corresponds to $\hat{\gamma}_{A}^{\textrm{\tiny \,dis}}$ (the hemisphere).
The red curve and the blue curve correspond to $\hat{\gamma}_{A}^{\textrm{\tiny \,con}}$
and they have been obtained through the analytic results discussed in Sec.\;\ref{sec disk profiles} and in Appendix\;\ref{app:disk}.
The red curve provides the global minimum in this case. 
Right: Extremal surfaces $\hat{\gamma}_{A}^{\textrm{\tiny \,con}}$ having $R_\circ/R_{\mathcal{Q}}\simeq 0.85$ 
for  different values of $\alpha$: 
$\alpha=\pi/3$ (red), $\alpha=\pi/2.5$ (magenta), $\alpha=\pi/2$ (green), $\alpha=2\pi/3$ (blue) and $\alpha=3\pi/4$ (black). 
The dashed curves are the profiles of the auxiliary surfaces $\hat{\gamma}_{A, \textrm{\tiny \,aux}}^{\textrm{\tiny \,con}}$, with the same color code. 
All the profiles correspond to the smaller value of $k$ whenever two surfaces $\hat{\gamma}_{A}^{\textrm{\tiny \,con}}$  exists.
All the curves except for the red one provide the global minimum of the corresponding configuration. 
}
\end{figure}

Adopting the coordinate system $(\rho, \phi, z)$ introduced in Sec.\,\ref{sec circular bdy},
the invariance under rotations around the $z$-axis for this configuration in the $z=0$ plane 
implies that the local extrema of the area functional are described by the profiles of their sections at $\phi =\textrm{const}$.

For a given $A$, an extremal surface is the hemisphere anchored to the circle $\partial A$.
Since it does not intersect $\mathcal{Q}$, this solution will be denoted by  $\hat{\gamma}_{A}^{\textrm{\tiny \,dis}}$,
while we will refer to the extremal surfaces that intersect $\mathcal{Q}$ orthogonally as $\hat{\gamma}_{A}^{\textrm{\tiny \,con}}$.
The holographic entanglement entropy of $A$ is provided by the surface corresponding to the global minimum of the area.
Let us anticipate that we find at most two solutions $\hat{\gamma}_{A}^{\textrm{\tiny \,con}}$;
hence we have at most three local extrema for a given disk $A$.
The number of solutions depends on the value of $\alpha$, as we will discuss in the following. 
By employing the analytic result that will be presented below,
in the left panel of Fig.\;\ref{fig_3profiles} 
we show the three profiles corresponding to  $\hat{\gamma}_{A}^{\textrm{\tiny \,dis}}$ (black curve) and $\hat{\gamma}_{A}^{\textrm{\tiny \,con}}$ (blue and red curve)
in an explicit case.
The red curve provides the holographic entanglement entropy in this example. 

We find it worth introducing an auxiliary surface that allows to relate our problem to the one
of finding the extremal surfaces in $\mathbb{H}_3$ anchored to an annulus, which has been already addressed in the literature.
Given $\hat{\gamma}_{A}^{\textrm{\tiny \,con}}$, let us consider its unique surface $\hat{\gamma}_{A, \textrm{\tiny \,aux}}^{\textrm{\tiny \,con}}$ in the whole $\mathbb{H}_3$
such that $\hat{\gamma}_{A}^{\textrm{\tiny \,con}} \cup \,\hat{\gamma}_{A, \textrm{\tiny \,aux}}^{\textrm{\tiny \,con}}$ is an extremal area surface in $\mathbb{H}_3$
anchored to the annulus whose boundary is made by the two concentric circles with radii $R_\circ$ and $R_{\textrm{\tiny aux}} > R_\circ$.
Thus, $\hat{\gamma}_{A}^{\textrm{\tiny \,con}}$ can be viewed as part of an extremal surface anchored to a proper annulus
whose boundary are the union of two circles, one of which is $\partial A$.
By using the solution that will be discussed in the following, 
in the right panel of Fig.\;\ref{fig_3profiles} we fix $A$ and we show the profiles associated to $\hat{\gamma}_{A}^{\textrm{\tiny \,con}}$ (solid curves) for various $\alpha$ 
and the ones for the corresponding extensions $\hat{\gamma}_{A, \textrm{\tiny \,aux}}^{\textrm{\tiny \,con}}$ (dashed curves).
Other examples are shown in Fig.\;\ref{fig_profiles_evolver}.

The profile of a section of $\hat{\gamma}_{A}^{\textrm{\tiny \,con}}$ at fixed $\phi$ 
can be written as $(\rho, z)= ( \rho_\gamma(\theta), \rho_\gamma(\theta) \tan\theta)$,
where  the angular variable is defined as $\zeta \equiv \tan \theta = z/\rho$ (see Sec.\,\ref{sec circular bdy}).
Considering the construction of the extremal surfaces in $\mathbb{H}_3$ anchored to an annulus reported in \cite{Fonda:2014cca},
we have that the curve $\rho_\gamma(\theta) $ can be written by introducing two branches as follows
\be
\label{annulus branches profile}
\rho_\gamma(\theta) 
\,=\,
\Bigg\{\begin{array}{l}
R_\circ \, e^{-q_{-, k}(\zeta)}
\\
\rule{0pt}{.5cm}
R_{\textrm{\tiny aux}} \, e^{-q_{+, k}(\zeta)}
\end{array}
\ee
with $R_{\textrm{\tiny aux}} > R_{\circ}$.  
The functions $q_{\pm, k}(\zeta)$ are defined as
\be
\label{q function pm}
q_{\pm, k}(\zeta)
\equiv
\int_0^{\zeta} 
\frac{\lambda}{1+\lambda^2} \,
\bigg(1\pm \frac{\lambda}{\sqrt{k \,(1+\lambda^2)-\lambda^4}}\bigg)
\, d\lambda
\;\;\qquad\;\;
0 \leqslant \zeta \leqslant \zeta_m
\ee
being $k>0$ and $\zeta_m^2 \equiv \big(k+\sqrt{k(k+4)}\,\big) / 2$
the unique admissible root of the biquadratic equation
coming from the expression under the square root in (\ref{q function pm}).
Since $q_{\pm, k}(0) = 0$, the two branches in (\ref{annulus branches profile}) 
give $\rho_\gamma = R_\circ$ and $\rho_\gamma = R_{\textrm{\tiny aux}} $ when $z=0$.

The two branches characterised by $q_{\pm, k}(\zeta)$ in (\ref{annulus branches profile})
match at the point $P_m=(\rho_m, \zeta_m)$ associated to the maximum value of $\theta$.
The coordinates of $P_m$ read (see also Appendix\;\ref{app:disk})
\be
\label{Pm coords}
\zeta_m^2
=\frac{k+\sqrt{k (k+4)}}{2}
\;\;\qquad\;\;
\rho_m
=
R_\circ \, e^{-q_{-, k}(\zeta_m)}
= 
R_{\textrm{\tiny aux}} \, e^{-q_{+, k}(\zeta_m)}
\ee
The last equality in the second expression follows from the continuity of the profile \eqref{annulus branches profile} and it gives 
\be
\label{R_ratio_chi}
\frac{R_\circ}{R_{\textrm{\tiny aux}}} 
=\, 
e^{q_{-, k}(\zeta_m)- q_{+, k}(\zeta_m)}
\ee
which will be denoted by $\chi(\zeta_m)$ in the following.
Being $\zeta_m$ given by the first expression in (\ref{Pm coords}),
from (\ref{R_ratio_chi}) we observe that the ratio $R_\circ / R_{\textrm{\tiny aux}}$ is a function of the parameter $k>0$.
Moreover, by employing (\ref{q function pm}) in (\ref{R_ratio_chi}), it is straightforward to observe that
$R_\circ / R_{\textrm{\tiny aux}}<1$.

The integral in (\ref{q function pm}) can be computed analytically, finding that $q_{\pm,k} (\zeta) $
can be written in terms of the incomplete elliptic integrals of the first and third kind as follows
\be
\label{q_+-_explicit}
q_{\pm,k} (\zeta) =
\frac{1}{2} \log (1+\zeta^2)
\pm
\kappa \, 
\sqrt{\frac{1-2\kappa^2}{\kappa^2-1}}\,
\Big[\,
\Pi \big(1-\kappa^2, \Omega(\zeta) | \kappa^2 \big)
-
\mathbb{F}\big(\Omega(\zeta) | \kappa^2\big)
\Big]
\ee
 where  
\be
\Omega(\zeta)
\,\equiv\, 
\arcsin \bigg(
\frac{\zeta/\zeta_m}{\sqrt{1+\kappa^2({\zeta^2/\zeta^2_m-1)}}}
\bigg)
\qquad 
\kappa \equiv
\sqrt{\frac{1+\zeta_m^2}{2+\zeta_m^2}}
\ee

Let us remark that the above expressions depend on the positive parameters $R_\circ $ and $k$.
The dependence on the parameters $R_{\mathcal{Q}}$ and $\alpha$ characterising the boundary occurs through 
the requirement that  $\hat{\gamma}_{A}^{\textrm{\tiny \,con}} \perp \mathcal{Q}$.

Denoting by  $P_\ast = (\rho_\ast, z_\ast)$ the point in the radial profile
corresponding to the intersection between $\hat{\gamma}_{A}^{\textrm{\tiny \,con}}$ and $\mathcal{Q}$,
in Appendix\;\ref{app:disk} we have found that
\be
\label{star coords}
\zeta_\ast^2
=\frac{k+\sqrt{k (k+4(\sin \alpha)^2)}}{2}
\;\;\qquad\;\;
\rho_\ast
=
R_\mathcal{Q}\,
\frac{\sqrt{\zeta_\ast^2 +(\sin \alpha)^2}
+ \zeta_\ast \cos \alpha }{\left(\zeta_\ast^2+1\right) \sin\alpha}
\ee
where the first expression has been obtained by imposing that 
$\hat{\gamma}_{A}^{\textrm{\tiny \,con}}$ intersects $\mathcal{Q}$ orthogonally at $P_\ast$,
while the second one comes from (\ref{Qbrane circ bdy}).
In Appendix\;\ref{app profiles} (see below (\ref{orthogonality_branches})) we have also remarked that
the orthogonality condition also implies that $P_\ast$ belongs to the branch described $q_{-,k}$ when $\alpha \geqslant \pi/2$,
while it belongs to the branch characterised by $q_{+,k}$ when $\alpha \leqslant \pi/2$.
This observation and  (\ref{annulus branches profile}) specialised to $P_\ast$  lead to 
\begin{equation}
\label{Rcirc from star}
R_\circ
= 
\rho_\ast \left(
\frac{1+\eta_\alpha}{2} \; e^{q_{-,k}(\zeta_\ast)}+\frac{1-\eta_\alpha}{2}\; \chi(\zeta_m)\,  e^{q_{+,k}(\zeta_\ast)}
\right)
\end{equation}
where $\eta_\alpha \equiv -\,\text{sign}(\cot\alpha)$
and  $ \chi(\zeta_m)$ denotes the ratio in (\ref{R_ratio_chi}).

Notice that $e^{q_{-,k}(\zeta_\ast)} =  \chi(\zeta_m)\,  e^{q_{+,k}(\zeta_\ast)}$ for $\alpha =\pi/2$.
Moreover, if we employ this observation into the second expression of (\ref{Pm coords}), 
we find that  $P_\ast = P_m$ when $\alpha =\pi/2$.

\begin{figure}[t] 
\vspace{-.6cm}
\hspace{.2cm}
\includegraphics[width=.9\textwidth]{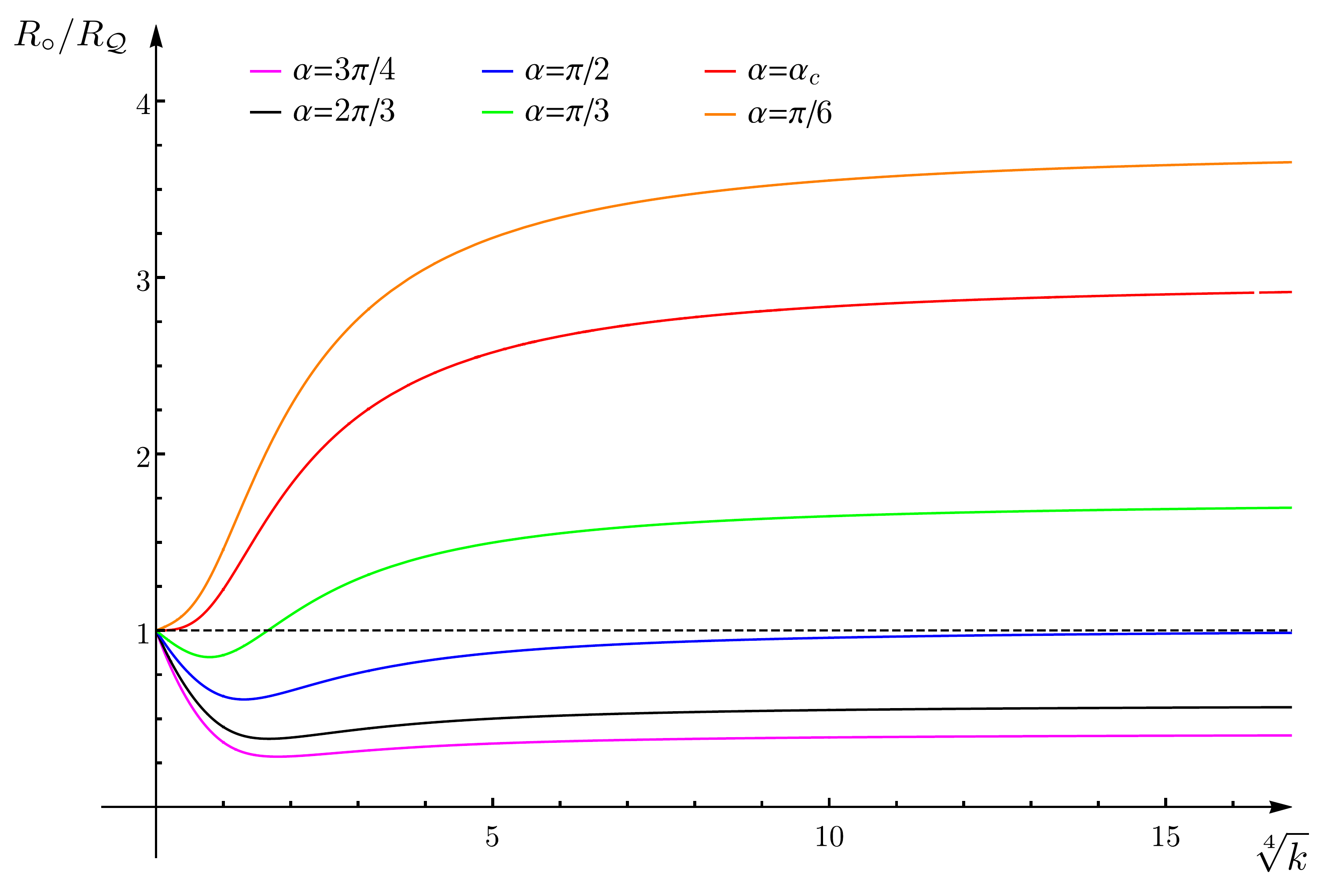}
\vspace{-.0cm}
\caption{\label{fig_RFk}
\small
The ratio $R_\circ/R_{\mathcal{Q}}$ providing $\hat \gamma_A^{\,\text{\tiny con}}$ as a function of $\sqrt[4]{k}$ from \eqref{initial ratio from k} for different values of $\alpha$. 
The allowed configurations have $R_\circ/R_{\mathcal{Q}} < 1$ and the black dashed line corresponds to the limiting value  $R_\circ/R_{\mathcal{Q}}=1$. 
The asymptotic behaviours of these curves for $k\to 0$ and $k \to \infty$ are given by (\ref{R_ratio_exp}) and (\ref{Rratio-k-inf}) respectively. 
For fixed values of $\alpha > \alpha_c$ and $R_\circ/R_{\mathcal{Q}}<1$, 
the number of extremal solutions $\hat \gamma_A^{\,\text{\tiny con}}$ is given by the number of intersections between 
the curve corresponding to $\alpha$ and the horizontal line characterised by the given value of $R_\circ/R_{\mathcal{Q}}$.
}
\end{figure}

\begin{figure}[t] 
\vspace{-.9cm}
\hspace{-.9cm}
\includegraphics[width=1.1\textwidth]{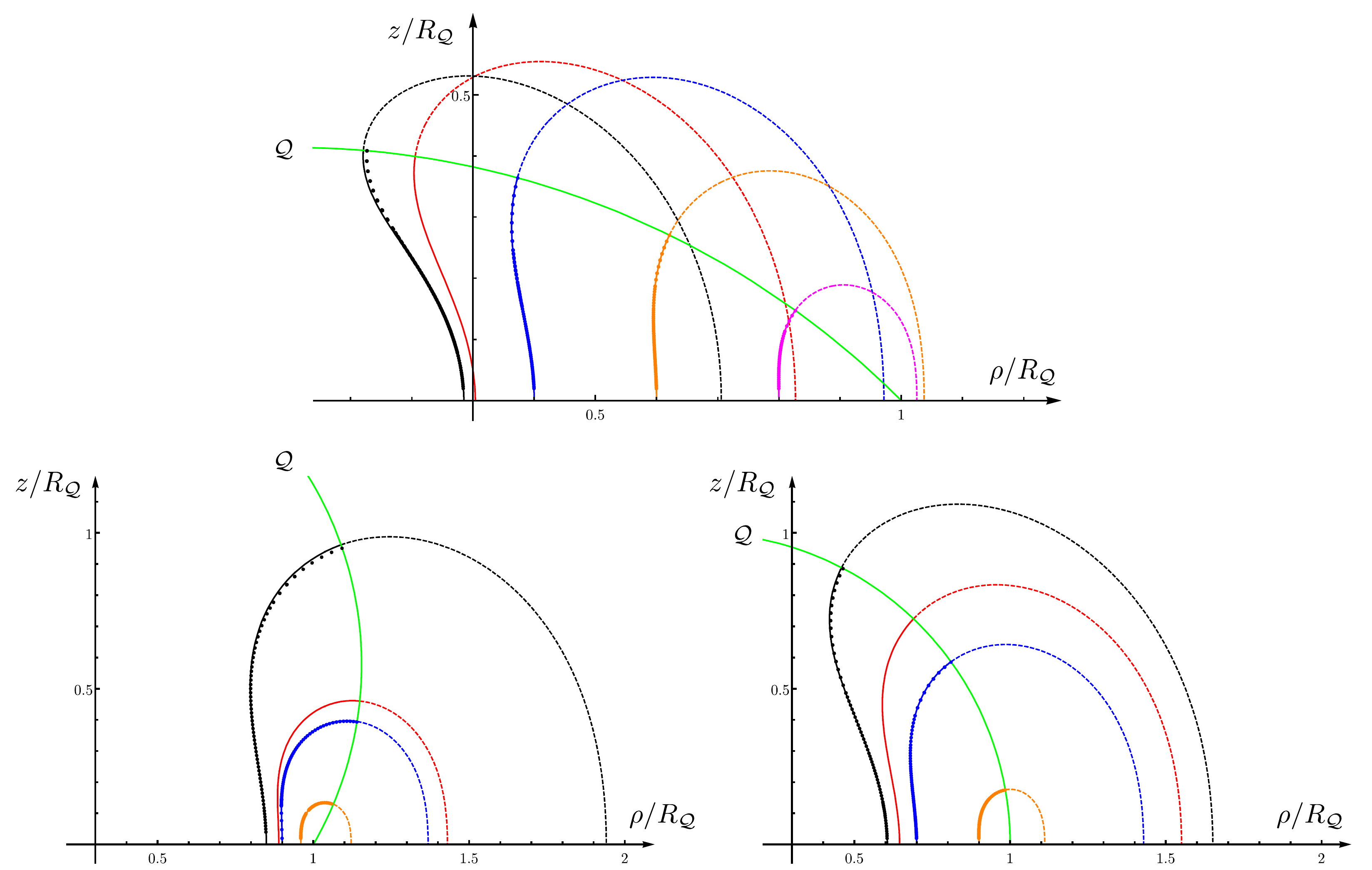}
\vspace{-.4cm}
\caption{\label{fig_profiles_evolver}
\small
Radial profiles of extremal surfaces $\hat{\gamma}_{A}^{\textrm{\tiny \,con}}$ intersecting $\mathcal{Q}$ (green curve) orthogonally
and anchored to a disk $A$ of radius $R_\circ$ concentric to a circular boundary with radius $R_{\mathcal{Q}}$ (see Sec.\;\ref{sec disk profiles}).
The value of $\alpha$ in the three panels is $\alpha = 3\pi/4$ (top), $\alpha = \pi/2$ (bottom, right) and $\alpha = \pi/3$ (bottom, left).
The solid lines give $\hat{\gamma}_{A}^{\textrm{\tiny \,con}}$, while the dashed ones (with the same colour) give the corresponding auxiliary surface $\hat{\gamma}_{A, \textrm{\tiny \,aux}}^{\textrm{\tiny \,con}}$.
The value of $k$ associated to all the shown profiles is the minimum one, whenever two solutions occur (see Fig.\;\ref{fig_RFk}). 
All the profiles except for the black one correspond to the global minimum.
The red curves correspond to the critical value of the ratio $R_\circ / R_{\mathcal{Q}}$ 
where the area of the extremal surface $\hat{\gamma}_{A}^{\textrm{\tiny \,dis}}$ is equal to the minimum of the area of the extremal surfaces $\hat{\gamma}_{A}^{\textrm{\tiny \,con}}$.
The points have been found by taking the $\phi = \textrm{const}$ section of the extremal surfaces constructed by Surface Evolver
and they nicely agree with the corresponding analytic solutions. 
}
\end{figure}

By using the expression of $\rho_\ast$ in (\ref{star coords}) into (\ref{Rcirc from star}),
we get the following relation
\be
\label{initial ratio from k}
\frac{R_\circ}{R_\mathcal{Q}}
= 
\frac{\sqrt{\zeta_\ast^2 +(\sin \alpha)^2}
+ \zeta_\ast \cos \alpha }{\left(\zeta_\ast^2+1\right) \sin\alpha}
\left(
\frac{1+\eta_\alpha}{2} \; e^{q_{-,k}(\zeta_\ast)}+\frac{1-\eta_\alpha}{2}\; \chi(\zeta_m)\,  e^{q_{+,k}(\zeta_\ast)}
\right)
\ee
where $\zeta_\ast$ is the function of $k$ and $\alpha$ given by the first formula in (\ref{star coords}).
The expression (\ref{initial ratio from k}) tells us that $R_\circ / R_\mathcal{Q}$ is a function of $k$ and $\alpha$.
In Fig.\;\ref{fig_RFk} we plot this function by employing $\sqrt[4]{k}$ as the independent variable and $\alpha$ as parameter. 
Since the disk $A$ is a spatial subsystem of the disk with radius $R_\mathcal{Q}$, the admissible configurations have $R_\circ / R_\mathcal{Q} < 1$.

We find it worth discussing the behaviour of the curves $R_\circ / R_\mathcal{Q}$ in (\ref{initial ratio from k}) parameterised by $\alpha$ 
in the limiting regimes given by $k \to 0$ and $k \to \infty$.
The technical details of this analysis have been reported in Appendix\;\ref{sec-app-limits}.

The expansion of \eqref{initial ratio from k} for small $k$ reads
\be
\label{R_ratio_exp}
\frac{R_\circ}{R_\mathcal{Q}}
\,=\,
1-\mathfrak{g}(\alpha) \,\sqrt[4]{k} + \frac{\mathfrak{g}(\alpha)^2}{2}\, \sqrt{k} +o\big(\sqrt{k}\,\big)
\ee
where $\mathfrak{g}(\alpha) $ has been defined in (\ref{g function def main}).
Since $\mathfrak{g}(\alpha)  > 0$ only for $\alpha>\alpha_c$,
being $\alpha_c$ the unique zero of $\mathfrak{g}(\alpha) $ introduced in Sec.\;\ref{sec strip adjacent},
the expansion (\ref{R_ratio_exp}) tells us that,
in the regime of small $k$,
an extremal surface $\hat{\gamma}_{A}^{\textrm{\tiny \,con}}$ can be found  only when $\alpha>\alpha_c$ because $R_\circ / R_\mathcal{Q} < 1$.
From Fig.\;\ref{fig_RFk} we notice that this observation can be extended to the entire regime of $k$.
Indeed, since $R_\circ / R_\mathcal{Q} \geqslant 1$ for the curves with $\alpha \leqslant \alpha_c$,
we have that $\hat{\gamma}_{A}^{\textrm{\tiny \,con}}$ does not exist in this range of  $\alpha$.

In Appendix\;\ref{sec-app-limits} also the limit of \eqref{initial ratio from k} for large $k$ has been discussed, finding that for any
$\alpha \in (0,\pi)$ it reads
\be
\label{Rratio-k-inf}
\lim_{k \,\to\,\infty}\,
 \frac{R_\circ}{R_\mathcal{Q}} 
\,= \,
\cot(\alpha/2)
\ee
which gives the asymptotic value of the curves in Fig.\;\ref{fig_RFk} for large $k$.

When $\alpha > \alpha_c$ the curve $R_\circ / R_\mathcal{Q} $ has only one local minimum (see Fig.\;\ref{fig_RFk}).
Denoting by $k_{\circ,\textrm{\tiny \,min}}$ and $r_{\circ,\textrm{\tiny \,min}}$ the values of $k$ and $R_\circ / R_\mathcal{Q} $ characterising this point,
we have  that $r_{\circ,\textrm{\tiny \,min}} < \cot(\alpha/2)$.
The plot of $r_{\circ,\textrm{\tiny \,min}}$ in terms of $\alpha > \alpha_c$ has been reported in Fig.\;\ref{fig_alpha_c} (black solid curve)
where $\cot(\alpha/2)$ corresponds to the dashed blue curve.

These observations about the limits of $R_\circ / R_\mathcal{Q} $ 
and the numerical analysis of Fig.\;\ref{fig_RFk} allow to discuss the number of extremal surfaces $\hat{\gamma}_{A}^{\textrm{\tiny \,con}}$ 
in the various regimes of the parameters.
When $\alpha \leqslant \alpha_c$ the solutions $\hat{\gamma}_{A}^{\textrm{\tiny \,con}}$  do not exist because $R_\circ / R_\mathcal{Q}  \geqslant 1$.
When $\alpha > \alpha_c$ also the global minimum $r_{\circ,\textrm{\tiny \,min}}$ of $R_\circ / R_\mathcal{Q} $ is an important parameter to consider. 
Indeed, for $\alpha_c <\alpha \leqslant \pi/2$ (see e.g. the green curve in Fig.\;\ref{fig_RFk})
one has two distinct extremal surfaces $\hat{\gamma}_{A}^{\textrm{\tiny \,con}}$
when $ r_{\circ,\textrm{\tiny \,min}} < R_\circ / R_\mathcal{Q} < 1$,
one extremal surface when $R_\circ / R_\mathcal{Q} = r_{\circ,\textrm{\tiny \,min}}$ 
and none of them when $R_\circ / R_\mathcal{Q} < r_{\circ,\textrm{\tiny \,min}}$.
For $\alpha > \pi/2$ also the asymptotic value (\ref{Rratio-k-inf}) plays an important role. 
Indeed, when $\cot(\alpha/2) \leqslant  R_\circ / R_\mathcal{Q} < 1$ we can find only one extremal surface $\hat{\gamma}_{A}^{\textrm{\tiny \,con}}$,
when  $r_{\circ,\textrm{\tiny \,min}} <  R_\circ / R_\mathcal{Q} < \cot(\alpha/2)$ there are two solutions $\hat{\gamma}_{A}^{\textrm{\tiny \,con}}$,
when $r_{\circ,\textrm{\tiny \,min}} =  R_\circ / R_\mathcal{Q} $ we have again only one solution,
while $\hat{\gamma}_{A}^{\textrm{\tiny \,con}}$ do not exist when $ R_\circ / R_\mathcal{Q} < r_{\circ,\textrm{\tiny \,min}} $.
Whenever two distinct solutions $\hat{\gamma}_{A}^{\textrm{\tiny \,con}}$ can be found,
considering their values $k_1 < k_2$ for the parameter $k$, 
we have that $k_1 < k_{\circ,\textrm{\tiny \,min}} < k_2$ because $R_\circ / R_\mathcal{Q} $ has at most one local minimum for $k>0$.

As for the extremal surface $\hat{\gamma}_{A}^{\textrm{\tiny \,dis}}$, which does not intersect $\mathcal{Q}$, 
its existence depends on the value of $\alpha$ because the condition that $\hat{\gamma}_{A}^{\textrm{\tiny \,dis}}$ 
does not intersect $\mathcal{Q}$ provides a non trivial constraint when $\alpha < \pi/2$.
In order to write this constraint, one first evaluates  the $z$ coordinate $z_\mathcal{Q}$ of the tip of $\mathcal{Q}$ by setting $\rho=0$ in (\ref{Qbrane disk}),
finding that $ z_\mathcal{Q}/ R_\mathcal{Q} = \cot(\alpha/2)$.
Then, being $\hat{\gamma}_{A}^{\textrm{\tiny \,dis}}$ a hemisphere, we must impose that $R_\circ \leqslant z_\mathcal{Q}$ and this leads to $R_\circ / R_\mathcal{Q} \leqslant \cot(\alpha/2)$.

Focusing on the regimes where at least one extremal surface $\hat{\gamma}_{A}^{\textrm{\tiny \,con}}$ exists
and employing the above observations,
we can plot the profile given by the section of $\hat{\gamma}_{A}^{\textrm{\tiny \,con}}$ at $\phi = \textrm{const}$ 
by using (\ref{annulus branches profile}) and the related expressions.
In Fig.\;\ref{fig_profiles_evolver} we show some radial profiles of $\hat{\gamma}_{A}^{\textrm{\tiny \,con}}$ (solid lines) 
and of the corresponding auxiliary surfaces $\hat{\gamma}_{A, \textrm{\tiny \,aux}}^{\textrm{\tiny \,con}}$ (dashed lines)
obtained from the analytic expressions discussed above. 
These analytic results have been also checked numerically by employing Surface Evolver as done in  \cite{Fonda:2014cca,Fonda:2015nma,Seminara:2017hhh} for other configurations. 
The data points in Fig.\;\ref{fig_profiles_evolver} correspond to the $\phi = \textrm{const}$ section of the extremal surfaces 
obtained numerically with Surface Evolver.
The nice agreement between the solid curves and the data points provides a highly non trivial check of our analytic results. 
We remark that Surface Evolver constructs also extremal surfaces that are not the global minimum corresponding to a given configuration.

 A detailed discussion about the position of the auxiliary circle with respect to the circular boundary
 has been reported in Appendix\;\ref{app aux_domains}.
 Here let us notice that  in the top panel, where $\alpha = 3\pi/4$, for the black curve and the blue curve we have $R_{\textrm{\tiny aux}} < R_{\mathcal{Q}}$.

In the above analysis we have considered the case of a disk concentric to a circular boundary.
Nonetheless, we can also study the case of a disk whose center does not coincide with the center of the circular boundary
by combining the analytic expressions obtained for this configuration and the mapping discussed in Appendix\;\ref{app:mapping}.

\subsubsection{Area}
\label{sec area disk concentric}

Given a configuration characterised by a disk $A$ of radius $R_\circ < R_\mathcal{Q}$ concentric to the spatial disk of radius $R_\mathcal{Q}$ and the value $\alpha$ for $\mathcal{Q}$,
in Sec.\,\ref{sec disk profiles} we have seen that
we can find at most three local extrema of the area functional among the surfaces anchored to $A$:
the hemisphere $\hat{\gamma}_{A}^{\textrm{\tiny \,dis}}$ and at most two surfaces $\hat{\gamma}_{A}^{\textrm{\tiny \,con}} \perp \mathcal{Q}$.
Since for these three surfaces the expansion of the regularised area is given by the r.h.s. of (\ref{hee bdy intro}) with $P_{A,B} = P_A = 2\pi R_\circ$,
the holographic entanglement entropy of $A$ can be found by comparing their subleading terms $F_A$.
Let us denote by $ F_{\textrm{\tiny con}} $ the subleading term for the surfaces intersecting $\mathcal{Q}$ orthogonally 
discussed in Sec.\,\ref{sec disk profiles}.
Since $F_A = 2\pi$  for the hemisphere \cite{RT, Berenstein:1998ij}, 
the holographic entanglement entropy of $A$ is given by 
\begin{equation}
\label{area_erik}
\mathcal{A}[\hat{\gamma}_\varepsilon]
\,=\, 
 \frac{2\pi R_\circ}{\varepsilon} 
 - \textrm{max}\big(2\pi,  \widehat{F}_{\textrm{\tiny con}}   \big) +\mathcal{O}(\varepsilon)
\end{equation}
where we have denoted by $ \widehat{F}_{\textrm{\tiny con}} $ the maximum between the (at most) two values 
taken by $F_{\textrm{\tiny con}}$ for the values of $k$
corresponding to the local extrema $\hat{\gamma}_{A}^{\textrm{\tiny \,con}}$.

In Appendix\;\ref{app_area}, we have computed $F_{\textrm{\tiny con}}$ by employing two methods:
a straightforward evaluation of the integral coming from the area functional and 
the general expression (\ref{F_A willmore ads-finite}) specialized to the extremal surfaces $\hat{\gamma}_{A}^{\textrm{\tiny \,con}}$ of these configurations. 
Both these approaches lead to the following result
\be
\label{F_A disk circ bdy}
F_{\textrm{\tiny con}} \,=\,
 2\pi \left[ \,
 \frac{1+\eta_\alpha}{2} \;\mathcal{F}_k(\zeta_\ast)
 + \frac{1-\eta_\alpha}{2} \Big( 2\, \mathcal{F}_k(\zeta_m)   -  \mathcal{F}_k(\zeta_\ast) \Big) 
 \, \right]  
\ee
where 
\be
\label{F_A int_evaluated}
	\mathcal{F}_k(\zeta)
	\,\equiv \,
	\frac{\sqrt{k(1+\zeta^2) - \zeta^4}}{\sqrt{k}\,\zeta}
	\,-\,
	\frac{\mathbb{F} ( \arcsin ( \zeta/\zeta_m )\,|-\zeta_m^2-1 )
		-
		\mathbb{E} ( \arcsin ( \zeta/\zeta_m )\,|-\zeta_m^2-1 )
	}{
		\zeta_m}
\ee
and we recall that $\zeta_m$ and $\zeta_\ast$ are the values of $\zeta$ corresponding to the points $P_m$ and $P_\ast$  respectively
(see Sec.\;\ref{sec disk profiles}).
For $\zeta=\zeta_m$, we have
\begin{equation}
\label{Fcal-zetam}
\mathcal{F}_k(\zeta_m)=\frac{\mathbb{E} ( -\zeta_m^2-1 )-\mathbb{K} ( -\zeta_m^2-1 )}{\zeta_m}
\end{equation}
where $\mathbb{K}$ and $\mathbb{E}$ are the complete elliptic integral of the first and second kind respectively.
Since $\zeta_m$ is a function of $k$ (see (\ref{Pm coords})), the r.h.s. of (\ref{Fcal-zetam}) depends only on this parameter. 
Instead, since $\zeta_\ast$ depends on both $k$ and $\alpha$ (see the first expression in (\ref{star coords})),
we have that (\ref{F_A disk circ bdy}) defines a family of functions of $k$ parameterised by $\alpha \in (0,\pi)$.

\begin{figure}[t] 
\vspace{-1.2cm}
\hspace{.85cm}
\includegraphics[width=.9\textwidth]{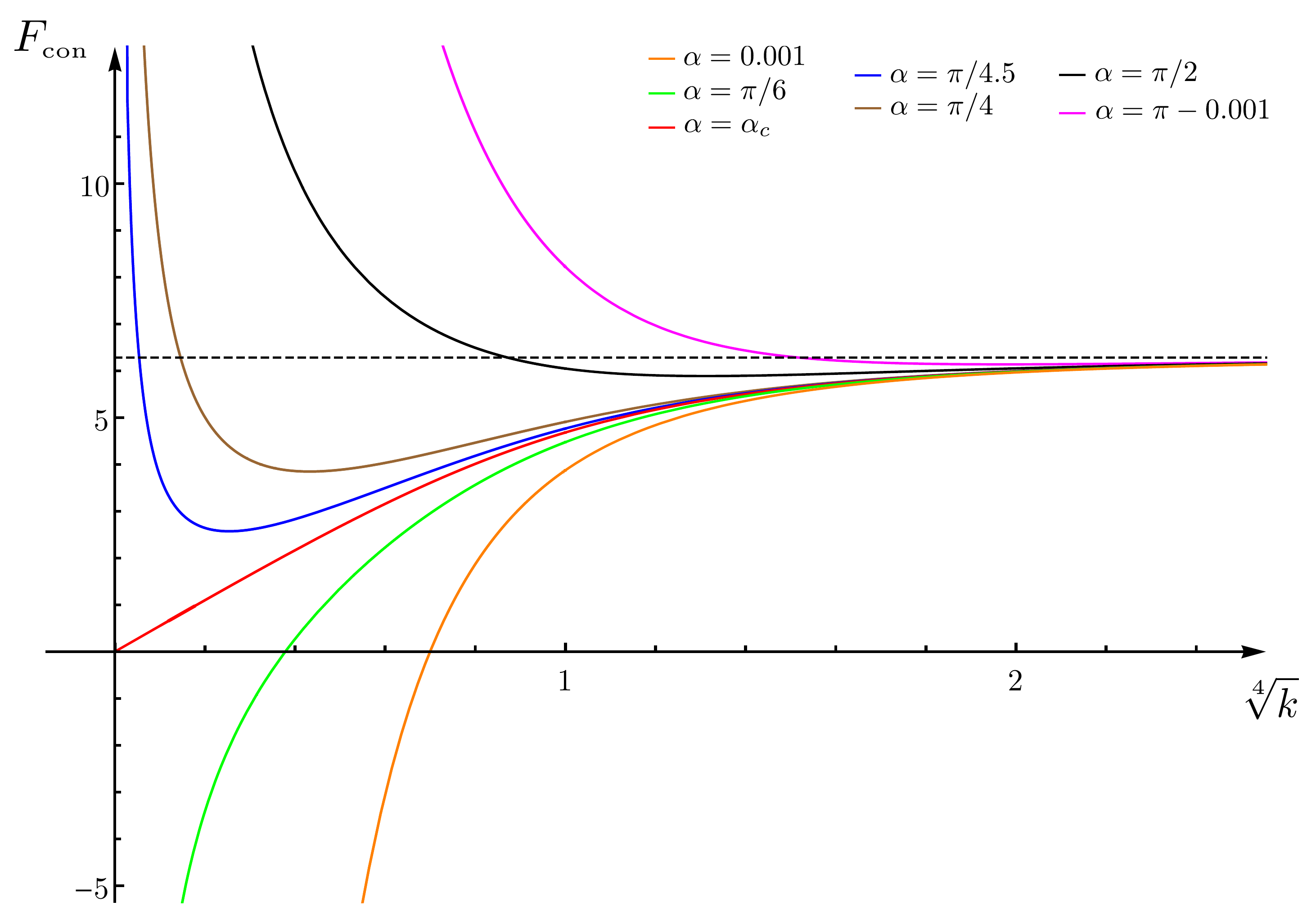}
\vspace{-.1cm}
\caption{\label{fig_Fk}
\small
The subleading term $F_{\text{\tiny{con}}}$ for the extremal surfaces $\hat{\gamma}_{A}^{\textrm{\tiny \,con}}$ 
which intersect $\mathcal{Q}$ orthogonally  as a function of $\sqrt[4]{k}$ (see  (\ref{F_A disk circ bdy})).
The horizontal dashed line corresponds to $2\pi$, i.e. the value of $F_A$ for the hemisphere $\hat{\gamma}_{A}^{\textrm{\tiny \,dis}}$,
and it provides the asymptotic limit at large $k$ for any value of $\alpha$.
The asymptotic behaviour for $k \to 0$ is given by (\ref{FA ann K}).
The curve with $\alpha=\alpha_c$ vanishes as $k \to 0$ and the slope of its tangent at $k=0$ is given by the coefficient 
of the $O(  \sqrt[4]{k}   \,) $ term in (\ref{FA ann K}).
We numerically observe that, for $\alpha \geqslant \alpha_c$, 
the values of $k$ corresponding to the local minima coincide with the values of $k$ of the local minima in Fig.\;\ref{fig_RFk}.
}
\end{figure}

\begin{figure}[t] 
\vspace{-1.2cm}
\hspace{.5cm}
\includegraphics[width=.9\textwidth]{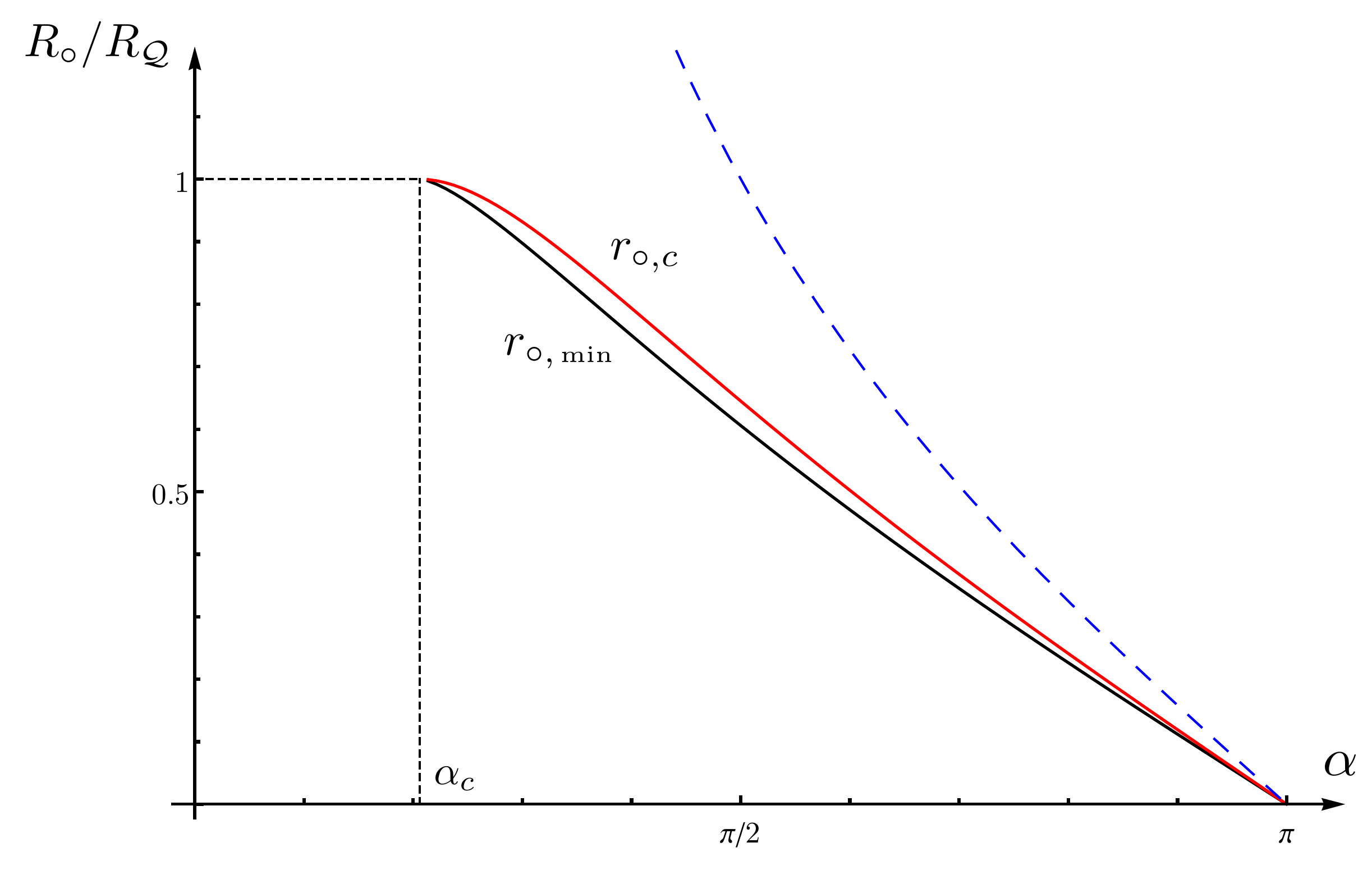}
\vspace{-.2cm}
\caption{\label{fig_alpha_c}
\small
The solid black curve is the minimal value $r_{\circ,\textrm{\tiny \,min}} $ of $R_{\circ}/R_{\mathcal{Q}}$, below which the local solutions 
$\hat{\gamma}_A^{\,\text{\tiny con}}$ intersecting $\mathcal{Q}$ orthogonally do not exist (see also Fig.\;\ref{fig_RFk}), in terms of $\alpha > \alpha_c$.
The solid red curve gives the value $r_{\circ, c} > r_{\circ,\textrm{\tiny \,min}} $ of $R_{\circ}/R_{\mathcal{Q}}$ for $\alpha > \alpha_c$
corresponding to the critical configuration where $\hat{\gamma}_A^{\,\text{\tiny con}}$ and $\hat{\gamma}_A^{\,\text{\tiny dis}}$ 
provide the same finite term $F_A$ of the holographic entanglement entropy. 
The dashed blue curve is the asymptotic value (\ref{Rratio-k-inf}).
}
\end{figure}

\begin{figure}[t] 
\vspace{-1.2cm}
\hspace{.6cm}
\includegraphics[width=.9\textwidth]{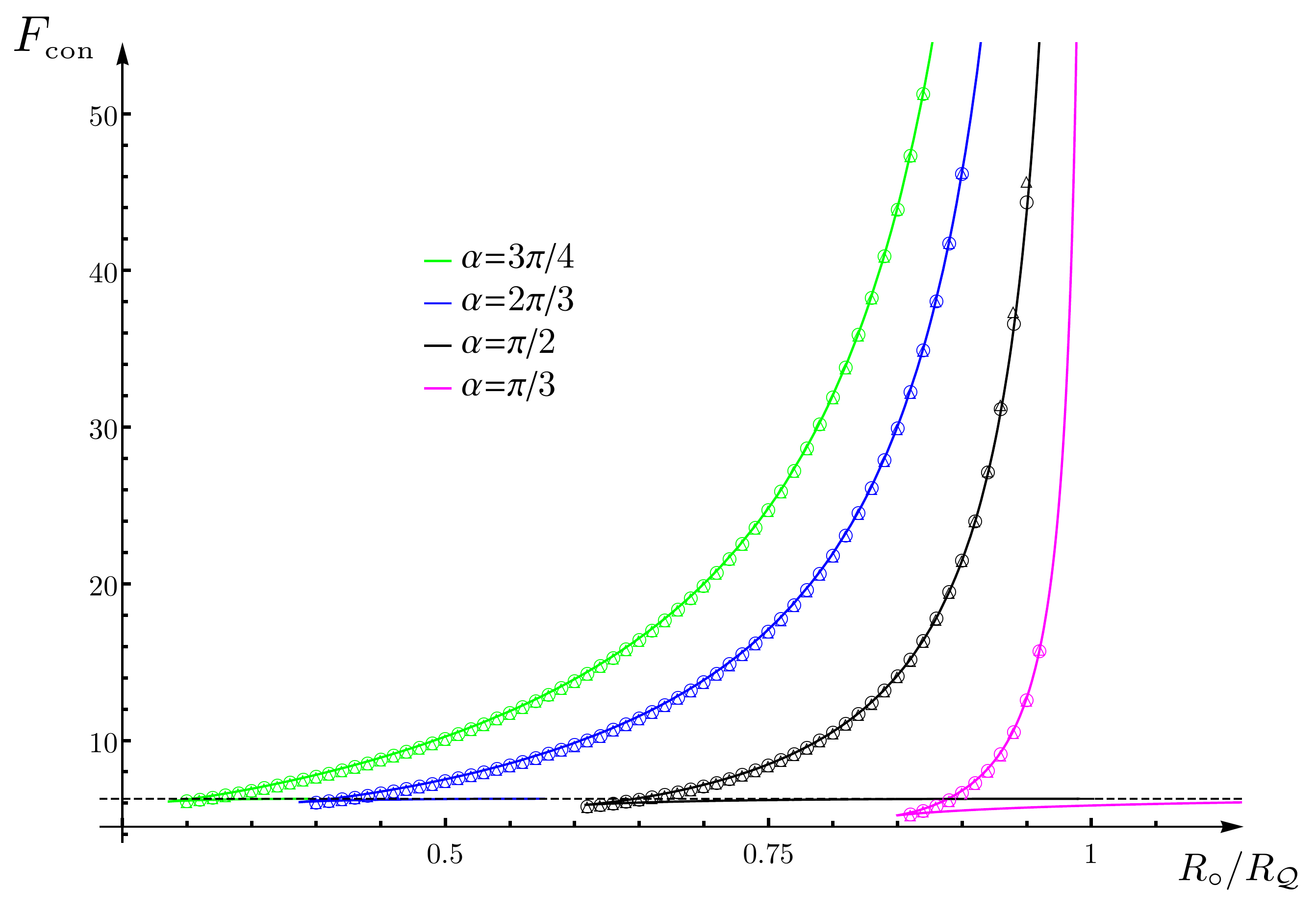}
\vspace{-.2cm}
\caption{\label{fig_R0RQ}
\small
The subleading term $F_{\text{\tiny con}}$ for the extremal surfaces $\hat{\gamma}_{A}^{\textrm{\tiny \,con}}$ 
intersecting orthogonally $\mathcal{Q}$ in terms of the ratio $R_\circ/R_\mathcal{Q}$,
for some values of $\alpha$.
The allowed configurations have $R_\circ/R_\mathcal{Q} <1$.
The solid curves have been obtained by combining the analytic expressions \eqref{initial ratio from k} and \eqref{F_A disk circ bdy}.
The horizontal dashed line corresponds to the value of the subleading term of the hemisphere $\hat{\gamma}_{A}^{\textrm{\tiny \,dis}}$, 
i.e.  $F_A = 2\pi$.
The data points are the numerical values obtained through Surface Evolver. 
The ones below the horizontal dashed line correspond to extremal surfaces that are not global minima.
Different kind of markers are associated to the two different ways employed to extract $F_{\text{\tiny con}}$ from the numerical data provided by Surface Evolver:
either by subtracting the area law term from the area of the entire extremal surface (empty circles)
or by applying the general formula \eqref{F_A willmore ads} (empty triangles).
}
\end{figure}

We find it worth discussing the limiting regimes of $F_{\textrm{\tiny con}}$ in (\ref{F_A disk circ bdy})  for small and large values of $k$
(the technical details of this analysis have been reported in Appendix\;\ref{sec-app-limits}).

In the limit $k \to 0$, which corresponds to $R_\circ \to R_{\mathcal{Q}}$ (see (\ref{R_ratio_exp}) and Fig.\;\ref{fig_RFk}), the expansion of $F_{\textrm{\tiny con}}$
reads
	\be
	\label{FA ann K}
	F_{\textrm{\tiny con}}
	=
	\frac{ 2 \pi \, \mathfrak{g}(\alpha)}{\sqrt[4]{k} }
	+ \frac{\pi}{2} \left( 
	\frac{\cot\alpha}{\sqrt{\sin\alpha}}+\mathbb{F}\big( \pi/4-\alpha/2 \, |\, 2 \big) +\frac{\Gamma^2\left(\frac{1}{4} \right)}{4 \sqrt{2\pi}}
	\right) \sqrt[4]{k} 
	+o\big(  \sqrt[4]{k}   \,\big) 
	\ee
Since the coefficient of the leading term is positive when $\alpha > \alpha_c$, negative when $\alpha < \alpha_c$ and zero when $\alpha = \alpha_c$,
different qualitative behaviours are observed when $k \to 0$.
In particular, for $\alpha = \alpha_c$ the subleading term is $o(1)$; therefore $F_{\textrm{\tiny con}} \to 0$.

By using (\ref{R_ratio_exp}), the expansion (\ref{FA ann K}) can be written also as an expansion for $R_\circ /R_\mathcal{Q} \to 1$,  finding that
	\be
	\label{FA ann K radius}
	F_{\textrm{\tiny con}}
	=
	\frac{2 \pi \, \mathfrak{g}(\alpha)^2}{1- R_\circ /R_\mathcal{Q}}-\pi \, \mathfrak{g}(\alpha)^2 +\mathcal{O}(1-R_\circ /R_\mathcal{Q})
	\ee

In the limit $k \to \infty$ we have seen that (\ref{Rratio-k-inf}) 
and in Appendix\;\ref{sec-app-limits} we find that $F_{\textrm{\tiny con}} \to (2\pi)^-$ for every $\alpha$.

In Fig.\;\ref{fig_Fk} we show $F_{\textrm{\tiny con}} $ in terms of $\sqrt[4]{k}$ for different values of $\alpha$.
The horizontal dashed line corresponds to $2\pi$, which is the value of the subleading term in the expansion of the area
of the hemisphere $\hat{\gamma}_{A}^{\textrm{\tiny \,dis}}$. 
This value provides the asymptotic limit of all the curves, confirming the result obtained in Appendix\;\ref{sec-app-limits}.

When $\alpha \leqslant \alpha_c$, from Fig.\;\ref{fig_Fk} we observe that  $F_{\textrm{\tiny con}} < 2\pi$ for all values of $k$.
Since in Sec.\,\ref{sec disk profiles} we have shown that the local solutions $\hat{\gamma}_{A}^{\textrm{\tiny \,con}}$ do not exist in this regime,
the curves $F_{\textrm{\tiny con}}$ having $\alpha \leqslant \alpha_c$ do not occur in the computation of holographic entanglement entropy. 
Thus, for $\alpha \leqslant \alpha_c$ the holographic entanglement entropy is given by $\hat{\gamma}_{A}^{\textrm{\tiny \,dis}}$.

When $\alpha > \alpha_c$ we have that $F_{\textrm{\tiny con}} \to +\infty$ for $k \to 0$ and $F_{\textrm{\tiny con}} \to (2\pi)^-$ for $k \to \infty$.
This implies that at least a local minimum exists. 
We observe numerically that $F_{\textrm{\tiny con}}$ has only one local extremum
for $k=k_{\circ,\textrm{\tiny \,min}} $, i.e. the same value for $k$ corresponding to the minimum of the ratio $R_\circ / R_{\mathcal{Q}}$.
This observation and the fact that, whenever two solutions $\hat{\gamma}_{A}^{\textrm{\tiny \,con}}$ can be found, 
for their values $k_1 < k_2$ of $k$ we have $k_1 < k_{\circ,\textrm{\tiny \,min}} < k_2$
lead to conclude that $F_{\textrm{\tiny con}}(k_2) < 2\pi$.
Hence, the holographic entanglement entropy is obtained by comparing $2\pi$ with $F_{\textrm{\tiny con}}$ evaluated on $k_1$. 
When $\alpha > \alpha_c$, let us denote with $k=k_c$  the solution of  $F_{\textrm{\tiny con}} = 2\pi$,
which can be found numerically and characterises the configuration where the subleading terms for
$\hat{\gamma}_{A}^{\textrm{\tiny \,con}}$ and $\hat{\gamma}_{A}^{\textrm{\tiny \,dis}}$ take the same value. 
Since $k_c < k_{\circ,\textrm{\tiny \,min}} $, 
the minimal surface providing the holographic entanglement entropy is $\hat{\gamma}_{A}^{\textrm{\tiny \,con}}$ if $k_1 < k_c$
and $\hat{\gamma}_{A}^{\textrm{\tiny \,dis}}$ if $k_1 > k_c\,$.
Denoting by $r_{\circ, c}$ the value of the ratio $R_\circ / R_\mathcal{Q}$ for the critical configuration having $k=k_c$,
in Fig.\;\ref{fig_alpha_c} we show $r_{\circ,\textrm{\tiny \,min}}  < r_{\circ, c}$ in terms of $\alpha \in (\alpha_c, \pi)$.

The solid curves in Fig.\;\ref{fig_R0RQ}, which are parameterised by $\alpha$, 
have been obtained by combining (\ref{initial ratio from k}) and (\ref{F_A disk circ bdy}) through a parametric plot.
The allowed configurations have $R_\circ / R_\mathcal{Q} < 1$.
A vertical line having $R_\circ / R_\mathcal{Q} < 1$ can intersect twice a solid curve corresponding to a fixed value of  $\alpha>\alpha_c$.
These two intersection points provide the values of $F_{\textrm{\tiny con}}$ (see Fig.\;\ref{fig_Fk}) 
obtained from the two values of $k$ given by the intersection of the horizontal line $R_\circ / R_\mathcal{Q}$ with the curve 
in Fig.\;\ref{fig_RFk} having the same $\alpha$.

In Fig.\;\ref{fig_R0RQ}, the value of $R_\circ / R_\mathcal{Q}$ corresponding to the intersection between 
$F_{\textrm{\tiny con}}$ for a given $\alpha$ and the horizontal dashed line
(whose height is $2\pi$) is $r_{\circ, c}$ (see the red line in Fig.\;\ref{fig_alpha_c}),
while $r_{\circ,\textrm{\tiny \,min}} $ is the value of $R_\circ / R_\mathcal{Q}$ corresponding to the cusp.

The analytic expression for $F_{\textrm{\tiny con}}$ has been checked numerically with Surface Evolver,
by adapting the method discussed in \cite{Seminara:2017hhh} to the configurations considered in this manuscript. 
The numerical results are the data points in Fig.\;\ref{fig_R0RQ}, where
the two different kind of markers (the empty circles and the empty triangles) correspond to two different ways to obtain the numerical value of 
$F_{\textrm{\tiny con}}$ from the numerical data about the extremal surface $\hat{\gamma}_{A}^{\textrm{\tiny \,con}}$.
One way is to evaluate $\hat{\mathcal{A}}^{\textrm{\tiny SE}}_\varepsilon - 2\pi R_\circ / \varepsilon $,
being $\hat{\mathcal{A}}^{\textrm{\tiny SE}}_\varepsilon$ the numerical value of the  area of  the extremal surface $\hat{\gamma}_{A}^{\textrm{\tiny \,con}}$.
The other method consists in finding $F_{\textrm{\tiny con}}$ by plugging into \eqref{F_A willmore ads} 
the geometrical quantities about $\hat{\gamma}_{A}^{\textrm{\tiny \,con}}$ required to employ this formula,
which are also given by Surface Evolver.

Notice that Fig.\;\ref{fig_R0RQ} shows that the extremal surfaces $\hat{\gamma}_{A}^{\textrm{\tiny \,con}}$ do not exist when $R_\circ / R_\mathcal{Q} \to 0$.
This means that the hemisphere $\hat{\gamma}_{A}^{\textrm{\tiny \,dis}}$ provides the holographic entanglement entropy in this regime, as expected.

The agreement between the solid curves and the data points in Fig.\;\ref{fig_R0RQ} provides
a highly non trivial confirmation of the analytic expressions obtained above.

The formula (\ref{F_A disk circ bdy}) can be found also 
by specialising the general result \eqref{F_A willmore ads-finite} to the extremal surfaces $\hat{\gamma}_{A}^{\textrm{\tiny \,con}}$ 
for the disks $A$ that we are considering. 
The details of this computation have been reported in Appendix\;\ref{app_area} and in the following we report only the main results. 
For the surface integral in \eqref{F_A willmore ads-finite} we find
\be
\label{disk_surface_term_main}
\int_{\hat{\gamma}_A}  \, \frac{(\tilde n^z)^2}{z^2} \, d\tilde{\mathcal{A}} 
\,=\, 
2\pi 
\left( \,
\frac{1+\eta_\alpha}{2} \, \mathcal{F}_{k,-} (\zeta_*)
+ \frac{1-\eta_\alpha}{2}\,
 \Big[ \mathcal{F}_{k,+} (\zeta_m) +\mathcal{F}_{k,-} (\zeta_m) -\mathcal{F}_{k,+} (\zeta_*) \Big]
 \,\right)
\ee
where the functions $\mathcal{F}_{k,\pm}$ can be written in terms of the function $\mathcal{F}_k$ introduced in (\ref{F_A int_evaluated}) as follows
(the derivation of this identity is briefly discussed in Appendix\;\ref{app_area}) 
\be
\label{F_k cal def}
\mathcal{F}_{k,\pm} (\zeta) 
  \,=\,
\mathcal{F}_k(\zeta)
- 
\frac{\sqrt{k (\zeta^2+1 )-\zeta^4}}{\sqrt{k}\; \zeta \big(\zeta^2+1\big)}
  \pm \frac{\zeta^2}{\sqrt{k} \,(\zeta^2+1)}
\ee
Since for $\zeta = \zeta_m$ the expression under the square root in (\ref{F_k cal def}) vanishes, 
it is straightforward to observe that, by plugging (\ref{F_k cal def}) into (\ref{disk_surface_term_main}), 
one obtains (\ref{F_A disk circ bdy}) and an additive contribution which depends on $\zeta_\ast$ but that does not contain $\zeta_m$.
This additive contribution is cancelled by the integral over the line $\partial \hat{\gamma}_{\mathcal{Q}} = \hat{\gamma}_{A}^{\textrm{\tiny \,con}} \cap \mathcal{Q}$
in \eqref{F_A willmore ads-finite}, which gives
\be
\label{disk_bound_term_main}
\int_{\partial \hat{\gamma}_{\mathcal{Q}}} 
\frac{\tilde{b}^z}{z} \, d\tilde{s}  
\,= \,
2\pi \,
\frac{ \sqrt{\zeta^2_* +(\sin \alpha)^2}  \; \zeta_* -\cos\alpha }{\zeta_* \left(\zeta^2_*+1\right) }
\ee

This concludes our analysis of the disk concentric to a circular boundary. 
We remark that we can easily study disks which are not concentric to the circular boundary
by combining the analytic expressions presented above with the mapping discussed in Appendix\;\ref{app:mapping}.

\subsection{Disk disjoint from a flat boundary}
\label{subsec:disk disjoint}

In the final part of this section we consider a disk $A$ of radius $R$ at finite distance $d$ from a flat boundary, 
in the AdS$_4$/BCFT$_3$ setup described in Sec.\;\ref{sec flat bdy}.
By combining the results presented in Sec.\;\ref{sec disk disjoint circ} with the mapping (\ref{mapping}) discussed in Appendix\;\ref{app:mapping},
one can easily obtain the analytic expressions for the extremal surfaces anchored to $\partial A$ and for 
the corresponding subleading term in the expansion of the area as $\varepsilon \to 0$.

The values of $R$ and $d$ are related to the parameters $R_\circ$ and $R_{\mathcal{Q}}$ 
characterising the configuration considered in Sec.\;\ref{sec disk profiles} and Sec.\;\ref{sec area disk concentric}
as follows
\be
R=\frac{R_\circ \, R_\mathcal{Q}^2}{R_\mathcal{Q}^2-R_\circ^2}
\hspace{.5cm} \qquad \hspace{.5cm}
d=\frac{R_\mathcal{Q} (R_\mathcal{Q} - R_\circ )}{2(R_\mathcal{Q}+R_\circ)}
\label{R and d}
\ee
From these expressions it is straightforward to find that
\begin{equation}
\label{delta}
\frac{d}{R} \,=\,
\frac{( R_\circ / R_\mathcal{Q} -1)^2}{2\,R_\circ / R_\mathcal{Q}}
\;\; \qquad \;\;
\frac{R_\circ}{R_\mathcal{Q}}
\,=\,
\frac{d}{R}  +1-\sqrt{\frac{d}{R}\bigg(\frac{d}{R}+2\bigg)}
\end{equation}

Since the extremal surfaces anchored to a 
disk disjoint from the flat boundary in the setup of Sec.\;\ref{sec flat bdy}
are obtained by mapping the extremal surfaces described in Sec.\;\ref{sec disk profiles} through (\ref{mapping}),
also for this configuration we have at most three local extrema of the area functional, depending on the ratio $d/R$:
the hemisphere $\hat{\gamma}_{A}^{\textrm{\tiny \,dis}}$ and at most two solutions $\hat{\gamma}_{A}^{\textrm{\tiny \,con}}$ 
intersecting the half plane $\mathcal{Q}$ orthogonally.

\begin{figure}[t] 
\vspace{-.6cm}
\hspace{-1.3cm}
\includegraphics[width=1.14\textwidth]{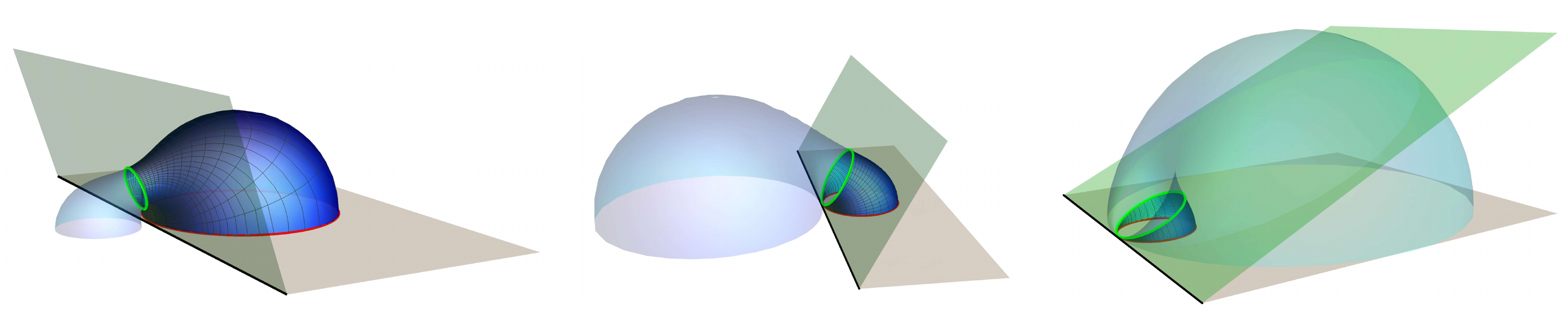}
\vspace{-.4cm}
\caption{\label{fig_shadow_alpha}
\small
Extremal surfaces $\hat\gamma_{A}^{\text{\tiny \,con}}$ anchored to a disk of radius $R$ (bounded by the red circle) 
at finite distance $d$ from the flat boundary
(see Sec.\;\ref{subsec:disk disjoint}).
Here $d/R\sim 0.042$ is fixed and different values of $\alpha$ are considered: 
$\alpha = \pi/2.5$ (left), $\alpha = 2\pi/3$ (middle) and $\alpha = 2.7$ (right).
The surface $\hat\gamma_{A}^{\text{\tiny \,con}}$ intersects the green half plane $\mathcal{Q}$ orthogonally  along the green circle
$\partial \hat{\gamma}_{\mathcal{Q}}$.
The shaded surfaces correspond to the auxiliary surfaces $\hat{\gamma}_{A, \textrm{\tiny \,aux}}^{\textrm{\tiny \,con}}$ (see also Appendix\;\ref{app aux_domains}).
The extremal surface $\hat\gamma_{A}^{\text{\tiny \,con}}$ is the global minimum when the corresponding $F_A$ is larger than $2\pi$.
Here $F_A = 5.6$ (left), $F_A = 17.1$ (middle) and $F_A = 47.1$ (right).
The surface in the left panel has the smallest area among the two solutions $\hat\gamma_{A}^{\text{\tiny \,con}}$ but the global minimum is 
the hemisphere $\hat\gamma_{A}^{\text{\tiny \,dis}}$ in this case. 
}
\end{figure}

\begin{figure}[t] 
\vspace{-.6cm}
\hspace{-1cm}
\includegraphics[width=1.14\textwidth]{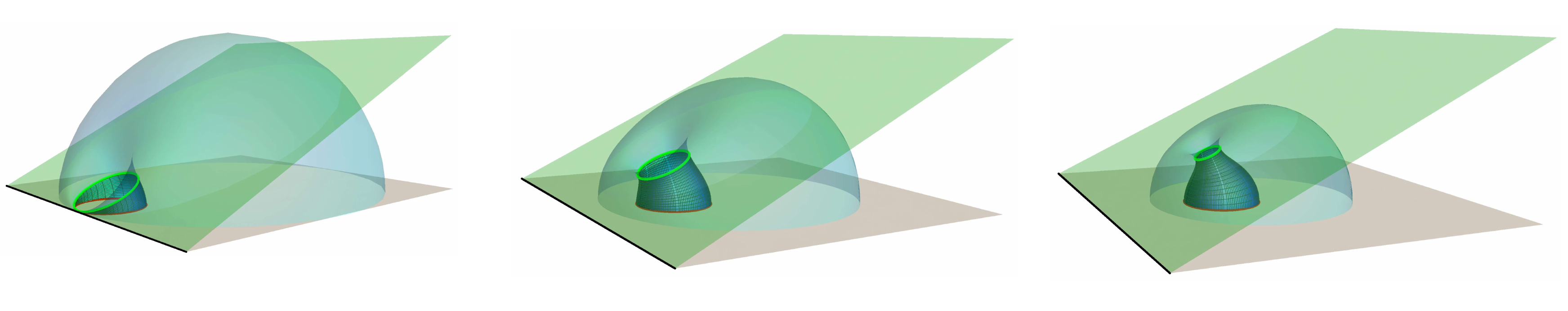}
\vspace{-.4cm}
\caption{\label{fig_shadow_dist}
\small
Extremal surfaces $\hat\gamma_{A}^{\text{\tiny \,con}}$ anchored to a disk (bounded by the red circle) 
of radius $R$ at finite distance $d$ from the flat boundary, like in Fig.\;\ref{fig_shadow_alpha}.
Here $\alpha=2.7$ is fixed (like in the right panel of Fig.\;\ref{fig_shadow_alpha}) and the different values of $d/R$ are considered:
$d/R\sim 0.042$ (left), $d/R\sim 1.6$ (middle) and $d/R\sim 2.243$ (right). 
The shaded surfaces correspond to $\hat{\gamma}_{A, \textrm{\tiny \,aux}}^{\textrm{\tiny \,con}}$ and for all the 
configurations of this figure part of $\hat{\gamma}_{A, \textrm{\tiny \,aux}}^{\textrm{\tiny \,con}}$ belongs to the gravitational spacetime (\ref{AdS4 regions})
(see also Appendix\;\ref{app aux_domains}).
The extremal surface $\hat{\gamma}_{A, \textrm{\tiny \,aux}}^{\textrm{\tiny \,con}}$ is a global minimum when its $F_A$ is larger than $2\pi$.
The configuration in the left panel is the same shown in the right panel of Fig.\;\ref{fig_shadow_alpha}.
In the remaining panels $F_A = 6.95$ (middle) and $F_A = 6.13$ (right).
}
\end{figure}

In Fig\;\ref{fig_shadow_alpha} we show some examples of $\hat\gamma_{A}^{\text{\tiny \,con}}$ 
 for a fixed configuration of the disk $A$ and three different slopes of $\mathcal{Q}$ (the green half plane).
In each panel, the shaded surface is the auxiliary surface $\hat{\gamma}_{A, \textrm{\tiny \,aux}}^{\textrm{\tiny \,con}}$ corresponding to $\hat\gamma_{A}^{\text{\tiny \,con}}$, 
which intersects orthogonally $\mathcal{Q}$ along $\partial \hat{\gamma}_{\mathcal{Q}}$ and is such that 
$\hat\gamma_{A}^{\text{\tiny \,con}} \cup \hat{\gamma}_{A, \textrm{\tiny \,aux}}^{\textrm{\tiny \,con}}$ is an extremal surface in $\mathbb{H}_3$
anchored to the two disjoint circles (one of them is $\partial A$).
In Fig.\;\ref{fig_shadow_dist} we show $\hat\gamma_{A}^{\text{\tiny \,con}}$  and the corresponding $\hat{\gamma}_{A, \textrm{\tiny \,aux}}^{\textrm{\tiny \,con}}$
for a fixed value of $\alpha$ and three different values of $d/R$.
Notice that for some configurations $\hat{\gamma}_{A, \textrm{\tiny \,aux}}^{\textrm{\tiny \,con}}$ lies entirely outside the gravitational spacetime 
(\ref{AdS4 regions}) (see e.g. the left panel and the middle panel of Fig\;\ref{fig_shadow_alpha}), 
while for other ones part of $\hat{\gamma}_{A, \textrm{\tiny \,aux}}^{\textrm{\tiny \,con}}$ belongs to it. 
The latter case occurs when the auxiliary region $A_\textrm{\tiny \,aux}$ is a subset of the half plane $x>0$, where also $A$ is defined.

For the extremal surfaces that we are considering, 
the leading term of $\mathcal{A}[\hat{\gamma}_\varepsilon]$ as $\varepsilon\to 0$  is the area law term $2\pi R / \varepsilon$
and the subleading finite term is $ - \, \textrm{max}(2\pi,  \widehat{F}_{\textrm{\tiny con}})$, like in (\ref{area_erik}),
where $\widehat{F}_{\textrm{\tiny con}}$ corresponds to the maximum between the values of $F_{\textrm{\tiny con}}$
evaluated for the extrema $\hat{\gamma}_{A}^{\textrm{\tiny \,con}}$.
The analytic expression of $F_{\textrm{\tiny con}}$ as function of $d/R$ can be obtained 
through a parametric plot involving $F_{\textrm{\tiny con}}$ in (\ref{F_A disk circ bdy}), 
$d/R$ in (\ref{delta})
and $R_\circ / R_\mathcal{Q}$ in (\ref{initial ratio from k}).
This procedure has been employed to find the solid black curves in Fig.\;\ref{fig_ellipses_F}, which correspond to a disk.

From (\ref{delta}), it is straightforward to observe that 
$d/R \to \infty$ corresponds to $R_\circ / R_\mathcal{Q} \to 0$,
and $d/R \to 0$ to $R_\circ / R_\mathcal{Q} \to 1$.
Thus, when $d/R \to \infty$ the hemisphere $\hat\gamma_{A}^{\text{\tiny \,dis}}$ is the minimal surface providing the 
holographic entanglement entropy (see also Sec.\;\ref{sec area disk concentric}).
In the opposite limiting regime $d/R \to 0$, the second expression in (\ref{delta}) implies that
$R_\circ / R_\mathcal{Q}
= 1-  \sqrt{2\, d/R}+ d/R +O((d/R) ^{3/2}) $.
Hence, from the expansion (\ref{FA ann K radius}), it is straightforward to obtain that
$F_{\textrm{\tiny con}} = 2\pi\,\mathfrak{g}(\alpha)^2 / \sqrt{2d /R} +\mathcal{O}\big(\sqrt{d/R}\big)$
at leading order.

\section{On smooth domains disjoint from the boundary}
\label{sec domain disjoint}

\begin{figure}[t] 
\vspace{-.8cm}
\hspace{-.8cm}
\includegraphics[width=1.1\textwidth]{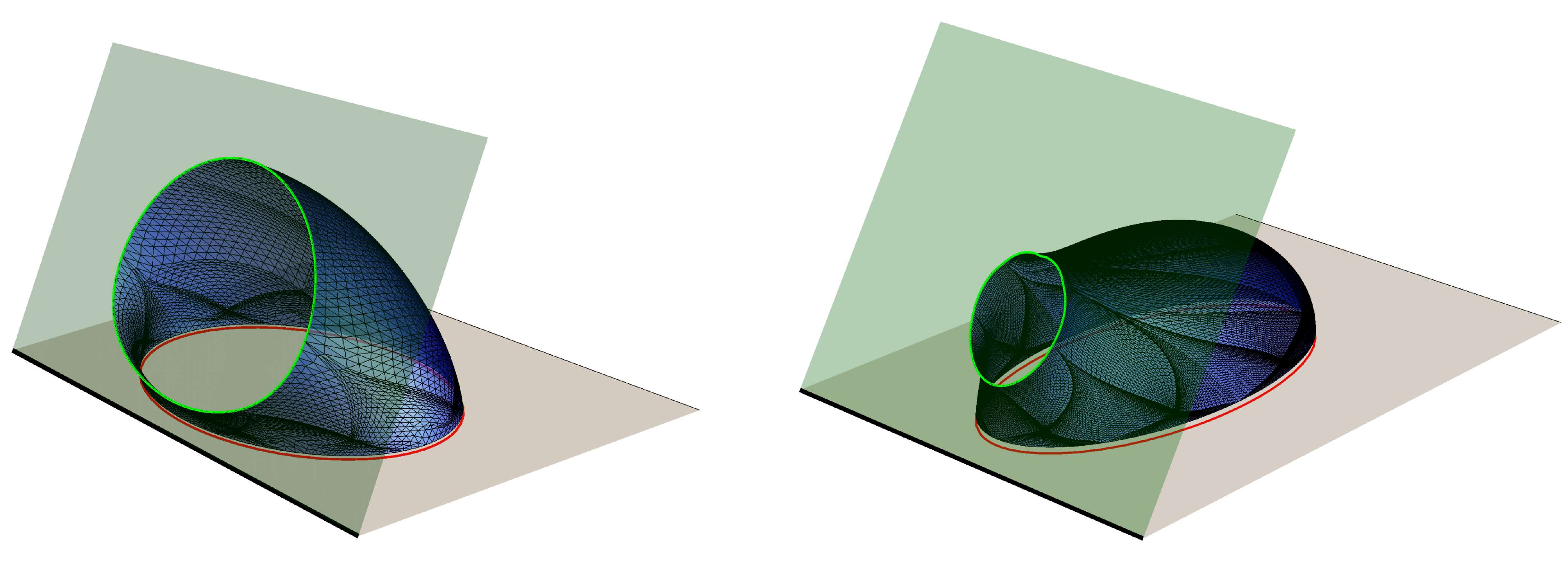}
\vspace{-.4cm}
\caption{\label{fig_ellipses3d}
\small
Extremal surfaces $\hat \gamma_A^{\text{\tiny con}}$ found with Surface Evolver 
in the gravitational setup described in Sec.\;\ref{sec flat bdy}.
The extremal surfaces are anchored to the boundary of two different ellipses $A$ (red curves) 
and intersect orthogonally the half plane $\mathcal{Q}$ with $\alpha=2\pi/3$ (green half plane). 
Here $\varepsilon=0.03$.
Denoting by $R_\perp$ and $R_\parallel$ the lengths of the semiaxis which are respectively  orthogonal and parallel to the flat boundary, 
and by $d$ the distance of $\partial A$ from the flat boundary, 
we have $d/R_\perp=0.2$ in both the panels.
Instead, $R_{\parallel}=2 R_\perp$ in the left panel 
and $R_{\parallel}=0.5 R_\perp$ in the right panel. 
}
\end{figure}

Analytic expressions for the subleading term $F_A$ in (\ref{hee bdy intro}) 
can be obtained for configurations which are particularly simple or highly symmetric.
Two important cases have been discussed in Sec.\;\ref{sec strip adjacent} and Sec.\;\ref{sec disk}.
In order to find analytic solutions for an extremal surface anchored to a generic entangling curve,
typically a partial differential equation must be solved, which is usually a difficult task. 
Thus, it is useful to develop efficient numerical methods that allow us to study the shape dependence of $F_A$.

The crucial tool of our numerical analysis is Surface Evolver, which 
has been already employed to study the holographic entanglement entropy in AdS$_4$/CFT$_3$ \cite{Fonda:2014cca, Fonda:2015nma}
and to check the corner functions in AdS$_4$/BCFT$_3$ \cite{Seminara:2017hhh}.
In this manuscript we consider some regions disjoint from the boundary in AdS$_4$/BCFT$_3$.
In Sec.\;\ref{sec disk disjoint circ} Surface Evolver has been used to check numerically the analytic expressions
of the extremal surfaces and of $F_A$ for a disk concentric to a circular boundary 
(see Fig.\;\ref{fig_profiles_evolver} and Fig.\;\ref{fig_R0RQ} respectively).
In this section we use  Surface Evolver to study 
the extremal surfaces $\hat{\gamma}_A$ and the corresponding $F_A$ for some simple domains
which cannot be treated through analytic methods.

\begin{figure}[t] 
\vspace{-1.2cm}
\hspace{-1cm}
\includegraphics[width=1.1\textwidth]{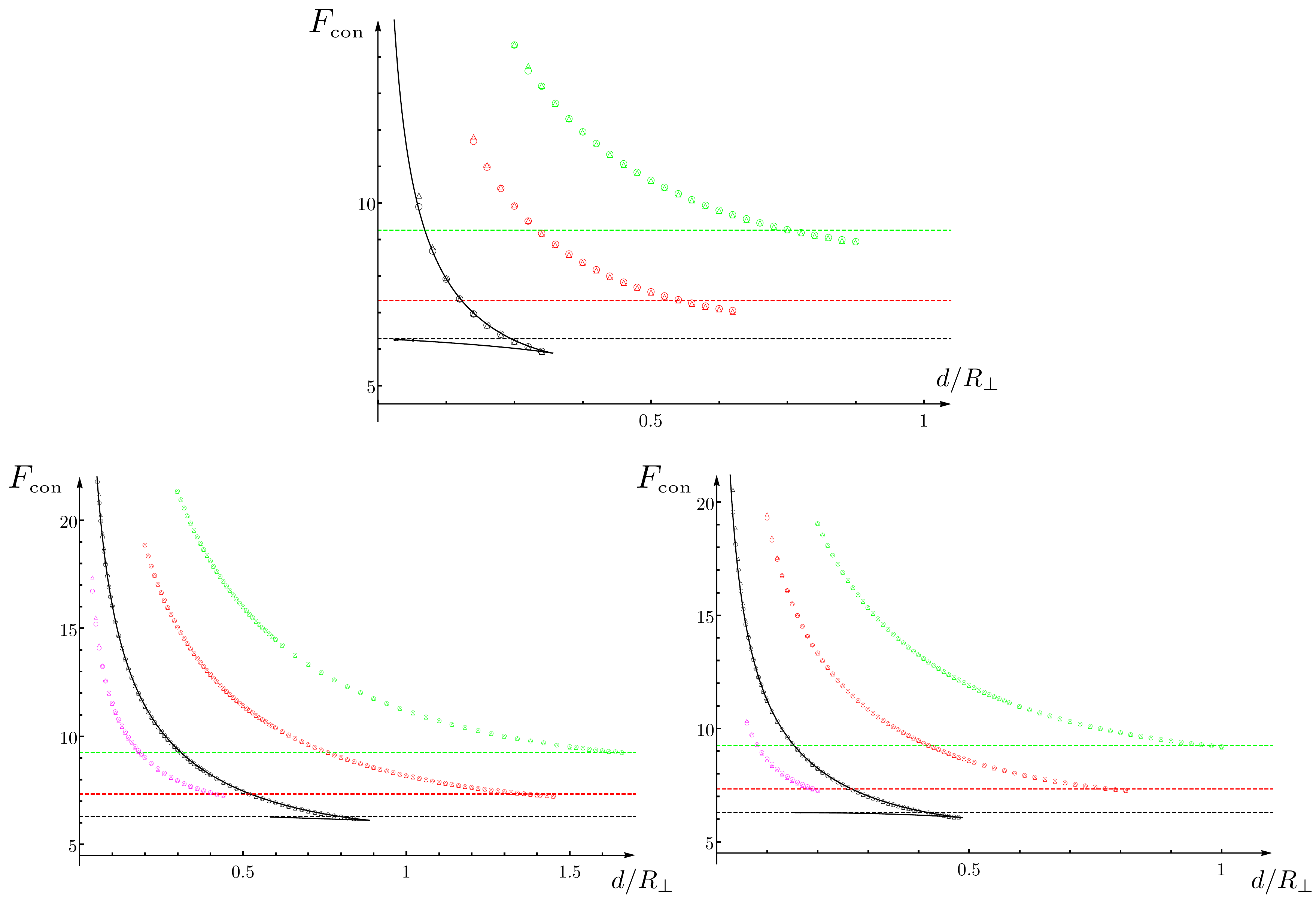}
\vspace{-.4cm}
\caption{\label{fig_ellipses_F}
\small
The subleading term $F_{\text{\tiny con}}$ for the extremal surfaces $\hat{\gamma}_{A}^{\textrm{\tiny \,con}}$ intersecting orthogonally the half plane $\mathcal{Q}$ 
and anchored to ellipses at distance $d$ from a flat boundary (see Fig.\;\ref{fig_ellipses3d}).
A semiaxes of the ellipse is orthogonal to the flat boundary and its length is $R_\perp$, while $R_{\parallel}$ is the length of the other one. 
The three panels are characterised by three diverse values of the slope $\alpha$ for the half plane $\mathcal{Q}$ (see Fig.\;\ref{fig_ellipses3d}):
$\alpha = \pi/2$ (top), $\alpha = 2\pi/3$ (bottom right) and $\alpha = 3\pi/4$ (bottom left).
Different colours correspond to different eccentricities:
$R_{\parallel}=3 R_\perp$ (green), $R_{\parallel}=2 R_\perp$ (red), $R_{\parallel}=R_\perp$ (black) and $R_{\parallel}=0.5 R_\perp$ (magenta).
The solid black curves correspond to the analytic expressions obtained in Sec.\;\ref{subsec:disk disjoint} for disks. 
The dashed horizontal lines provide the value $F_A = F_{\text{\tiny dis}}$ for the extremal surfaces disconnected from $\mathcal{Q}$.
In particular, $F_{\text{\tiny dis}} = 9.25$ (green),  $F_{\text{\tiny dis}} = 2\pi$ (black) and $F_{\text{\tiny dis}} = 7.33$ (red and magenta).
}			
\end{figure}

Considering the simple AdS$_4$/BCFT$_3$ setup described in Sec.\;\ref{sec flat bdy},
in Fig.\;\ref{fig_advertising} we showed the extremal surface corresponding to a region $A$ with a complicated shape
(the entangling curve is the red curve in the inset) which has been constructed by using Surface Evolver
and which is very difficult to describe analytically.

In the same setup, let us consider, for simplicity, regions $A$ delimited by ellipses
at distance $d$ from the flat boundary with one of the semiaxis parallel to the flat boundary. 
These regions are given by the points $(x,y) \in \mathbb{R}^2$ with $x>0$
such that $(x-d-R_\perp)^2/R^2_{\parallel} + y^2/R^2_{\perp} \leqslant 1$,
where $R_\perp$ and $R_{\parallel}$ are the lengths of the semiaxis which are respectively
orthogonal and parallel to the flat boundary $x=0$.
As for the extremal surfaces anchored to the entangling curve $\partial A$, 
either they are disconnected from the half plane $\mathcal{Q}$ or they intersect it orthogonally.
The occurrence of these different kind of extremal surfaces 
and which of them gives the global minimum 
depend on the values of $\alpha$, of the ratio $d/R_\perp$ and of the eccentricity of $A$.
For some configurations only the solutions disconnected from $\mathcal{Q}$ are allowed,
while for other configurations only the extremal surfaces intersecting $\mathcal{Q}$ exist,
as discussed in a specific example in the final part of Sec.\;\ref{sec disk profiles}. 
In Fig.\;\ref{fig_ellipses3d} we show two examples of extremal surfaces anchored to ellipses in the $z=0$ half plane (the red curves)
which intersect $\mathcal{Q}$ orthogonally along the green line $\partial \hat{\gamma}_{\mathcal{Q}}$. 

In Fig.\;\ref{fig_ellipses_F} the values of the subleading term for extremal surfaces 
intersecting $\mathcal{Q}$ and anchored to various ellipses are plotted in terms of the ratio $d/R_\perp$.
These data points have been obtained through Surface Evolver by first 
constructing the extremal surface $\hat{\gamma}_\varepsilon^{\textrm{\tiny SE}}$ anchored to the ellipses defined at $z=\varepsilon$
and then employing the information about $\hat{\gamma}_\varepsilon^{\textrm{\tiny SE}}$ provided by the code 
(in particular its area $\mathcal{A}[\hat{\gamma}_\varepsilon^{\textrm{\tiny SE}}]$ and its normal vectors)
in two different ways. 
One way to extract the subleading term is to compute $P_A / \varepsilon - \mathcal{A}[\hat{\gamma}_\varepsilon^{\textrm{\tiny SE}}]$ 
(empty circles in Fig.\;\ref{fig_ellipses_F}).
Another way is to evaluate \eqref{F_A willmore ads Q-plane} from the unit vector $\tilde{n}^\mu$ normal to $\hat{\gamma}_\varepsilon^{\textrm{\tiny SE}}$
(empty triangles  in Fig.\;\ref{fig_ellipses_F}).
The agreement between these two approaches provides a strong numerical evidence that \eqref{F_A willmore ads Q-plane} is correct.
The numerical analysis has been performed by adapting the method discussed in \cite{Seminara:2017hhh} to the configurations considered here.

The horizontal dashed lines in Fig.\;\ref{fig_ellipses_F} correspond to the extremal surfaces that do not intersect $\mathcal{Q}$.
Denoting by $F_{\text{\tiny dis}}$ the subleading term in the expansion of $\mathcal{A}[\hat{\gamma}_\varepsilon^{\textrm{\tiny SE}}]$ for these surfaces, 
we have that $F_A$ in (\ref{hee bdy intro}) is finite and given by $F_A = \textrm{max}(F_{\text{\tiny con}}, F_{\text{\tiny dis}})$.
The relation $F_{\text{\tiny con}} = F_{\text{\tiny dis}}$
provides the critical value of $d/R_\perp$  characterising the transition  in the holographic entanglement entropy
between the surfaces connected to $\mathcal{Q}$ and the ones disjoint from $\mathcal{Q}$
(see the intersection between the curve identified by the data points and the horizontal dashed line having the same colour  in Fig.\;\ref{fig_ellipses_F}, except for the magenta points,
that must be compared with the red dashed line).

The black points in Fig.\;\ref{fig_ellipses_F} correspond to disks disjoint from a flat boundary and the solid black curves 
have been obtained through the analytic expressions discussed in Sec.\;\ref{sec disk} (see \eqref{F_A disk circ bdy} and \eqref{delta}).
The nice agreement with the data points found with Surface Evolver is a strong check for the analytic expressions.

In Sec.\;\ref{sec disk} we have found that the critical value $\alpha_c$ (defined as the unique zero of (\ref{g function def main})) 
for the slope of $\mathcal{Q}$ in the AdS$_4$/BCFT$_3$  setup of Sec.\;\ref{sec flat bdy} is such that 
extremal surfaces anchored to a disk $A$ disjoint from the flat boundary and intersecting $\mathcal{Q}$ orthogonally
do not exist for $\alpha \leqslant \alpha_c $.
We find it reasonable to conjecture the validity of this property (with same $\alpha_c$) for any smooth region $A$ disjoint from the boundary
in the AdS$_4$/BCFT$_3$  setups described in Sec.\;\ref{sec flat bdy} and Sec.\;\ref{sec circular bdy}.

We find it worth exploring the existence of bounds on the subleading term $F_A$.
In the AdS$_4$/CFT$_3$ duality when the dual gravitational background is AdS$_4$,
by employing a well known bound for the Willmore functional in $\mathbb{R}^3$,
it has been shown that $F_A \geqslant 2\pi$ for any kind of spatial region, including the ones with singular $\partial A$ and the ones made by disjoint components \cite{Fonda:2015nma}.

In the remaining part of this section we discuss that, in the context of AdS$_4$/BCFT$_3$ and 
when the gravitational dual is the part of AdS$_4$ delimited by $\mathcal{Q}$ and the conformal boundary,
for any kind of spatial region $A$ disjoint from the boundary we have
\be
\label{bound}
F_A \geqslant 2\pi
\ee

If $A$ contains at least one corner, this bound is trivially satisfied because $F_A$ diverges logarithmically 
and the coefficient of this divergence is positive, being determined by the corner function of \cite{Drukker:1999zq}.

For regions $A$ with smooth $\partial A$, the subleading term $F_A$ in (\ref{hee bdy intro}) is finite
and the corresponding minimal surface $\hat{\gamma}_A$ is such that either 
$\hat{\gamma}_A \cap \mathcal{Q} = \emptyset$ or $\hat{\gamma}_A \cap \mathcal{Q} \neq \emptyset$.
In the former case $\hat{\gamma}_A$ is also a minimal surface in $\mathbb{H}_3$, therefore we can employ the observation 
made in \cite{Fonda:2015nma} for AdS$_4$/CFT$_3$ and conclude that (\ref{bound}) holds.

If $\hat{\gamma}_A \cap \mathcal{Q} \neq \emptyset$, let us denote by $F_A = F_{\text{\tiny con}}$ the value of the 
subleading term corresponding to $\hat{\gamma}_A $.
In these cases, we have two possibilities:
either another extremal surface $\hat{\gamma}_A^{\,\textrm{\tiny dis}}$ such that $\hat{\gamma}_A^{\,\textrm{\tiny dis}} \cap \mathcal{Q} = \emptyset$ exists or not. 
In the former case, being $\hat{\gamma}_A $ the global minimum, we have that $F_{\text{\tiny con}} \geqslant F_{\text{\tiny dis}} \geqslant 2\pi$,
where the last inequality is obtained from the observation of \cite{Fonda:2015nma}, as above.

The remaining configurations are the ones such that only the extremal surface $\hat{\gamma}_A $ with $\hat{\gamma}_A \cap \mathcal{Q} \neq \emptyset$ exists
(see e.g. the explicit case discussed in the final part of Sec.\;\ref{sec disk profiles}).
In these cases $\hat{\gamma}_A^{\,\textrm{\tiny dis}}$ does not occur because,
by introducing the extremal surface $\hat{\gamma}_A^{\textrm{\tiny $(0)$}}$ in $\mathbb{H}_3$ anchored to $\partial A$,
we have that $\hat{\gamma}_A^{\textrm{\tiny $(0)$}} \cap \mathcal{Q} \neq \emptyset$.
Let us consider the part $\hat{\gamma}_A^{\textrm{\tiny $\,\angle$}}  \subset \hat{\gamma}_A^{\textrm{\tiny $(0)$}} $
of $\hat{\gamma}_A^{\textrm{\tiny $(0)$}} $ belonging to the region of AdS$_4$ delimited by $\mathcal{Q}$ and the conformal boundary.
We remark that $\hat{\gamma}_\varepsilon^{\textrm{\tiny $\,\angle$}}$ intersects $\mathcal{Q}$ but, 
typically,  they are not orthogonal along their intersection.
Restricting both $\hat{\gamma}_A^{\textrm{\tiny $(0)$}}$ and $\hat{\gamma}_A^{\textrm{\tiny $\,\angle$}}$ to $z\geqslant \varepsilon$, 
for the resulting surfaces $\hat{\gamma}_\varepsilon^{\textrm{\tiny $(0)$}}$ and $\hat{\gamma}_\varepsilon^{\textrm{\tiny $\,\angle$}}$
the expansion  (\ref{area expansion intro}) holds with the same $P_A$ but different $O(1)$ terms, 
that will be denoted by $F_A^{\textrm{\tiny $(0)$}}$ and $F_A^{\textrm{\tiny $\,\angle$}}$ respectively. 
Notice that the observation of  \cite{Fonda:2015nma} here gives $F_A^{\textrm{\tiny $(0)$}} \geqslant 2\pi$.
Since $\hat{\gamma}_A^{\textrm{\tiny $\,\angle$}}  \subset \hat{\gamma}_A^{\textrm{\tiny $(0)$}} $,
we have $ \mathcal{A}[ \hat{\gamma}^{\textrm{\tiny $(0)$}} _\varepsilon ] \geqslant \mathcal{A}[ \hat{\gamma}^{\textrm{\tiny $\,\angle$}} _\varepsilon ]$,
which implies $F_A^{\textrm{\tiny $(0)$}} \leqslant  F_A^{\textrm{\tiny $\,\angle$}}$,
being $P_A$ the same for $\hat{\gamma}_\varepsilon^{\textrm{\tiny $(0)$}}$ and $\hat{\gamma}_\varepsilon^{\textrm{\tiny $\,\angle$}}$.
Since $F_{\text{\tiny con}}$ corresponds to an extremal surface and $\hat{\gamma}_\varepsilon^{\textrm{\tiny $\,\angle$}}$
is not extremal, we can conclude that $F_{\text{\tiny con}} \geqslant  F_A^{\textrm{\tiny $\,\angle$}}$. 
Collecting these observations, we find that $F_{\text{\tiny con}} \geqslant  F_A^{\textrm{\tiny $\,\angle$}} \geqslant F_A^{\textrm{\tiny $(0)$}} \geqslant 2\pi$.

This completes our discussion about the validity of the inequality (\ref{bound})  for any spatial region $A$ disjoint from the boundary, 
including the ones having singular $\partial A$ or  that are made by disjoint connected components.
We find it worth remarking that the bound (\ref{bound}) does not hold in general when $A$ is adjacent to the boundary because the corner function 
is negative for some configurations \cite{Seminara:2017hhh}.
\\

\section{Domains with corners adjacent to the boundary}
\label{sec corners}

The holographic entanglement entropy of domains $A$ with corners whose tip is on the boundary
contains a subleading logarithmic divergence whose coefficient
is determined  by  a model dependent corner function which depends
also on the boundary conditions. 
In the setups of AdS$_4$/BCFT$_3$ of sec. \ref{sec flat bdy} , the analytic expression of the corner function 
$F_\alpha(\omega)$ has been found in \cite{Seminara:2017hhh} from a direct evaluation of the area of
the minimal surface corresponding to an infinite wedge adjacent to the flat boundary (see (\ref{area wedge intro exp})).

Below, we show that the corner function  $F_\alpha(\omega)$ can be also obtained also from (\ref{F_A willmore ads Q-plane}).
In Sec.\;\ref{sec app half disk} we focus on the simplest configuration given by  a half disk centered on the flat boundary,
while in Sec.\;\ref{sec wedge} we discuss the most general case of an infinite wedge adjacent to the flat boundary
with generic opening angle.

\subsection{Half disk centered on the boundary}
\label{sec app half disk}

\begin{figure}[t] 
\vspace{-.8cm}
\hspace{-.8cm}
\includegraphics[width=1.1\textwidth]{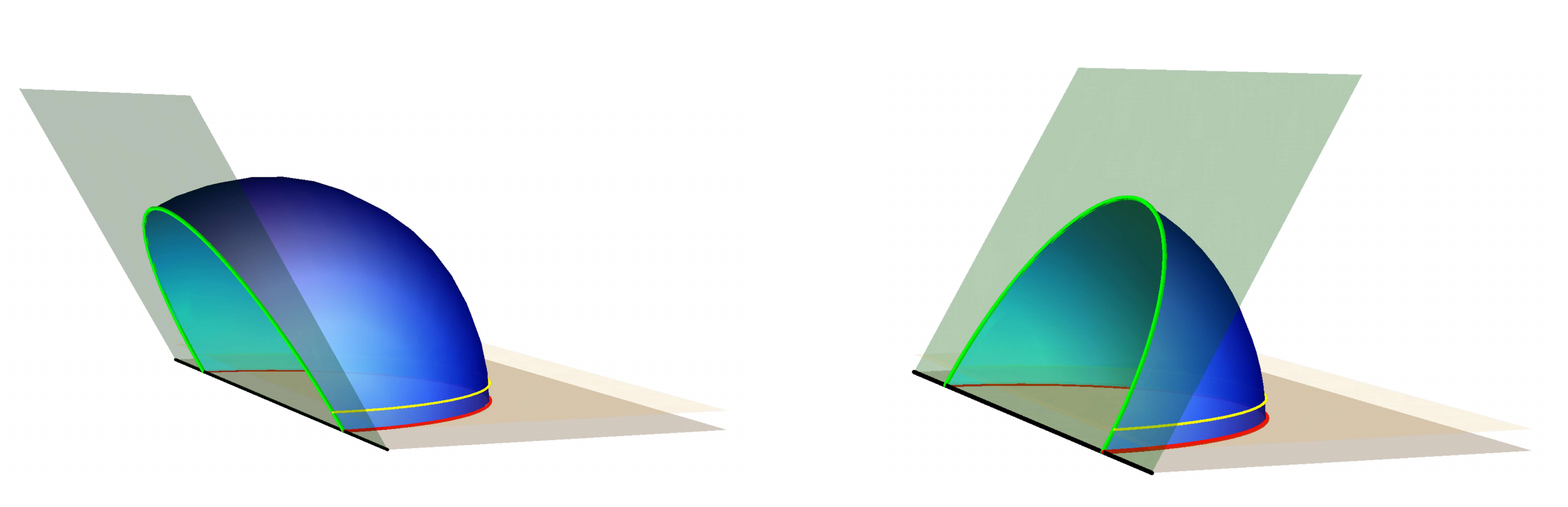}
\vspace{-.5cm}
\caption{\label{fig_3d_half_disk}
\small
Minimal surfaces  $\hat \gamma_A$ anchored to the entangling curve corresponding to a half disk $A$ centered on the flat boundary. 
The slope of the green half plane $\mathcal{Q}$ is $\alpha=\pi/3$ in the left panel and $\alpha=2\pi/3$ in the right panel. 
The yellow half plane has $z=\varepsilon$ and its intersection with  $\hat \gamma_A$ is the yellow curve.
The green curve corresponds to $\partial \hat{\gamma}_{\mathcal{Q}} $ in (\ref{interQ half-disk}).
}
\end{figure}

In the gravitational setup described in Sec.\;\ref{sec flat bdy}, let us consider the half disk $A$ of radius $R$
centered in the origin, which belongs to the flat boundary, i.e.
$A=\{ (x,y) \in \mathbb{R}^2 \, | \,x^2+y^2\leqslant R^2 ,  \,x\geqslant 0\}$. 
The minimal surface corresponding to this configuration is simply given by the part of the hemisphere anchored to the entire circle 
centered in the origin which satisfies the constraint (\ref{sec flat bdy}).
In Fig.\;\ref{fig_3d_half_disk} the minimal surface $\hat \gamma_A$ 
is shown for two different values of $\alpha$.
When $\alpha \neq \pi/2$, a non trivial logarithmic divergence occurs in the expansion of the area
$\mathcal{A} [\hat{\gamma}_\varepsilon ] $. In particular, it has been found that  \cite{Seminara:2017hhh}
\be
\label{area half-disk final main}
\mathcal{A} [\hat{\gamma}_\varepsilon ] 
\,=\, 
\frac{\pi R}{\varepsilon} + 2 \cot \alpha\, \log(R/\varepsilon) + O(1) 
\ee
which tells us that $F_{\alpha}(\pi/2)= - \cot \alpha\,$ for the corner function introduced in (\ref{area wedge intro exp}), 
being the factor of $2$ due to the fact that $A$ has two corners adjacent to the boundary. 
The expression of $F_{\alpha}(\pi/2)$ has been first obtained in \cite{FarajiAstaneh:2017hqv} by considering 
the equal bipartition of the half plane where the entangling curve is the half line orthogonal to the flat boundary.

It is instructive to show that the general formula \eqref{F_A willmore ads Q-plane} is able to reproduce 
the logarithmic term occurring in \eqref{area half-disk final main}.  
Let us observe that the integral over $\hat{\gamma}_\varepsilon$ in (\ref{F_A willmore ads Q-plane}) provides a finite result as $\varepsilon \to 0$ because
$\hat{\gamma}_\varepsilon$ is part of the hemisphere $\hat{\gamma}_{A} \cup \,\hat{\gamma}_{A, \textrm{\tiny \,aux}}$ and,
being the integrand positive, the integral over $\hat{\gamma}_\varepsilon$ is smaller than the integral over
the entire hemisphere $\hat{\gamma}_{A} \cup \,\hat{\gamma}_{A, \textrm{\tiny \,aux}}\,$, which gives $2\pi$.

The intersection between $\hat{\gamma}_{A} $ and $\mathcal{Q}$ is given by the following semi-circle
\be
\label{interQ half-disk}
\partial \hat{\gamma}_{\mathcal{Q}} : \;
\left\{ \begin{array}{l}
x^2+y^2+z^2 = R^2
\\
\rule{0pt}{.4cm}
z =  - \, x \tan \alpha
\end{array}
\right.
\ee
By employing the  spherical coordinates
\begin{equation}
\label{polar coords spike}
z= R\, \sin\theta\, \cos \phi
\qquad
x= -\, R\, \sin\theta \, \sin\phi
\qquad
y= R\, \cos\theta
\end{equation}
one finds the following parametric representation of $\partial \hat{\gamma}_{\mathcal{Q}}$
\be
\label{polar coords spike willmore}
\partial \hat{\gamma}_{\mathcal{Q}}: \;\;
(z,x,y) = R\big(
\sin\theta\, \cos(\pi/2-\alpha) \,,\,
- \sin\theta \, \sin(\pi/2-\alpha) \,,\,
\cos\theta
\big)
\qquad
\theta_\varepsilon \leqslant  \theta \leqslant \pi -  \theta_\varepsilon
\ee
The angle $\theta_\varepsilon$ is given by the intersection of 
$\partial \hat{\gamma}_{\mathcal{Q}}$ with the cutoff $z=\varepsilon$;
therefore it can be found from the condition $\varepsilon = R\sin\theta_\varepsilon \cos (\pi/2-\alpha)$.  
Since the line element is $d\tilde{s} = R \,d\theta$, from \eqref{polar coords spike willmore} we easily obtain the following result for the 
line integral over $\partial \hat{\gamma}_{\mathcal{Q}}$ in (\ref{F_A willmore ads}) and (\ref{F_A willmore ads Q-plane}) for this configuration
\be
  \int_{\partial \hat{\gamma}_{\mathcal{Q}}}  \frac{\tilde{b}^z}{z} \, d\tilde{s}
\,=\,
-\cot\alpha 
\int_{{\theta}_\varepsilon}^{\pi- \theta_{\varepsilon}}\frac{1}{\sin\theta} \, d\theta
\,=\,
- \cot\alpha \, \log \big[ \tan(\theta / 2 )\big]
\Big|_{\theta_\varepsilon}^{\pi-\theta_\varepsilon}
\ee
As $\varepsilon\to 0$, at the leading order  we obtain
\be
\label{willmore-final-pi/2}
 \int_{\partial \hat{\gamma}_{\mathcal{Q}}}  \frac{\tilde{b}^z}{z} \, d\tilde{s}
\,=\,
-\,2 \cot\alpha \, \log(R/ \varepsilon ) + O(1)
\ee

Thus, the logarithmic divergence and its coefficient in \eqref{area half-disk final main} have been recovered by specifying the general formula (\ref{F_A willmore ads Q-plane}) to this configuration,
finding that they come from the line integral over $\partial \hat{\gamma}_{\mathcal{Q}}$.

\subsection{Infinite wedge adjacent to the boundary}
\label{sec wedge}

\begin{figure}[t] 
\vspace{-.8cm}
\hspace{-.8cm}
\includegraphics[width=1.1\textwidth]{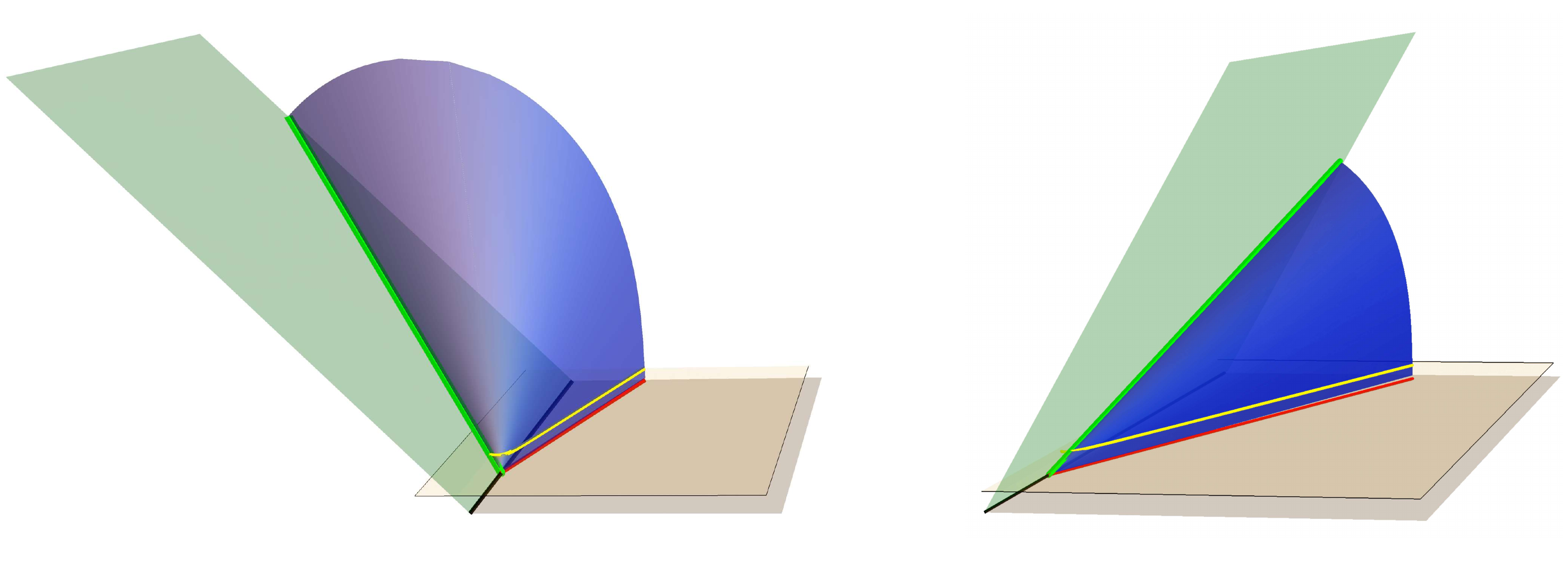}
\vspace{-.3cm}
\caption{\label{fig_3d_wedge}
\small
Minimal surface $\hat{\gamma}_A$ anchored to the entangling edge (the red line) of an infinite wedge $A$ adjacent to the flat boundary, in the gravitational setup of Sec.\;\ref{sec flat bdy}.
The analytic results of \cite{Seminara:2017hhh} have been employed to plot these surfaces (see Sec.\;\ref{sec wedge}).
The green half plane is $\mathcal{Q}$ and the yellow plane corresponds to $z = \varepsilon$,
and they intersect $\hat{\gamma}_A$ along the green  line and the yellow curve  respectively.
Left: $\omega=0.2$ and $\alpha=\pi/4$.
Right: $\omega=0.3$ and $\alpha=2\pi/3$.
}
\end{figure}

In the gravitational setup described in Sec.\;\ref{sec flat bdy}, let us consider the following infinite wedge $A$ adjacent to the flat boundary
\be
\label{domain wedge bdy}
A = \big\{ (\rho, \phi) \,\big|\, 0\leqslant \phi \leqslant \omega \,, \,\rho \leqslant L \big\}
\;\;\qquad\;\;
L \gg \varepsilon
\ee
where $\omega$ is the opening angle of the wedge and we have adopted the polar coordinates $(\rho,\phi)$ for the spatial section of the conformal boundary
such that $\phi =0$ corresponds to the positive $y$ semiaxis, which are related to the usual Cartesian coordinates  as $x=\rho\sin\phi$ and $y = \rho\cos\phi$.

The  minimal surface $\hat{\gamma}_A$  has been found analytically in \cite{Seminara:2017hhh}.
In Fig.\;\ref{fig_3d_wedge} we show two examples of $\hat{\gamma}_A$ corresponding to the same $A$ and to different slopes for $\mathcal{Q}$.
In \cite{Seminara:2017hhh} the area of the corresponding regularised surface $\hat{\gamma}_\varepsilon = \hat{\gamma}_A \cap \{z \geqslant \varepsilon \}$
has been computed, finding (\ref{area wedge intro exp}) and an explicit expression for the corner function $F_\alpha(\omega)$.

The parametric form of the minimal surface $\hat{\gamma}_A$ can be written in cylindrical coordinates $(z,\rho, \phi)$ by introducing the following ansatz
\begin{equation}
\label{cusp_parametric_form}
(z,\rho,\phi)=\left( \,\frac{\rho}{q(	\phi)} \,,\, \rho\,, \,\phi \right)
\qquad
\rho \in (0,L), 
\qquad
\phi\in (\phi_*\, , \omega)
\end{equation}
where $\phi_*$ corresponds to the value of $\phi$ characterising the line $\partial \hat{\gamma}_\mathcal{Q}$ along which $\hat{\gamma}_A\perp \mathcal{Q}$
(the green line in Fig.\;\ref{fig_3d_wedge}).

The function $q(\phi)$, which provides the minimal surface, can be implicitly obtained from
(see (F.9) and (F.10) in \cite{Seminara:2017hhh}) 
\begin{equation}
\big|\phi- \omega +P_0(q_0)\big|
=
P(q,q_0)
\end{equation}
where 
\begin{eqnarray}
\label{P0 expression v2}
P_0(q_0) 
 &\equiv&
\frac{1}{q_0}\;
\bigg\{
(1+Q_0^2) \; \Pi\left( -1/Q_0^2 \, ,  - Q_0^2 \right) 
- Q_0^2\;
\mathbb{K}\left( -Q_0^2 \right)
\bigg\} \\
\rule{0pt}{.8cm}
\label{Pfunction def main}
P(q,q_0) 
&\equiv&
\frac{1}{q_0}\,
\bigg\{
(1+Q_0^2) \; \Pi\big( -1/Q_0^2\, , \, \sigma(q,q_0)  \, \big|  - Q_0^2\big) 
- Q_0^2\;
\mathbb{F}\big( \sigma(q,q_0)\, \big| - Q_0^2 \big)
\bigg\}
\end{eqnarray}
with
\begin{equation}
\label{Q0^2 def main}
\sigma(q,q_0) \equiv \,\arctan \sqrt{\frac{q^2-q_0^2}{1+2q_0^2}}
\hspace{.4cm} \qquad \hspace{.4cm} 
Q_0^2 \equiv \frac{q_0^2}{1+q_0^2} \in (0,1)
\end{equation}
being $\mathbb{F}(\phi | m)$ and $\Pi(n,\phi | m)$ the incomplete elliptic integrals of the first and third kind respectively,
while $\mathbb{K}(x)$ is the complete elliptic integral of the first kind.
Here $q_0$ is the minimum value of $q$.  
Given the opening angle $\omega$ of the wedge and the slope $\alpha$ of $\mathcal{Q}$,
the values of $q_0$ and $\phi_*$ are obtained  by inverting the following transcendental equations
\begin{equation}
\label{phi_ast q0 main}
\phi_\ast(\alpha,q_0) =     \, \eta_\alpha \arcsin [  s_\ast(\alpha, q_0) ]
\qquad
\omega
\,=\,
P_0(q_0) + 
 \phi_\ast(\alpha,q_0)-  \eta_\alpha \,P\big(q_\ast(\alpha,q_0) ,q_0\big)
\end{equation}
where we have introduced
\be
\label{s_ast def main}
s_\ast(\alpha,q_0) 
\equiv\,
-\,\eta_\alpha  \frac{\cot \alpha}{\sqrt{2}}   
\left\{ \frac{\sqrt{1+4 (\sin\alpha)^2 (q_0^4+q_0^2) } - \cos (2 \alpha) }{(\cos\alpha)^2 + q_0^4+q_0^2}
\right\}^{\frac{1}{2}} 
\qquad 
q_\ast(\alpha,q_0) = \frac{| \cot\alpha \,|}{s_\ast(\alpha,q_0)}
\ee

The expansion of the area of the minimal surface $\hat{\gamma}_\varepsilon $ as $\varepsilon \to 0$ is given by (\ref{area wedge intro exp}).
The analytic expression of the corner function reads \cite{Seminara:2017hhh} 
\begin{equation}
\label{total corner func bdy main}
F_\alpha(\omega)
=
F(q_0) + \eta_\alpha\, \mathcal{G}\big( q_\ast(\alpha,q_0) ,q_0\big)
\end{equation}
where
\be
\label{F_q0}
F(q_0) \, \equiv \,\frac{\mathbb{E}(\tilde q_0^2)-(1-\tilde q_0^2)\,\mathbb{K}(\tilde q_0^2)}{\sqrt{1-2 \tilde q_0^2}}
\;\;\qquad\;\;
\tilde q_0 = \frac{q_0^2}{1+2 q_0^2}
\ee
and
\be
\mathcal{G}(q,q_0) 
\, \equiv \,
\sqrt{1+q_0^2} \,
\left\{
\mathbb{F}\big( \sigma(q,q_0)\, \big| - Q_0^2 \big)
-
\mathbb{E}\big( \sigma(q,q_0)\, \big| - Q_0^2 \big)
+
\sqrt{\frac{(q^2+1)(q^2-q_0^2)}{(q_0^2+1)(q^2+q_0^2+1)}}
\;\right\} 
\label{mathcalG def}
\ee

The goal of this section is to show that (\ref{total corner func bdy main}) can be recovered also from the general expression \eqref{F_A willmore ads Q-plane}.
In Appendix\;\ref{app: wedge adj}, we discuss the details of this computation, while in the following we only report the main intermediate steps. 
Let us remark that, while for the half disk centered on the flat boundary the logarithmic divergence in the expansion of $\mathcal{A}[\hat{\gamma}_\varepsilon]$ 
comes only from the line integral over $\partial \hat{\gamma}_{\mathcal{Q}}$ (see Sec.\;\ref{sec app half disk}),
for the wedge adjacent to the boundary both the surface integral over $\hat{\gamma}_\varepsilon$ and the line integral over $\partial \hat{\gamma}_{\mathcal{Q}}$ provide a logarithmic divergence.
In particular,  for the line integral over $\partial\hat \gamma_\mathcal{Q}$ we find 
\begin{equation}
\label{linecon}
\int_{\partial\hat \gamma_\mathcal{Q}} \frac{\tilde b^z}{z} \,d\tilde s 
\,=\,
-\cot\alpha \, \sqrt{1+(\cos\alpha \,\cot\phi_*)^2} \; \log (L / \varepsilon)+\mathcal{O}(1)
\end{equation}
Notice that, since for the half disk centered on the flat boundary $\phi_*= \eta_\alpha \, \pi/2$,
the expression (\ref{linecon}) is consistent with (\ref{willmore-final-pi/2})
(where we recall that the factor of $2$ occurs because the half disk contains two corners).

The evaluation of the surface integral over $\hat{\gamma}_\varepsilon$ in  \eqref{F_A willmore ads Q-plane} is less straightforward than (\ref{linecon})
and it provides the following logarithmic divergence
\begin{equation}
\label{bulkcon}
\int_{\hat{\gamma}_\varepsilon} 
\frac{(\tilde{n}^{z})^2}{z^2} 
\, d\tilde{\mathcal{A}} 
\, = \,
\mathcal{I} (q_*,q_0) \log (L / \varepsilon) + O(1)
\end{equation}
whose coefficient is given by 
\be
\label{coeff curly-I}
\mathcal{I} (q_*,q_0) 
\equiv
F(q_0)
-\eta_\alpha \Bigg( S(q_*,q_0) + \sqrt{\frac{(q_*-q_0) \, (q_*+q_0 ) \,(q_*^2+q_0^2+1)}{q_*^2+1}} \;\Bigg)
\ee
where 
	\be
	\label{Sfunc qqstar}
	S(q_*,q_0)
	\equiv 
	\sqrt{q_0^2+1}
	\left[ \,
	\mathbb{E}
	\left(i\, \textrm{arccsch}\, \frac{q_0}{\sqrt{q^2+1}} \;
	\bigg\vert\,
	\frac{-\, q_0^2}{q_0^2+1} \right) 
	-\,
	\mathbb{F}
	\left(i\, \textrm{arccsch}\, \frac{q_0}{\sqrt{q^2+1}} \;
	\bigg\vert\,
	\frac{-\, q_0^2}{q_0^2+1} \right) 
	\right]
	\Bigg\vert_{q_0}^{q_*}
	\ee

By combining (\ref{linecon}) and (\ref{bulkcon}) as prescribed by the general formula (\ref{F_A willmore ads})
(see the Appendix\;\ref{app: wedge adj} for some technical details), 
we recover the expression (\ref{total corner func bdy main}) for the corner function.

\section{Conclusions}

Understanding the gauge/gravity correspondence when the dual conformal field theory 
 has a physical boundary is an important question.
 
In this manuscript we studied the 
holographic entanglement entropy in AdS$_4$/BCFT$_3$
for spatial regions having arbitrary shapes, along the lines of 
\cite{Takayanagi:2011zk, Fujita:2011fp, Nozaki:2012qd, Miao:2017gyt, Astaneh:2017ghi, FarajiAstaneh:2017hqv, Nagasaki:2011ue, Seminara:2017hhh}.
Considering the expansion of the holographic entanglement entropy as the UV cutoff vanishes 
(see (\ref{hee bdy intro}) and (\ref{RT formula bdy intro})),
our main result is the analytic formula  (\ref{F_A generic}) for the subleading term $F_A$,
that can be applied for any spatial region and any static gravitational background. 
Known analytic expressions corresponding to some particular configurations such as
an infinite strip parallel to a flat boundary \cite{Nagasaki:2011ue, Miao:2017gyt, Seminara:2017hhh} 
or an infinite wedge adjacent to a flat boundary \cite{Seminara:2017hhh}
have been recovered through (\ref{F_A generic}).

The  second result is the analytic study of the extremal surfaces anchored to a disk
disjoint from a boundary which is either flat or circular,
when the gravitational background is a part of $\mathbb{H}_3$.
The corresponding expression for the subleading term $F_A$ has been obtained
both by evaluating the area in the standard way and by specialising (\ref{F_A generic})
to this configuration.
Furthermore, when the spatial section of the gravitational spacetime is 
a part of $\mathbb{H}_3$, we found the bound $F_A \geqslant 2\pi$ for any region $A$ 
that does not intersect the boundary.

The numerical analysis of the holographic entanglement entropy in AdS$_4$/BCFT$_3$ 
performed in this manuscript is based on Surface Evolver, which has been previously employed 
to study the holographic corner functions in AdS$_4$/BCFT$_3$ \cite{Seminara:2017hhh}
and the holographic entanglement entropy in AdS$_4$/CFT$_3$ for regions with arbitrary shape \cite{Fonda:2014cca, Fonda:2015nma}.

Many interesting directions can be explored in the future. 
In the AdS/BCFT construction, it is important to identify the possible relation occurring between 
the geometrical parameter $\alpha$ in the bulk and the allowed boundary conditions for the dual BCFT$_3$.
As for the holographic entanglement entropy in AdS$_4$/BCFT$_3$,
gravitational backgrounds dual to a BCFT$_3$ at finite temperature or to a boundary RG flows could be considered. 
The expression (\ref{F_A generic}) found in this manuscript holds also in these cases;
nonetheless, it would be interesting to find explicit analytic expressions in some simple setups. 
An interesting direction to address involves time-dependent gravitational backgrounds. 

The results and the methods discussed in this manuscript could be useful also in the context of the 
gauge/gravity correspondence in the presence of defects (AdS/dCFT)
\cite{Nagasaki:2011ue, Kristjansen:2012tn}.

\subsection*{Acknowledgments}

It is our pleasure to thank Matthew Headrick, Luca Heltai, Alberto Sartori, Michael Smolkin, Tadashi Takayanagi 
and in particular Jonas Hirsch and Martina Teruzzi for useful discussions. 
JS and ET are grateful to the Instituto Balseiro, Bariloche, 
for hospitality and the stimulating environment enjoyed during the {\it It from Qubit} workshop/school.
ET thanks the Hebrew University for the hospitality during part of this work.
We are grateful to the Galileo Galilei Institute for Theoretical Physics for the hospitality during the program {\it Entanglement in Quantum Systems}
and the INFN for partial support during the final stage of this work.

\appendix

\section{Useful mappings}
\label{app:mapping}

In this appendix we discuss two useful transformations employed in Sec.\;\ref{sec ads4} and Sec.\;\ref{sec disk}.

Let us consider the map $(x,y,z) \to (X,Y,Z)$ with $z>0$ and $Z>0$ defined by \cite{Berenstein:1998ij}
\begin{equation}
\left\{\;
\begin{split}
X &=
\lambda \; 
\frac{x - a_x + c_x \big[ (\boldsymbol{x}-\boldsymbol{a})^2+ z^2\big]
}{
	1+2 \,\boldsymbol{c} \cdot (\boldsymbol{x}-\boldsymbol{a})+\boldsymbol{c}^2 \big[(\boldsymbol{x}-\boldsymbol{a})^2+z^2\big]
}
\\
\rule{0pt}{.8cm}
Y &=
\lambda \; 
\frac{y - a_y + c_y \big[ (\boldsymbol{x}-\boldsymbol{a})^2+ z^2\big]
}{
	1+2 \,\boldsymbol{c} \cdot (\boldsymbol{x}-\boldsymbol{a})+\boldsymbol{c}^2 \big[(\boldsymbol{x}-\boldsymbol{a})^2+z^2\big]
}
\\
\rule{0pt}{.6cm}
Z &=
\lambda\; 
\frac{z}{1+2 \, \boldsymbol{c} \cdot (\boldsymbol{x}-\boldsymbol{a})+\boldsymbol{c}^2 \big[(\boldsymbol{x}-\boldsymbol{a})^2+z^2\big]}
\end{split}
\right.
\label{general_mapping}
\end{equation}
where $\lambda >0$, the vectors $\boldsymbol{x} = (x, y)$, $\boldsymbol{a} = (a_x, a_y)$ and $\boldsymbol{c} = (c_x, c_y)$ belong to $\mathbb{R}^{2}$
and $\cdot$ denotes the standard scalar product between vectors in $\mathbb{R}^2$. 
The transformation (\ref{general_mapping}) leaves the metric \eqref{H3 metric} invariant up to a conformal factor.
On the conformal boundary, given by $Z=z=0$, the map (\ref{general_mapping}) becomes a special conformal transformation.

The first special case of (\ref{general_mapping}) that we need is the map 
sending the right half plane $\{(x,y) \in \mathbb{R}^2, x\geqslant 0\}$ at $z=0$
into the disk {$\{ (X,Y) \in \mathbb{R}^2 , X^2+Y^2 \leqslant R^2_{\mathcal{Q}} \}$} of radius $R_{\mathcal{Q}}$ at $Z=0$.
Since this transformation must send the straight line $(x,y,z)=(0,y,0)$ 
into the circle $\mathcal{C}_\mathcal{Q}$ given by ${(X,Y,Z)=(R_\mathcal{Q} \cos\phi, R_\mathcal{Q} \sin\phi,0)}$
with $\phi \in [0, 2\pi)$,
it can be constructed by first setting ${a_y=a_z=0}$ and $x=z=0$ in \eqref{general_mapping},
and then imposing $X^2+Y^2=R^2_{\mathcal{Q}}$.
This leads to
\begin{equation}
\frac{ \lambda^2\, (a_x^2+y^2)}{\left(a_x^2 + y^2 \right)
	\left(c_x^2+c_y^2\right)-2 a_x c_x+2 c_y y+1}-R^2_\mathcal{Q}=0 
	\qquad
	\forall y \in \mathbb{R}
\end{equation}
which can be written as a quadratic equation in $y$ that must hold $\forall y \in \mathbb{R}$; 
therefore we have to impose the vanishing of its coefficients. 
This procedure gives $a_x=\pm \,R_\mathcal{Q}/(2\lambda)$ and $\boldsymbol{c}= (\pm \, \lambda/R_\mathcal{Q},0)$,
where the choice of the sign determines whether the right half plane $x\geqslant 0$ is mapped in the region
inside (positive sign) or outside (negative sign) the circle $\mathcal{C}_\mathcal{Q}$.
Considering the former option, we find that (\ref{general_mapping}) becomes
\be
\label{mapping}
\left\{\;
\begin{split}
X& 
=\frac{R_{\mathcal{Q}}\big[ 4 \lambda ^2 (x^2+y^2+z^2 )-R^2_\mathcal{Q} \big]}{R^2_\mathcal{Q}+ 4 \lambda^2  \left(x^2+y^2+z^2\right)+4 \lambda R_\mathcal{Q} \,x}
\\
\rule{0pt}{.7cm}
Y & 
=\frac{4 \lambda R^2_\mathcal{Q}  \, y  }{R^2_\mathcal{Q}+ 4 \lambda^2  \left(x^2+y^2+z^2\right)+4 \lambda R_\mathcal{Q} \,x}
\\
\rule{0pt}{.7cm}
Z & 
= \frac{4 \lambda R^2_\mathcal{Q}  \, z}{R^2_\mathcal{Q}+ 4 \lambda^2  \left(x^2+y^2+z^2\right)+4 \lambda R_\mathcal{Q}\, x}
\end{split}
\right.
\;\; \qquad \;\;
\left\{\;
\begin{split}
x& =    \frac{R_\mathcal{Q} \big(R^2_\mathcal{Q}-X^2-Y^2-Z^2\big)}{2\lambda \big[ (R_\mathcal{Q}-X)^2+Y^2+Z^2 \big]} 
\\
\rule{0pt}{.7cm}
y & =  \frac{R^2_\mathcal{Q}\, Y}{\lambda \big[ (R_\mathcal{Q}-X)^2+Y^2+Z^2 \big]} 
\\
\rule{0pt}{.7cm}
z & =  \frac{R^2_\mathcal{Q}\, Z}{\lambda \big[ (R_\mathcal{Q}-X)^2+Y^2+Z^2 \big]}
\end{split}
\right.\;\;
\ee
where also the inverse map has been reported.
The transformations in (\ref{mapping}) relate the setups described in Sec.\;\ref{sec flat bdy} and Sec.\;\ref{sec circular bdy}.
Since in (\ref{mapping}) the constant $\lambda$ can be reabsorbed  through the rescaling $(x,y,z) \to \lambda (x,y,z) $, which leaves $\mathbb{H}_3$ invariant,
we are allowed to set $\lambda =1 $ in (\ref{mapping}) without loss of generality. 
The first transformation in (\ref{mapping}) maps the half plane (\ref{brane-profile})
 into the following spherical cap \cite{Fujita:2011fp}
\begin{equation}
\label{brane_trasf}
X^2+Y^2 + \left(Z-R_\mathcal{Q}\cot\alpha \right)^2
= \frac{R^2_\mathcal{Q}}{\sin^2\alpha}
\;\;\qquad\;\;
Z>0
\end{equation}
which has been written also in (\ref{Qbrane disk}) by means of cylindrical coordinates. 
When $\alpha = \pi/2$, (\ref{brane_trasf}) reduces to the hemisphere of radius $R_\mathcal{Q}$.

The second map in (\ref{mapping}) has been used in Sec.\;\ref{subsec:disk disjoint} to obtain
the holographic entanglement entropy of a disk disjoint from a flat boundary 
starting from the holographic entanglement entropy of a disk concentric to a circular boundary computed in Sec.\;\ref{sec disk disjoint circ}.
Indeed, by considering the circle $(X,Y)=(b_\circ+ R_\circ \cos\phi , R_\circ \sin\phi)$ with $\phi \in [0, 2\pi)$ inside the disk delimited by $\mathcal{C}_\mathcal{Q}$,
its image through the second map in (\ref{mapping}) is the circle $(x,y)=(d+R+ R \cos\phi,R \sin\phi)$ in the right half plane at $z=0$,
which has radius $R$ and distance $d$ from the straight boundary at $x=0$.
We find that $(R_\circ, b_\circ)$ can be written in terms of $(R, d)$ as follows
\begin{eqnarray}
\label{R_circ} 
\frac{R_\circ}{R_{\mathcal{Q}} } 
& = & 
\frac{4  \, R/ R_{\mathcal{Q}}}{1 + 4  (d/ R_{\mathcal{Q}}+2R/ R_{\mathcal{Q}}) d/ R_{\mathcal{Q}}  +4  (d / R_{\mathcal{Q}} +R/ R_{\mathcal{Q}} ) } 
\\
\rule{0pt}{.8cm}
\label{b_circ eq}
\frac{b_\circ}{R_{\mathcal{Q}} } 
& = & 
1- \frac{2\big[ 1+2(d / R_{\mathcal{Q}} +R/ R_{\mathcal{Q}} )  \big]}{\big[ 1+ 2(d / R_{\mathcal{Q}} +2 R/ R_{\mathcal{Q}} ) \big]\, \big[ 1+ 2 d / R_{\mathcal{Q}}\big]}
\end{eqnarray}
where the r.h.s.'s depend only on the ratios $R/ R_{\mathcal{Q}}$ and $d / R_{\mathcal{Q}}$.
For a circle concentric to the circular boundary (considered e.g. in Sec.\;\ref{sec disk disjoint circ}), $b_\circ = 0$.
The expressions in \eqref{R and d} have been obtained by solving \eqref{R_circ} and \eqref{b_circ eq} in this special case. 

The second map in (\ref{mapping}) has been also employed to obtain the analytic expressions for the extremal surfaces shown in 
Fig.\;\ref{fig_shadow_alpha} and Fig.\;\ref{fig_shadow_dist}.

The second transformation coming from  (\ref{general_mapping}) that we consider is the one 
mapping the disk delimited by $\mathcal{C}_\mathcal{Q}$ into itself. 
Let us rename $(x,y,z)=(X',Y',Z')$ in (\ref{general_mapping}) for this case, where $Z=Z'=0$.
By imposing that the circle $\mathcal{C}_\mathcal{Q}$ is mapped into itself in the coordinates $(X',Y')$,
we find the following two options:
either ${\boldsymbol{a}=(\pm\,R_\mathcal{Q}\sqrt{(\lambda+1)/\lambda},0)}$ and ${\boldsymbol{c}= (\pm \sqrt{\lambda(1+\lambda)}/R_\mathcal{Q},0)}$ 
or $\boldsymbol{a}= (\pm R_\mathcal{Q}\sqrt{(\lambda-1)/\lambda},0)$ and $\boldsymbol{c}=(\mp\sqrt{\lambda(\lambda-1)}/R_\mathcal{Q},0)$ with $\lambda \geqslant 1$. 
Since the first option exchanges the interior and the exterior of the disk, we have to select the second one,
where the lower or upper choice of the signs move the center of the disk along either $X'>0$ or $X'<0$ respectively.
Being the disk invariant under a rotation of $\pi$ about the origin, we can choose one of these two options without loss of generality.
Considering e.g. $\boldsymbol{a}=- (R_\mathcal{Q}\sqrt{(\lambda-1)/\lambda},0)$ and $\boldsymbol{c}= (\sqrt{\lambda(\lambda-1)}/R_\mathcal{Q},0)$ with $\lambda \geqslant 1$,
the resulting transformation maps the circle $(X,Y)=(R_\circ \cos\phi , R_\circ \sin\phi)$ with $R_\circ  < R_\mathcal{Q}$ 
into the circle $(X',Y')=(b'_\circ+ R'_\circ \cos\phi , R'_\circ \sin\phi)$, where
\be
\label{mappp}
\frac{R'_\circ}{R_\mathcal{Q}} 
= \frac{R_\circ/ R_\mathcal{Q} }{\lambda\big[1-(R_\circ/ R_\mathcal{Q})^2\big]+(R_\circ/ R_\mathcal{Q})^2}
\;\;\qquad\;\;
\frac{b'_{\circ}}{R_{\mathcal{Q}}} = \frac{\sqrt{(\lambda-1)\lambda}\; \big[1-(R_\circ/ R_\mathcal{Q})^2\big]}{\lambda\big[1-(R_\circ/ R_\mathcal{Q})^2\big]+(R_\circ/ R_\mathcal{Q})^2}
\ee
By inverting these relations, 
one gets $R_\circ/ R_\mathcal{Q}$ and $\lambda$ in terms of $R'_\circ / R_\mathcal{Q}$ and $b'_{\circ} / R_{\mathcal{Q}}$.
We have checked that, under the transformation that we have constructed, 
the surface $\mathcal{Q}$ in (\ref{brane_trasf}) remains unchanged for any value of $\lambda \geqslant 1$.

The expression of $R_\circ/ R_\mathcal{Q}$ obtained in this way and (\ref{F_A disk circ bdy}) provide the finite term $F_A$ 
for the holographic entanglement entropy of a disk $A$ inside the disk delimited by $\mathcal{C}_\mathcal{Q}$
in the cases where these two disks are not concentric.

	\section{On the disk concentric to a circular boundary}
\label{app:disk}

In this appendix we provide some technical details underlying the derivation of the results reported in Sec.\;\ref{sec disk disjoint circ}.
Considering the setup introduced in Sec.\;\ref{sec circular bdy}, we are interested in
the extremal surfaces anchored to the boundary of a disk $A$ with radius $R_\circ $ 
concentric to the disk of radius $R_\mathcal{Q} > R_\circ $,
which corresponds to a spatial slice of the spacetime where the BCFT$_3$ is defined. 
In the following we will adapt to this case the analysis reported in Appendix\;D.2 of \cite{Fonda:2014cca}
about the extremal surfaces anchored to the boundary of an annulus in AdS$_4$/CFT$_3$
(see also \cite{Drukker:2005cu}).

\subsection{Extremal surfaces}	
\label{app profiles}

The invariance under rotations about the vertical axis $z$ of this configuration
significantly simplifies the analysis of the corresponding extremal surfaces. 
Indeed, by introducing the polar coordinates $(\rho, \phi)$ in the $z=0$ plane,
an extremal surface is determined by the curve $z=z(\rho)$ obtained by taking its section at a fixed angle $\phi$.
The area functional evaluated on these surfaces becomes
\be
\mathcal{A}
\,=\,
2\pi L^2_{\textrm{\tiny AdS}} \int d\rho \, \rho\;\frac{\sqrt{z'^2+1}}{z^2}
\label{disk_area_func}
\ee
The equation of motion coming from the extremization of this functional reads
\begin{equation}
z \,z''+\big(1+z'^2\big)\left( 2+\frac{z \,z'}{\rho} \right)=0
\label{eq_of_motion}
\end{equation}
By introducing the variable $u$ and the function $\zeta(\rho)$ as follows
\begin{equation}
z(\rho)=\rho \, \zeta(\rho)
\qquad
u= \log \rho
\qquad
\zeta_u= \partial_u \zeta
\label{disk_subst}
\end{equation}
the differential equation \eqref{eq_of_motion} becomes
\begin{equation}
\zeta \,\zeta_u (1+\partial_{\zeta}\zeta_u)+\big[1+(\zeta+\zeta_u)^2\big]  \big[ 2+\zeta(\zeta+\zeta_u) \big]
\,=\,0
\label{eq_diff_disk_zeta}
\end{equation}
Integrating this equation, one finds
\begin{equation}
\zeta_{u,\pm}=-\frac{1+\zeta^2}{\zeta} \left[ 1\pm \frac{\zeta}{\sqrt{k (1+\zeta^2)-\zeta^4}} \right]^{-1}
\qquad 
k>0
\label{zeta_u}
\end{equation}
where $k$ is the integration constant.
By employing that $du = d\zeta / \zeta_u$ and integrating (\ref{zeta_u}) starting from an arbitrary initial point, we get
\be
\log(\rho / \rho_{\textrm{\tiny in}})
=
\int_{u_{\textrm{in}}}^{u} d\tilde u 
\,=\,
-\int_{\zeta_{\textrm{\tiny in}}}^{\zeta } \frac{\lambda}{1+\lambda^2} \left[ 1\pm \frac{\lambda}{\sqrt{k(1+\lambda^2)-\lambda^4}} \right] d\lambda 
\label{integral_profile_circle}
\ee

Since the extremal surfaces are anchored to the boundary of the disk $A$ of radius $R_\circ$ at $z=0$, 
from (\ref{disk_subst}) we have $\zeta(R_\circ) =0$ and $u=\log R_\circ$ when $\rho = R_\circ$.
Choosing $\rho_{\textrm{\tiny in}} = R_\circ$ and the negative sign within the integrand in (\ref{integral_profile_circle}), one finds
the first equation in the r.h.s. of (\ref{annulus branches profile}), namely
\begin{equation}
\label{first_branch}
\log (\rho / R_\circ )\, =\,  -\,q_{-,k}(\zeta)
\end{equation}
where $q_{-,k}(\zeta)$ has been defined in \eqref{q function pm}.
The choice of the negative sign in (\ref{first_branch}) will be discussed at the end of this subsection.

The solution \eqref{first_branch} is well defined as long as the expression under the square root of \eqref{integral_profile_circle} is positive.
Such expression vanishes at the point $P_m=(\rho_m, \zeta_m)$, whose coordinates have been reported in  \eqref{Pm coords}.
Following the curve given by (\ref{first_branch}) starting from $(\rho, z) = (R_\circ, 0)$, if it intersects $\mathcal{Q}$ before reaching $P_m$,
then (\ref{first_branch}) fully describes the profile of $\hat{\gamma}_{A}^{\textrm{\tiny \,con}}$.
Otherwise, (\ref{first_branch}) provides the profile of $\hat{\gamma}_{A}^{\textrm{\tiny \,con}}$ until $P_m$ and for the part between $P_m$ 
and the point $P_\ast =(\rho_\ast, \zeta_\ast)$ 
(which fully  characterises the curve $\partial \hat{\gamma}_{\mathcal{Q}}=\hat{\gamma}_A \cap \mathcal{Q}$ in this case)
also the function defined by (\ref{integral_profile_circle}) with the positive sign must be employed.
In particular, the profile between $P_m$ and $P_\ast$ reads
\be
\label{second_branch}
\log (\rho / R_\circ )\, =\,  -\,q_{+,k}(\zeta) + q_{+,k}(\zeta_m) - q_{-,k}(\zeta_m)
\ee
which can be written also in the form given by the second expression in the r.h.s. of (\ref{annulus branches profile}),
once (\ref{R_ratio_chi}) has been used.

In order to justify  \eqref{Pm coords} for the coordinates of $P_m$,
let us consider the unit vectors $v^\mu_\pm$ tangent to the radial profile of $\hat \gamma_A^{\text{\tiny con}}$ along the two branches 
characterised by $q_{\pm,k}$. They read
\begin{equation}
\label{tang_surf_erik}
v^\mu_\pm =  
\big(v^\rho_\pm,v^z_\pm,v^\phi_\pm\big) 
=\,
\frac{\pm \, z}{\sqrt{(q'_{\pm,k})^2+ ( 1-\zeta \, q'_{\pm,k} )^2}}\, \big( \,q'_{\pm,k}\,, \zeta\,q'_{\pm,k} - 1\,,0 \, \big)
\end{equation}
where $\pm$ refer to the two different branches. 
At the matching point $P_m$, the tangent vector field defined by $v^\mu_{\pm}$ must be continuous,
hence a necessary condition is that $g_{\mu\nu} \, v_+^\mu v_-^\nu = 1$ at $P_m$.
From (\ref{tang_surf_erik}), one finds that this requirement gives $ \zeta^4 = k(1+\zeta^2) $,
whose only admissible solution is the first expression in  \eqref{Pm coords}.

The boundary condition along the curve $\partial \hat{\gamma}_{\mathcal{Q}}=\hat{\gamma}_A \cap \mathcal{Q}$ provides the parameter $k$. 
The condition to impose is that $\hat \gamma_A^{\textrm{\tiny \,con}}$ and $\mathcal{Q}$ intersects orthogonally along $\partial \hat{\gamma}_{\mathcal{Q}}$.
This requirement is equivalent to impose that the vector $v_\mu$ tangent to $\hat \gamma_A^{\textrm{\tiny \,con}}$ 
and the vector $u_\mu$ tangent to $\mathcal{Q}$ are orthogonal along $\partial \hat{\gamma}_{\mathcal{Q}}$.
From \eqref{Qbrane disk}, we find
\bea
u^\mu  = (u^\rho,u^z,u^\phi) =  \left(\cot\alpha- \rho \, \zeta / R_\mathcal{Q} \,, \rho/R_\mathcal{Q} \,,0 \right)
\label{tang_brane_erik}
\eea

By using \eqref{tang_surf_erik} and \eqref{tang_brane_erik}, we find that the orthogonality condition $v^\rho u^\rho + v^z u^z=0$
at the intersection between $\hat{\gamma}_{A}^{\textrm{\tiny \,con}}$ and $\mathcal{Q}$
gives
\begin{equation}
q'_{\pm,k}(\rho_*)= \frac{\rho_*}{R_Q} \tan \alpha
\label{f1_star}
\end{equation} 
where  $q_{\pm,k}'$ can be read from \eqref{q function pm} and
$\rho_*/R_{\mathcal{Q}}$ can be obtained by specializing \eqref{Qbrane circ bdy} to $P_\ast$.
This leads to 
\begin{equation}
\frac{\sqrt{\zeta_*^2+\sin^2\alpha}}{\cos\alpha}
\,=\,
\pm \,\frac{\zeta^2_*}{\sqrt{k(1+\zeta_*^2)-\zeta_*^4}}
\label{orthogonality_branches}
\end{equation} 
that allows us to write $\zeta_*$ as a function of $k$ and $\alpha$.
Indeed, the first expression of (\ref{star coords}) can be found by taking the square of (\ref{orthogonality_branches}).
The $\pm$ in the r.h.s. of (\ref{orthogonality_branches}) correspond to the same choice of sign occurring in (\ref{f1_star}).
From (\ref{orthogonality_branches}) and $\zeta_* \geqslant 0$, one observes that
the orthogonality condition can be satisfied only by $q_{+,k}$ when $\alpha \leqslant \pi/2$,
while for $\alpha \geqslant \pi/2$ the orthogonality condition leads to select $q_{-,k}$\,.
Consequently, $P_\ast$ belongs to the branch described $q_{-,k}$ for $\alpha \leqslant \pi/2$
and to the one characterised by $q_{+,k}$ for $\alpha \geqslant \pi/2$.
When $\alpha \to \pi/2$ the l.h.s. of (\ref{orthogonality_branches}) diverges; 
therefore the argument of the square root in the r.h.s. must vanish in this limit.
This means that $\zeta_*=\zeta_m$, being $\zeta_m$ given in (\ref{Pm coords}).
Thus, when $\alpha = \pi/2$, the extremal surface $\hat \gamma_A^{\,\text{\tiny con}}$ intersects $\mathcal{Q}$ at the matching point $P_m$ of the two branches characterised by $q_{\pm,k}$.

In order to justify the choice of $q_{-,k}$ in \eqref{first_branch}, in the following we show that a contradiction is obtained 
if $q_{+,k}$ is assumed in \eqref{first_branch} instead of $q_{-,k}\,$.
In this case the profile of $\hat \gamma_A$ can be obtained from \eqref{annulus branches profile} simply by exchanging the role of $R_\circ$ and $R_\text{\tiny aux}$, i.e. 
	\be
	\label{annulus branches modific}
	\rho_\gamma(\theta) 
	\,=\,
	\Bigg\{\begin{array}{l}
		R_\circ \, e^{-q_{+, k}(\zeta)}
		\\
		\rule{0pt}{.5cm}
		R_{\textrm{\tiny aux}}	 \, e^{-q_{-, k}(\zeta)}
	\end{array}
	\ee
	where now $R_\mathcal{Q}>R_\circ>R_{\text{\tiny aux}}$. 
	First, let us notice  that the maximum value of $z(\zeta)$ is realized in the $q_{+,k}$ branch
	because from \eqref{tang_surf_erik} we have that $v_{\pm}^z=0$ only for the $q_{+,k}$ branch (at $\zeta=\sqrt[4]{k}$). 
	Since $R_\mathcal{Q}>R_\circ>R_{\text{\tiny aux}}$, this observation leads to conclude that $\mathcal{Q}$ cannot intersect the $q_{-,k}$ branch 
	without intersecting the one described by $q_{+,k}$ (see e.g. the red and the black curves in the top panel of Fig.\;\ref{fig_profiles_evolver} as guidance). 
	Thus, the only possibility is that $\mathcal{Q}$ intersects orthogonally the branch described by $q_{+,k}\,$.
	In this case, the condition \eqref{orthogonality_branches} leads to $\alpha \leqslant \pi/2$. 
	In order to find a contradiction, let us compare the quantity $\rho^2+z^2$ for the branch $q_{+,k}$ with the one for $\mathcal{Q}$. 
	For $\mathcal{Q}$ in the range $\alpha \leqslant \pi/2$ we get 
	\be
	\rho^2+z^2 
	= R_\mathcal{Q}^2 \big(1+\zeta^2\big) Q_\alpha^2 
	= R^2_\mathcal{Q} 
	\frac{\big( \sqrt{\zeta^2 (\csc\alpha)^2 +1}
	+ \zeta \cot \alpha\big)^2}{\zeta^2+1}  
	\geqslant R_\mathcal{Q}^2
	\ee 
	being $Q_\alpha$ the function introduced in (\ref{Qbrane circ bdy}). As for the $q_{+,k}$ branch, from \eqref{annulus branches modific} and \eqref{disk_subst} we get $\rho_\gamma^2+z^2=(1+\zeta^2)\rho^2_\gamma=R_\circ^2\, e^{-2f_{+,k}} $ 
	where $f_{+,k} \equiv q_{+,k} - \log \sqrt{1+\zeta^2}$ (see \eqref{q_+-_explicit}).
	Since $f_{+,k}>0$ for any $\zeta$ and $R_\circ>R_\mathcal{Q}$, we have $\rho_\gamma^2+z^2 < R_\mathcal{Q}^2$. 
	This means that the branch described by $q_{+,k}$ cannot intersect $\mathcal{Q}$ in the whole range $\alpha \leqslant \pi/2$,
	ruling out the possibility that $\hat \gamma_{A}$ is described by the profile \eqref{annulus branches modific}.

\subsection{Area}
\label{app_area}

In this appendix we evaluate the area of $\hat \gamma_A^{\,\text{\tiny con}}$ in two ways:
by a direct computation of the integral \eqref{disk_area_func}
and by specialising the general formula \eqref{F_A willmore ads-finite} to the extremal surfaces $\hat \gamma_A^{\textrm{\tiny \,con}}$.

The analysis performed in Sec.\;\ref{app profiles} allows to write the area of $\hat \gamma_A^{\,\text{\tiny con}}$ 
from \eqref{disk_area_func} and \eqref{disk_subst} as follows
\begin{equation}
\mathcal{A} \,=\,	\left\{\; \begin{array}{ll}
\label{integral_for_area}
2\pi L^2_{\textrm{\tiny AdS}} \bigg( \,\displaystyle\int_{\varepsilon/R_\circ}^{\zeta_m} \, 
\frac{d\zeta}{\zeta^2 \sqrt{1+\zeta^2-\zeta^4/k}}
+ \displaystyle\int_{\zeta_*}^{\zeta_m} \, \frac{d\zeta}{\zeta^2 \sqrt{1+\zeta^2-\zeta^4/k}}\, \bigg)
& \hspace{.5cm}  0 < \alpha \leqslant \pi/2
\\
\rule{0pt}{.9cm}
2\pi L^2_{\textrm{\tiny AdS}} \displaystyle\int_{\varepsilon/R_\circ}^{\zeta_*} \, 
\frac{d\zeta}{\zeta^2 \sqrt{1+\zeta^2-\zeta^4/k}}
& \hspace{.5cm} \pi/2 \leqslant \alpha <\pi 
\end{array}\right.
\end{equation}
where the UV cutoff $\varepsilon$ has been introduced to regularise $\mathcal{A}$, which is a divergent quantity as $\varepsilon \to 0$.
Le us recall that $\zeta_*=\zeta_m$ for $\alpha = \pi/2$.
The integrals in (\ref{integral_for_area}) can be explicitly written by using that
\begin{equation}
\label{F_torto_k}
\int \frac{d\zeta}{\zeta^2 \sqrt{1+\zeta^2-\zeta^4/k}} \,=\,-\,\mathcal{F}_k(\zeta) + \textrm{const}
\end{equation}
where $\mathcal{F}_k(\zeta)$ has been introduced in \eqref{F_A int_evaluated}.
The expression \eqref{F_A disk circ bdy} for $F_{ \textrm{\tiny con}}$ can be found from (\ref{integral_for_area})
by employing the expansions of $\mathcal{F}_k(\zeta)$ as $\zeta \rightarrow 0^+$, which reads
\begin{equation}
\label{total_area_circ}
\mathcal{F}_k(\zeta) =\frac{1}{\zeta}+\frac{\zeta}{2}+O\left(\zeta^3\right)
\end{equation}

In the remaining part of this appendix we show that the analytic expression for $F_{\textrm{\tiny con}}$ given in (\ref{F_A disk circ bdy}) 
can be obtained also by applying the general formula \eqref{F_A willmore ads-finite} in the special cases of the extremal surfaces  $\hat \gamma_A^{\textrm{\tiny \,con}}$.

In order to evaluate the surface integral over $\hat{\gamma}_A$ in (\ref{F_A willmore ads-finite}), 
we need the normal vector $\tilde n_\mu$
and the area element $d \tilde{\mathcal{A}}$, which are given respectively by
\begin{equation}
\label{ntile_mu comp}
\tilde n^\mu =(n^\rho,n^z,n^\phi)=\frac{1}{\sqrt{1+z'^2}} \left( z',-1,0 \right) 
\qquad
d \tilde{\mathcal{A}}\,=\, \sqrt{z'^2+1}\;\rho\, d\rho \, d\phi
\end{equation}
The evaluation of the surface integral over $\hat{\gamma}_A$ in (\ref{F_A willmore ads-finite}) can be performed
by using (\ref{disk_subst}) and (\ref{ntile_mu comp}), finding
\begin{equation}
\label{wilmore_circ_integral}
\int \, \frac{(\tilde n^z)^2}{z^2} \, d\tilde{\mathcal{A}}  \,=\,	
\left\{\; \begin{array}{ll}
2\pi  \Big( 	\mathcal{F}_{k,-}(\zeta_m) + 	\mathcal{F}_{k,+}(\zeta_m) -  	\mathcal{F}_{k,+}(\zeta_\ast)  \Big)
& \hspace{.8cm}  0 < \alpha \leqslant \pi/2
\\
\rule{0pt}{.6cm}
\displaystyle 2\pi 	\, \mathcal{F}_{k,-}(\zeta_\ast) 
& \hspace{.8cm} \pi/2 \leqslant \alpha <\pi 
\end{array}
\right.
\end{equation}
(which can be written as reported in (\ref{disk_surface_term_main}))
where we have introduced the following functions
\begin{equation}
\label{Fcal deff0}
\mathcal{F}_{k,\pm}(\zeta)
\equiv 
\frac{1}{\sqrt{k}} 
\int_0^\zeta
\frac{\big(\sqrt{k(1+\xi^2) -\xi^4}\pm \xi \big)^2}{\left(\xi^2+1\right)^2 \sqrt{k(1+\xi^2)-\xi^4}}
\,d\xi		
\end{equation}
which can be written in terms of $\mathcal{F}_{k}(\zeta)$ (see \eqref{F_k cal def}).
The relation  \eqref{F_k cal def} has been found by integrating the following identity
\begin{equation}		
	\frac{\big(\sqrt{k \left(\zeta^2+1\right)-\zeta^4}\pm \zeta \big)^2}{\sqrt{k}\left(\zeta^2+1\right)^2 \sqrt{k
			\left(\zeta^2+1\right)-\zeta^4}}
	+
	\frac{1}{\sqrt{k}}\,
	\frac{\partial}{\partial \zeta}
	\bigg(\frac{\sqrt{k \left(\zeta^2+1\right)-\zeta^4} \pm \zeta }{\zeta(\zeta^2+1)}\bigg)
	=
	-\, \frac{1}{\zeta^2 \sqrt{\zeta^2+1-\zeta^4/k}}
\end{equation}
The result of this indefinite integration contains an arbitrary integration constant which can be fixed by taking $\zeta \to 0$
 and imposing that both sides of the equation are consistent in this limit
(also (\ref{total_area_circ}) is useful in this computation).

In order to facilitate the recovering of the expression (\ref{F_A disk circ bdy}) for $F_{\textrm{\tiny con}}$,
let us observe that, by employing \eqref{F_k cal def}, the expression \eqref{disk_surface_term_main} 
can be written as follows
\be
\label{surf_term_disk}
\int_{\hat{\gamma}}  \, \frac{(\tilde n^z)^2}{z^2} \, d\tilde{\mathcal{A}} 
\,=\,
F_{\textrm{\tiny con}}
-2\pi \, \frac{\zeta^3_\ast + \eta_\alpha \sqrt{k \left(\zeta_*^2+1\right)-\zeta_*^4} }{\sqrt{k}\, \zeta_\ast (\zeta^2_*+1)} 
\,=\,
F_{\textrm{\tiny con}}
-2\pi\, \frac{\zeta^3_\ast - \sqrt{k}\, \cos\alpha }{\sqrt{k}\, \zeta_\ast (\zeta^2_*+1)} 
\ee
where in the last step we used the identity 
$\sqrt{k \left(\zeta_*^2+1\right)-\zeta_*^4} =  -\, \sqrt{k}\,\eta_\alpha \cos\alpha$, 
which comes from the explicit form of $\zeta_\ast$ given in the first expression of \eqref{star coords}.

As for  the boundary term in \eqref{F_A willmore ads-finite}, 
the vector $\tilde b^\mu$ can be obtained from the vector which is tangent to $\mathcal{Q}$ given in \eqref{tang_brane_erik}, 
finding
\begin{equation}
\label{btilde comps}
\tilde b^\mu=\big( \,\tilde b^\rho,\tilde b^z,\tilde b^\phi\, \big) 
= 
\Bigg(
\sqrt{ 1 -\left(\frac{\rho_*}{R_\mathcal{Q}} \zeta_* \sin\alpha -\cos\alpha \right)^2} , 
\frac{\rho_*}{R_\mathcal{Q}} \,\zeta_* \sin\alpha -\cos\alpha 
\,,0 
\Bigg)
\end{equation}
that coincides with (\ref{tang_surf_erik})  evaluated at $P_\ast$.
From the component $\tilde b^z$ in (\ref{btilde comps}) 
and the fact that $d\tilde{s} = \rho_\ast d\phi $ along $\partial \hat{\gamma}_{\mathcal{Q}}$,
we find that the boundary contribution in \eqref{F_A willmore ads-finite} becomes
\begin{equation}
\label{bound_term_disk}
\int_{\partial \hat{\gamma}_{\mathcal{Q}}} 
\frac{\tilde{b}^z}{z} \, d\tilde{s}  
\,=\,
2\pi\,\frac{\tilde{b}^z}{\zeta_*}
\,=\,
2\pi \left(\frac{\rho_*}{R_\mathcal{Q}} \sin\alpha-\frac{\cos\alpha}{\zeta_*} \right)
\end{equation}
which reduces to \eqref{disk_bound_term_main}, once the second  expression  of \eqref{star coords} has been employed. 
Then, plugging  \eqref{surf_term_disk} and \eqref{disk_bound_term_main} into \eqref{F_A willmore ads-finite}, one obtains
\begin{equation}
	\label{contract}
	F_A 
	\,=\,
	F_{\textrm{\tiny con}} -
	2\pi \, \frac{\zeta^2_* - \sqrt{k\big[ \zeta^2_* +(\sin \alpha)^2 \big]} }{\sqrt{k} \left(\zeta^2_*+1\right)  } 
	\end{equation}
where, by using  (\ref{orthogonality_branches}) and the identity given in the text below (\ref{surf_term_disk}),
it is straightforward to observe that the numerator in the r.h.s. vanishes.

\subsection{Limiting regimes}
\label{sec-app-limits}

\begin{figure}[t] 
\vspace{-1.2cm}
\hspace{-.6cm}
\includegraphics[width=1.1\textwidth]{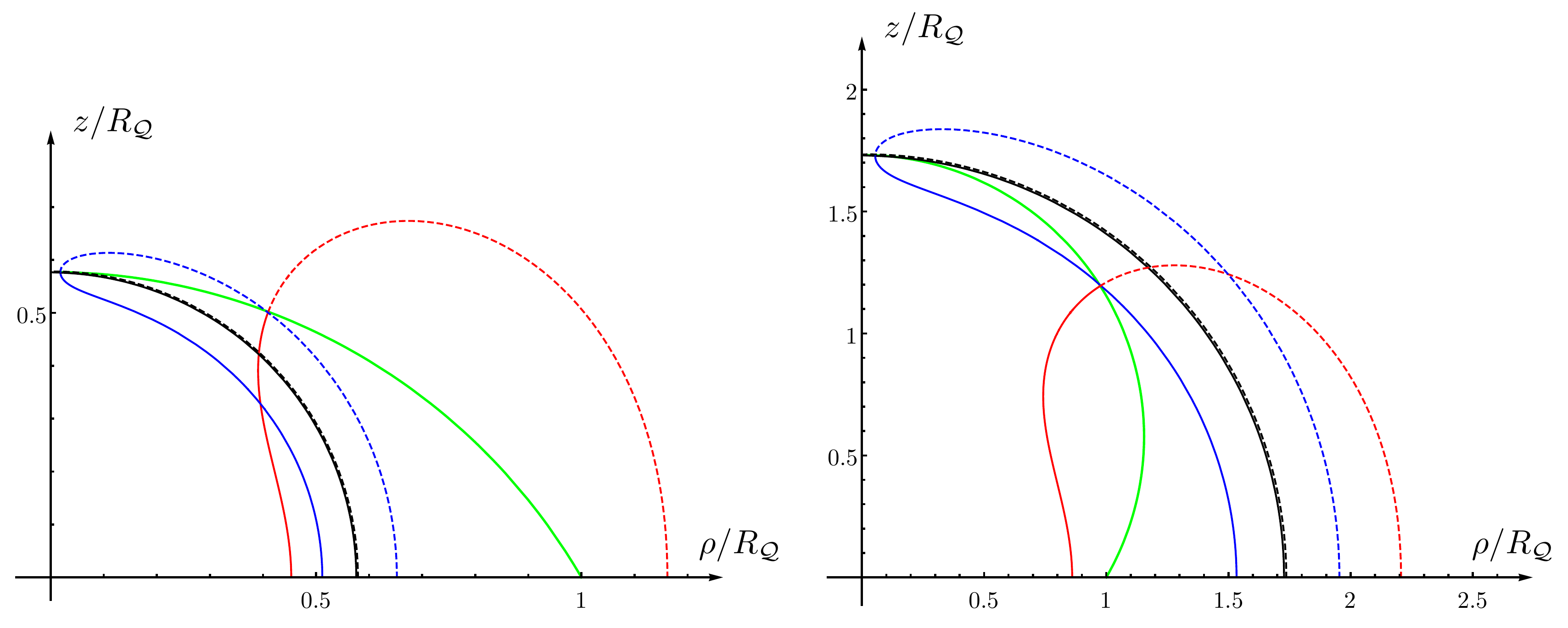}
\vspace{-.4cm}
\caption{\label{fig_app_limit}
\small
Radial profiles of extremal surfaces $\hat{\gamma}_{A}^{\textrm{\tiny \,con}}$ intersecting $\mathcal{Q}$ (green curve) orthogonally
and anchored to a disk $A$ of radius $R_\circ$ concentric to a circular boundary with radius $R_{\mathcal{Q}}$ (see Sec.\;\ref{sec disk profiles}).
Here $\alpha =2\pi/3$ (left panel) and $\alpha =\pi/3$ (right panel).
Any solid line provides $\hat{\gamma}_{A}^{\textrm{\tiny \,con}}$ and the dashed line with the same colour gives the radial profile of the 
corresponding auxiliary surface $\hat{\gamma}_{A, \textrm{\tiny \,aux}}^{\textrm{\tiny \,con}}$.
Here the values of $k$ associated to  $\hat{\gamma}_{A}^{\textrm{\tiny \,con}}$ (see Fig.\;\ref{fig_RFk})
are $k=1$ (red), $k=1000$ (blue) and $k=10^7$ (black).
For large $k$, both $\hat{\gamma}_{A}^{\textrm{\tiny \,con}}$ and the corresponding $\hat{\gamma}_{A, \textrm{\tiny \,aux}}^{\textrm{\tiny \,con}}$
tend to the hemisphere with radius $\cot(\alpha/2)$, which is tangent to $\mathcal{Q}$ at $\rho=0$.
}
\end{figure}

	In the remaining part of this appendix we provide some technical details about the limiting regimes $k \to 0$ and $k \to \infty$
	of the analytic expressions for $R_\circ / R_\mathcal{Q}$ and $F_{\textrm{\tiny con}}$
	(see (\ref{initial ratio from k}) and (\ref{F_A disk circ bdy}) respectively).
	The results of this analysis have been reported in (\ref{R_ratio_exp}), (\ref{Rratio-k-inf}) and (\ref{FA ann K}).

	As for the ratio $R_\circ / R_\mathcal{Q}$, whose analytic expression is (\ref{initial ratio from k}) with $\chi(\zeta_m)$ given by (\ref{R_ratio_chi}),
	 we have to study $q_{\pm,k}(\zeta_*)$ and $q_{\pm,k}(\zeta_m)$ in these limiting regimes. 
	 
	 In order to find $q_{\pm,k}(\zeta_*)$ for $k \to 0$, let us write $q_{\pm,k}(\zeta_*)$ 
	 from the integral (\ref{q function pm}) evaluated for $\zeta=\zeta_\ast$ (see \eqref{star coords}) 
	 and adopt $\zeta_* \lambda$ as integration variable because it leads us to a definite integral whose extrema are $0$ and $1$.
	 By first expanding the integrand of the resulting formula and then integrating separately the terms of the expansion, we find 
	\begin{equation}
	\label{q_pm zeta_star exp}
	q_{\pm,k}(\zeta_*)
	\,=\,	 
	\pm \Big[\,
	\mathbb{E}\big(\arcsin(\sqrt{\sin\alpha}\,) \big|-1\big)
	-\mathbb{F}\big(\arcsin(\sqrt{\sin\alpha}\,) \big|-1\big)
	\Big]
	\sqrt[4]{k}
	+
	\frac{\sin\alpha}{2} \,\sqrt{k}  
	+ 
	O\big(k^{3/4}\big)
	\end{equation} 
	Adapting this analysis to $q_{\pm,k}(\zeta_m) $,  we obtain
	\be
	\label{q_pm zeta_m exp} 
	q_{\pm,k}(\zeta_m) 
	\,=\,
	\pm \Big(
	\mathbb{E}(-1)-\mathbb{K}(-1) 
	\Big)
	\sqrt[4]{k}+ \frac{\sqrt{k}}{2} + \mathcal{O}(k^{3/4}) 	
	\ee
	By employing the expansions (\ref{q_pm zeta_star exp}) and (\ref{q_pm zeta_m exp}) 
	into (\ref{R_ratio_chi}) and (\ref{initial ratio from k}), one gets the result (\ref{R_ratio_exp}). 
	
	As for the $k \rightarrow \infty$ regime, for the integrals (\ref{q function pm}) we have
\be
\label{q_pm svil k infinity}
q_{\pm,k}(\zeta)=\frac{1}{2}\log ( 1+\zeta^2 ) + \mathcal{O}\big(1/\sqrt{k}\,\big)
\ee
Moreover, from \eqref{Pm coords} and \eqref{star coords} notice that both $\zeta_\ast$ and $\zeta_m$ diverge, with $\zeta_\ast/\zeta_m \to 1$.
Thus, being $\zeta=z/\rho$ with finite $z$ for the surfaces that we are considering, we have that $\rho_\ast \to 0$ and $\rho_m\to 0$.
These observations tell us that, in the regime of large $k$,  
the two branches in \eqref{annulus branches profile} become the same arc of circle from $\rho=R_\circ$ to $\rho=0$
(see the black curves in Fig.\;\ref{fig_app_limit}).
In particular, we have $R_{\text{\tiny aux}} \to R_\circ$.
By taking the limit of \eqref{Qbrane circ bdy} for large $\zeta$ and employing the identity $\cot\alpha +\csc\alpha = \cot(\alpha/2)$,
one finds that $P_*=P_m=R_{\mathcal{Q}} (0, \cot(\alpha/2))$ in this regime. 
Then, being the limiting curve a circle of radius $R_\circ$, we have that $R_{\mathcal{Q}} \cot(\alpha/2) = R_\circ$.
The latter relation provides (\ref{Rratio-k-inf}), which is the asymptotic behaviour of the curves in Fig.\;\ref{fig_RFk}.
In Fig.\;\ref{fig_app_limit} we show some examples of extremal surfaces (which are not necessarily the global minimum of the area) 
as $k$ increases for two fixed values of $\alpha$, highlighting the limit of large $k$, which corresponds to the black curves.

In order to study the subleading term of area of the extremal surfaces as $k \to 0$ or $k \to \infty$,
we find it convenient to employ the expressions \eqref{F_A willmore ads-finite}, (\ref{disk_surface_term_main}) and (\ref{disk_bound_term_main}).
Indeed, since $\mathcal{F}_{k,\pm}(\zeta_*)$ and $\mathcal{F}_{k,\pm}(\zeta_m)$ can be written through the integral representation (\ref{Fcal deff0}) 
of the functions $\mathcal{F}_{k,\pm}(\zeta)$, we can adapt the above analysis to this case
(e.g. for $\mathcal{F}_{k,\pm}(\zeta_*)$ one first introduces $\zeta_* \xi$ as integration variable, obtaining a definite integral between $0$ and $1$,
then expands the integrand of the resulting expression and finally integrates the various terms of the expansion), finding
	\bea
	\label{f_pm zeta_star exp} 
	\mathcal{F}_{k,\pm}(\zeta_*) 
	& = & 
	\frac{1}{\sqrt[4]{k}}\,
	\Big[\,
	\mathbb{E}\big(\arcsin(\sqrt{\sin\alpha}\,) \big|-1\big)
	-\mathbb{F}\big(\arcsin(\sqrt{\sin\alpha}\,) \big|-1\big)
	\Big]
	\pm \sin\alpha 
	\nonumber 
	\\
	\rule{0pt}{.7cm}
	& &
	+\, \bigg(\frac{1}{4} \, \mathbb{F}\big(\arcsin(\sqrt{\sin\alpha}\,) \big|-1\big)  -  \eta_\alpha \cos\alpha \, \sqrt{\sin\alpha}  \, \bigg) \sqrt[4]{k}
	+\mathcal{O}\big(\sqrt{k}\,\big)
	\eea
	and
	\be
	\label{f_pm zeta_m exp}
	\mathcal{F}_{k,\pm}(\zeta_m) 
	\,=\, 
	\frac{\mathbb{E}(-1)-\mathbb{K}(-1)}{\sqrt[4]{k}} \, \pm 1+\frac{\mathbb{K}(-1)}{4} \, \sqrt[4]{k} + \mathcal{O}(\sqrt{k}) 
	\ee
	By using these expansions into \eqref{disk_surface_term_main}, together with \eqref{disk_bound_term_main} into \eqref{F_A willmore ads-finite},
	the expansion \eqref{FA ann K} is obtained. 
	
	The asymptotic value $2\pi$ for large $k$ in Fig.\;\ref{fig_Fk} can be found by employing 
	that the profile of $\hat \gamma_A^{\text{\tiny con}}$ in this regime is the one of the hemisphere in $\mathbb{H}_3$ anchored to $R_\circ$ 
	(see also the Appendix\;D in \cite{Fonda:2014cca}).
	Since the finite term of the area for the hemisphere in $\mathbb{H}_3$ is $2\pi$, 
	we can easily conclude that the curves in Fig.\;\ref{fig_Fk} tend to this value as $k \to \infty$.

\section{On the infinite wedge adjacent to the boundary}
\label{app: wedge adj}

In the gravitational setup described in Sec.\;\ref{sec flat bdy}, let us consider an infinite wedge $A$ in (\ref{domain wedge bdy}),
which is adjacent to the flat boundary and whose opening angle is $\omega$.
As for the corresponding holographic entanglement entropy, 
by a direct evaluation of the area $\mathcal{A}[\hat{\gamma}_\varepsilon]$, 
it has been found that (\ref{area wedge intro exp}) holds and 
the analytic expression of the corner function $F_\alpha(\omega)$ has been found \cite{Seminara:2017hhh}.
In this appendix we provide some technical details underlying the discussion of Sec.\;\ref{sec wedge}, where
we have shown that the analytic expression for $F_\alpha(\omega)$ can be recovered also through  \eqref{F_A willmore ads Q-plane}.

Let us consider first the line integral over $\partial \hat \gamma_{\mathcal{Q}}$ occurring in \eqref{F_A willmore ads Q-plane}. 
The curve $\partial \hat \gamma_{\mathcal{Q}}$ is a line on $\mathcal{Q}$ which can be 
parameterised as follows \cite{Seminara:2017hhh}
\begin{equation}
\label{paramegammaQ}
\partial \hat{\gamma}_{\mathcal{Q}}: \;\;
(z,x,y) = \rho\, 
\big(
- \sin\phi_* \tan\alpha \,,\, \sin\phi_* \,,\,
\cos\phi_*
\big)
\qquad
0 \leqslant  \rho \leqslant L
\end{equation} 
where $\phi_*$ is the angular coordinate characterising the projection of $\partial \hat{\gamma}_{\mathcal{Q}}$ on the $z=0$ plane.
The line element $d\tilde s$ induced by the flat metric in \eqref{F_A willmore ads Q-plane}  reads
\begin{equation}
\label{line_element}
d\tilde s
=
\sqrt{x'^2+y^2+z'^2}\,d\rho 
=\frac{\sqrt{x'^2+\cos^2\alpha \, y'^2}\,d\rho }{|\cos \alpha|}
= 
-\frac{\eta_\alpha}{\cos \alpha}\,\sqrt{\sin^2\phi_*+\cos^2\alpha\,  \cos^2\phi_*} \,d\rho 
\end{equation}
By employing \eqref{paramegammaQ} and \eqref{line_element},  
the line integral over $\partial \hat \gamma_{\mathcal{Q}}$ in \eqref{F_A willmore ads Q-plane} becomes
\begin{equation}
\label{integral_brane}
- 
\cos \alpha 
\int_{\partial \hat{\gamma}_{\mathcal{Q}}} 
\frac{1}{z} \, d\tilde{s} 
\,=\,
-\, \cot\alpha \int_{\rho_\varepsilon}^L \frac{\sqrt{1+\cos^2\alpha\cot^2\phi_*}}{\rho}\,d\rho
\end{equation}
where $\text{sign}(\sin\phi_*)=\eta_\alpha$ has been used. 
The integral in the r.h.s. of (\ref{integral_brane}) has been regularised by introducing  the lower extremum $\rho_\varepsilon$,
which is  defined by the condition $\varepsilon= -\,\rho_\varepsilon \,\sin\phi_* \tan\alpha$,
obtained by intersecting $\partial \hat{\gamma}_{\mathcal{Q}}$ in (\ref{paramegammaQ}) with the plane $z=\varepsilon$.
The radial  integral  \eqref{integral_brane} can be easily evaluated, finding (\ref{linecon}) at leading order as $\varepsilon \to 0$.

In order to compute the surface integral over $\hat{\gamma}_\varepsilon$ in \eqref{F_A willmore ads Q-plane}, we need the unit normal vector $\tilde n_\nu$.
Up to a normalization factor, this vector is given by  the gradient of the equation  $\mathcal{C}=z-\rho/q(\phi)=0$, where $q(\phi)$ has been introduced in (\ref{cusp_parametric_form}) 
and characterises the minimal surface. 
By imposing the normalization condition $\tilde n_\mu \tilde n^\mu=1$,  we get
\begin{equation}
\label{n_cusp}
\tilde n_\mu=\frac{1}{\sqrt{q^4+q^2+q'^2}}\left( q^2,-q,q'\rho \right)
\end{equation}
where the index $\mu$ spans the cylindrical coordinates $(z,\rho,\phi)$ defined in Sec.\;\ref{sec wedge}. 
The first derivative $q'$ of $q$ with respect to $\phi$ can be  expressed in term of $q$ and $q_0$ with the help of the integral of motion associated to the 
cyclic coordinate $\phi$  \cite{Drukker:1999zq, Seminara:2017hhh}, finding that
\begin{equation}
\label{first-integral-squared}
\frac{(q')^2}{q^2} = (q^2+1) \left(  \frac{q^4 + q^2}{q_0^4 + q_0^2}  - 1\right)
\qquad
q \geqslant q_0
\end{equation}
By using \eqref{n_cusp} and \eqref{first-integral-squared} the integrand  of the integral over $\hat{\gamma}_\varepsilon$  in  \eqref{F_A willmore ads Q-plane}  can be written as
\begin{equation}
\frac{(\tilde n^z)^2}{z^2}
=\frac{q^6}{\rho^2 \left( q^4+q^2+q'^2 \right)}
= \frac{q^2 (q_0^4+q_0^2)}{(q^2+1)^2 \rho^2}
\label{wilmore_integrand}
\end{equation}
In terms of  the cylindrical coordinates introduced in Sec.\;\ref{sec wedge},
the  area element induced by the flat metric reads
\begin{equation}
d\tilde{\mathcal{A}}= \frac{\sqrt{q'^2+q^4+q^2}}{q^2}\,\rho \,d\rho \,d\phi
\,=\,
\frac{q^2+1}{\sqrt{q_0^4+q_0^2}}\,\rho \,d\rho \,d\phi
\label{element_area_wilmore}
\end{equation}
Plugging \eqref{wilmore_integrand} and \eqref{element_area_wilmore} into the surface integral over $\hat{\gamma}_\varepsilon$ 
in \eqref{F_A willmore ads Q-plane}, it reduces to the following double integral 
\begin{equation}
\label{double-int-surface}
\int_{\hat{\gamma}_\varepsilon} 
\frac{(\tilde{n}^{z})^2}{z^2} 
\, d\tilde{\mathcal{A}} 
\,=  
\int_{\rho_{\textrm{\tiny min}}}^{\rho_{\textrm{\tiny max}}} \frac{1}{\rho} \,d\rho\int_{\phi_*}^{\omega_\varepsilon} \frac{q^2 \sqrt{q_0^4+q_0^2}}{q^2+1} \,d\phi 
\end{equation}
The integration domain in the angular integral is defined by the angle $\phi_*$ characterising $\partial \hat{\gamma}_\mathcal{Q}$ and $\omega_\varepsilon\equiv\omega-\delta_\varepsilon$,
where $\delta_\varepsilon\sim 0$  is the angle between the border  of the wedge at $\phi=\omega$  
and the straight line in the $z = 0$ half plane connecting the tip of the wedge to the intersection point between the circle given by $\rho = \rho_{\textrm{\tiny max}}$ and the projection of $\hat \gamma_A \cap \{z = \varepsilon\}$ on the  half plane $z = 0$. In the radial direction
we have introduced the large cutoff $\rho_{\textrm{\tiny max}}$  to regulate the infrared divergences of this integral, 
while the lower extremum
$ \rho_{\textrm{\tiny min}}=q_0\,\varepsilon$ (being $q_0$ the minimum value of $q$)
controls the UV behaviour.
The cutoff $\rho_{\textrm{\tiny max}}$ is related to $L$ in \eqref{domain wedge bdy} and to $\delta_\varepsilon$ through the relation $L=\rho_{\textrm{\tiny max}}\cos\delta_\varepsilon\,$,
and to $\varepsilon$ through the condition
\be
\rho_{\textrm{\tiny max}}
\,=\,
\varepsilon \,q (\omega-\delta_\varepsilon)
\ee 
In order to perform the angular integration in (\ref{double-int-surface}), it is convenient to change the integration variable from $\phi$ to $q$. 
However, since $q$  is not monotonic as function of $\phi$ for some values of $\alpha$, 
we have to split the integral into two separate contributions (depending on the  sign of $\cot\alpha$) as follows
\begin{equation}
\label{splitted integral}
\int_{\hat{\gamma}} 
\frac{(\tilde{n}^{z})^2}{z^2} 
\, d\tilde{\mathcal{A}}    
=
\int_{\rho_{\textrm{\tiny min}}}^{\rho_{\textrm{\tiny max}}} \frac{d \rho}{\rho}
\left(\,
\int_{q_0}^{\rho/\varepsilon}  \frac{q^2 \sqrt{q_0^4+q_0^2}}{\left(q^2+1\right)q'} \,dq 
-\eta_\alpha  \int_{q_0}^{q_*} \frac{q^2 \sqrt{q_0^4+q_0^2}}{\left(q^2+1\right)q'} \,dq 
\right)
 \end{equation}
 where \eqref{first-integral-squared} can be used to express $q^\prime$. By introducing the integration variable $\tilde \rho=\rho /\varepsilon$ in the radial integration, we get
\begin{equation}
\label{I1_plus_I2}
\int_{\hat{\gamma}} 
\frac{(\tilde{n}^{z})^2}{z^2} 
\, d\tilde{\mathcal{A}} 
 \,= 
 \int_{\rho_{\textrm{\tiny min}}/\varepsilon}^{\rho_{\textrm{\tiny max}}/\varepsilon} \frac{d\tilde \rho}{\tilde \rho}
 \left(
 \int_{\rho_{\textrm{\tiny min}}/\varepsilon}^{\tilde \rho} \frac{q^2 \sqrt{q_0^4+q_0^2}}{\left(q^2+1\right)q'} \,dq -\eta_\alpha   \int_{q_0}^{q_*} \frac{q^2 \sqrt{q_0^4+q_0^2}}{\left(q^2+1\right)q'} \,dq \right)
 \equiv\, I_1-\eta_\alpha \,I_2
\end{equation}
where $I_2$ is defined as the integral multiplying $\eta_\alpha$, while $I_1$ is the remaining one. 

\noindent
Considering $I_1$ first, in order to single out the logarithmic divergence we exchange the order of integration  between $\tilde\rho$ and $q$, finding that
\be
I_1 
=
\int_{\rho_{\textrm{\tiny min}}/\varepsilon}^{\rho_{\textrm{\tiny max}}/\varepsilon}   \frac{q^2 \sqrt{q_0^4+q_0^2}}{\left(q^2+1\right)q'} \,dq
\int_{q}^{\rho_{\textrm{\tiny max}}/\varepsilon} \frac{d\tilde \rho}{\tilde \rho}
\ee
Now the integration over $\rho$ can be easily performed, obtaining
\be
\rule{0pt}{.8cm} 
I_1 = 
\int_{\rho_{\textrm{\tiny min}}/\varepsilon}^{\rho_{\textrm{\tiny max}}/\varepsilon} 
\sqrt{q_0^4+q_0^2}\left(  \frac{q^2 }{\left(q^2+1\right)q'}\,\log (\rho_{\textrm{\tiny max}} /\varepsilon)- \frac{q^2\log q}{\left(q^2+1\right)q'}\right) dq
\ee
Since  $L$ is large, the dominant contribution comes from the first integral 
(the second one is finite in this limit).
In particular, we find
\be
\label{log i1}
I_1
= 
\bigg( \,\int_{q_0}^{+\infty}  \frac{q^2 \sqrt{q_0^4+q_0^2}}{\left(q^2+1\right)q'}\, dq \, \bigg)
\log (L / \varepsilon) + \cdots
\ee
where the integral multiplying the logarithmic divergence provides an integral representation of the function $F(q_0)$ 
given in \eqref{F_q0} in terms of elliptic function, i.e.
\begin{equation}
\int_{q_0}^{+\infty}\frac{q^2 \sqrt{q_0^4+q_0^2}}{\left(q^2+1\right)q'} \, dq = F(q_0)
\end{equation}

\noindent
The second  integral $I_2$ in  \eqref{I1_plus_I2} can be also calculated in closed form  
in terms of elliptic functions. 
Expanding the result for large $L$,  one finds that the dominant contribution is the following logarithmic divergence
\begin{equation}
\label{I2}
I_2 
= 
\Bigg(
S(q_*,q_0)+\sqrt{\frac{\left(q_*^2-q_0^2\right)  \left(q_*^2+q_0^2+1\right)}{q_*^2+1}} 
\; \Bigg) 
\log(L / \varepsilon)+\cdots
\end{equation}
where $S(q_*,q_0)$ has been defined in \eqref{Sfunc qqstar}.

Combining \eqref{log i1} and \eqref{I2} into (\ref{I1_plus_I2}), we get the logarithmic divergence 
provided by the surface integral over $\hat{\gamma}_\varepsilon$ in  \eqref{F_A willmore ads Q-plane}, which is given by
(\ref{bulkcon}) and (\ref{coeff curly-I}).
By taking into account also the logarithmic divergence 
provided by the line integral over $\partial\hat \gamma_\mathcal{Q}$ (see (\ref{linecon})),
for the coefficient of $\log(L / \varepsilon)$ in the subleading term $F_A$ we find 
\bea
\label{quasi_final_coef_wilmore}
F_\alpha(q_0)
&=&
F(q_0)-\eta_\alpha S(q_*(\alpha,q_0),q_0)
\\
\rule{0pt}{.9cm}
& &
-\; \eta_\alpha \, \sqrt{\frac{\left(q^2_*-q^2_0\right)  \left(q_*^2+q_0^2+1\right)}{q_*^2+1}}
- \sqrt{1+\cos^2\alpha\cot^2\phi_*(\alpha,q_0)}\; \cot\alpha
\nonumber
\eea
where the last two terms in \eqref{quasi_final_coef_wilmore} cancel, 
once the  explicit expressions for $\phi_*(\alpha,q_0)$   and $q_*(\alpha,q_0)$ (see \eqref{phi_ast q0 main} and \eqref{s_ast def main}) have been used. 
Hence, $F_\alpha(q_0)$ simplifies to
\begin{equation}
F_\alpha(q_0)=F(q_0)-\eta_\alpha  \, S(q_*(\alpha,q_0),q_0)
\label{corner_wilmore}
\end{equation}
In order to show that  \eqref{corner_wilmore} coincides with \eqref{total corner func bdy main}, we have to prove that $S(q_*,q_0)= -\,\mathcal{G}(q_*,q_0)$.
This  follows from two observations that can be easily verified:
the function obtained by taking the derivative of \eqref{mathcalG def} with respect to $q$ and then evaluating it for $q=q_\ast$
is the opposite of the derivative of \eqref{Sfunc qqstar} with respect to $q_\ast$
and $S(q_0,q_0)= \mathcal{G}(q_0,q_0)=0$ for any $\alpha$.

\section{Auxiliary surfaces}
\label{app aux_domains}

In this appendix we discuss a way to relate an extremal surface $\hat{\gamma}_A$ anchored to the entangling curve of a region $A$
in AdS$_4$/BCFT$_3$ to an extremal surface in AdS$_4$/CFT$_3$ anchored to a corresponding entangling curve in 
$\mathbb{R}^2$, which is  the spatial slice of the CFT$_3$,
being the gravitational background the one obtained by removing $\mathcal{Q}$.
We will discuss only the simplest cases where a spatial section of the 
gravitational spacetimes is given by $\mathbb{H}_3$ or part of it.

In AdS$_4$/BCFT$_3$ setups of Sec.\;\ref{sec flat bdy} and Sec.\;\ref{sec circular bdy}, 
if the extremal surface  $\hat{\gamma}_A$ does not intersect the boundary $\mathcal{Q}$,
then it can be also seen as an extremal surface in $\mathbb{H}_3$.
Instead, when $\hat{\gamma}_A$ intersects orthogonally $\mathcal{Q}$ 
along some curve $\partial \hat{\gamma}_{\mathcal{Q}}$ 
(since we mainly consider extremal surfaces intersecting $\mathcal{Q}$ orthogonally, 
in this appendix we denote by $\hat{\gamma}_A$  the surfaces $\hat{\gamma}_{A}^{\textrm{\tiny \,con}}$ of Sec.\;\ref{sec disk profiles}), 
we can consider the unique auxiliary surface $\hat{\gamma}_{A, \textrm{\tiny \,aux}}$ 
such that $\hat{\gamma}_A\cup \hat{\gamma}_{A, \textrm{\tiny \,aux}}$ is
an extremal surface in $\mathbb{H}_3$
and $\hat{\gamma}_{A, \textrm{\tiny \,aux}}$ is orthogonal to $\mathcal{Q}$ along $\partial \hat{\gamma}_{\mathcal{Q}}$.
The extremal surface $\hat{\gamma}_A\cup \hat{\gamma}_{A, \textrm{\tiny \,aux}}$  in $\mathbb{H}_3$
 is anchored to $\partial A_{\textrm{\tiny \,aux}}$ of some auxiliary region $A_{\textrm{\tiny \,aux}}$ 
 in the plane $\mathbb{R}^2$ at $z=0$.

As first example, let us consider an infinite strip $A$ of width $\ell$ adjacent to the flat boundary in the 
setup of Sec.\;\ref{sec flat bdy}.
In this case, $A_{\textrm{\tiny \,aux}} $ is a strip whose width is \cite{Seminara:2017hhh}
	\begin{equation}
	\label{ell_aux strip ddim}
	\ell_{\textrm{\tiny aux}} = 
	\frac{2\,\sqrt{\pi} \; \Gamma(\tfrac{3}{4})}{\Gamma(\tfrac{1}{4})\, \mathfrak{g}(\alpha)}\; \ell 
	\end{equation}
where $\mathfrak{g}(\alpha)$ has been defined in \eqref{g function def main}.
We remark that the strip $A$ is not necessarily a subset of the $A_{\textrm{\tiny \,aux}} $.
Indeed, for $\alpha \leqslant \alpha_{c, \textrm{\tiny \,aux}}$ we have that $A \subseteq A_{\textrm{\tiny \,aux}} $,
while $A_{\textrm{\tiny \,aux}}  \subseteq  A$ when $\alpha \geqslant \alpha_{c, \textrm{\tiny \,aux}}$.
The value of $\alpha_{c, \textrm{\tiny \,aux}}$ is defined by imposing that $\ell_{\textrm{\tiny aux}} = \ell$,
which gives
$\mathfrak{g}(\alpha_{c, \textrm{\tiny \,aux}}) =	2 \,\sqrt{\pi} \, \Gamma(\tfrac{3}{4}) / \Gamma(\tfrac{1}{4})$.
From the latter result and \eqref{g function def main}, for $\alpha \in (0,\pi)$ we have 
\be
\mathfrak{g}(\pi -\alpha)
\,=\,
\mathfrak{g}(\alpha_{c, \textrm{\tiny \,aux}})  -\mathfrak{g}(\alpha)
\ee
By specifying this relation to $\alpha = \alpha_c$, 
the critical value of $\alpha$ defined in Sec.\;\ref{sec strip adjacent} as the zero of $\mathfrak{g}(\alpha)$,
one finds that $\alpha_{c, \textrm{\tiny \,aux}}=\pi-\alpha_c$.

Another interesting configuration is given by a disk $A$ disjoint from the boundary which is either flat or circular (see Sec.\;\ref{sec disk}).
In these cases the extremal surfaces $\hat{\gamma}_{A, \textrm{\tiny \,aux}} \cup \hat{\gamma}_A$ are anchored to a pair of circles
and they have been studied in  \cite{Drukker:2005cu, Fonda:2014cca} for the gravitational background given by $\mathbb{H}_3$.
In the setup of Sec.\;\ref{sec circular bdy}, considering a disk $A$ of radius $R_\circ$ concentric to a circular boundary of radius $R_{\mathcal{Q}}$
as in Sec.\;\ref{sec disk disjoint circ}, we have that $ \hat{\gamma}_A \cup \hat{\gamma}_{A, \textrm{\tiny \,aux}} $ is an extremal surface in $\mathbb{H}_3$
anchored to the boundary of an annulus characterised by the radii $R_\circ$ and $R_{\textrm{\tiny aux}} > R_{\circ}$ (see also (\ref{annulus branches profile})).
The ratio $R_{\circ} / R_{\textrm{\tiny aux}}$ is given by (\ref{R_ratio_chi}).

\begin{figure}[t]
\vspace{-1.2cm}
\hspace{.1cm}
\includegraphics[width=.92\textwidth]{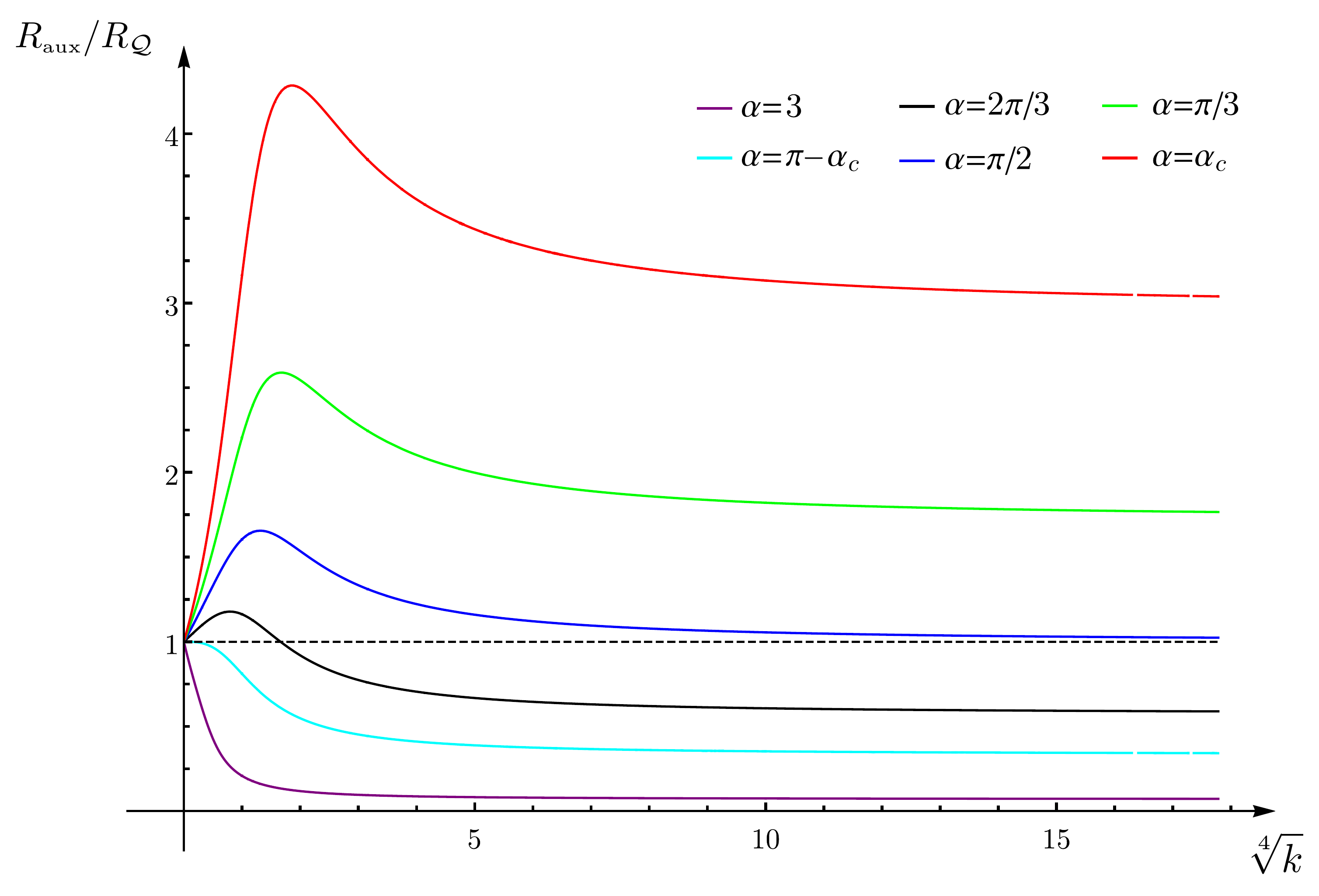}
\vspace{-.0cm}
	\caption{\label{fig:Raux_k}
		\small
		The ratio $R_{\text{\tiny \,aux}} / R_\mathcal{Q}$ for a disk $A$ concentric to a circular boundary of radius $R_\mathcal{Q}$ (see Sec.\;\ref{sec disk profiles} and Appendix\;\ref{app aux_domains})
		in terms of the parameter $k$, obtained by combining (\ref{R_ratio_chi}) and (\ref{initial ratio from k}).
		For $\alpha \geqslant \pi - \alpha_c$ we have that $R_{\text{\tiny \,aux}} \leqslant R_\mathcal{Q}$,
		therefore part of $\hat{\gamma}_{A, \textrm{\tiny \,aux}} $ belongs to the gravitational spacetime bounded by $\mathcal{Q}$.
	}
\end{figure}

Partitioning $\mathbb{H}_3$ into the part $\mathcal{M}_3$, introduced in Sec.\;\ref{sec general bg}, and its complement $\overline{\mathcal{M}}_3$,
we have that part of $\hat{\gamma}_{A, \textrm{\tiny \,aux}} $ belongs to $\overline{\mathcal{M}}_3$
because $\hat{\gamma}_{A, \textrm{\tiny \,aux}}  \perp \mathcal{Q}$ along $\partial \hat{\gamma}_{\mathcal{Q}}$.
It can happen that the intersection between $\hat{\gamma}_{A, \textrm{\tiny \,aux}} $ and $\mathcal{M}_3$ is non trivial (see e.g. the right panel in Fig.\;\ref{fig_shadow_alpha}).  
In Fig.\;\ref{fig:Raux_k} we show the ratio $R_{\textrm{\tiny aux}} / R_{\mathcal{Q}}$ as function of $k$ for some values of $\alpha$.
Let us introduce the critical value $\alpha_{c, \textrm{\tiny \,aux}}$ such that $R_{\textrm{\tiny aux}} / R_{\mathcal{Q}}<1$ for every $k$ at fixed  $\alpha > \alpha_{c, \textrm{\tiny \,aux}}$.
For this configuration we observe numerically that $\alpha_{c, \textrm{\tiny \,aux}}=\pi-\alpha_c$,
namely the same relation found above for the strip adjacent to the flat boundary.
Three qualitatively different situations are observed (see Fig.\;\ref{fig:Raux_k} ):
when $\alpha \leqslant \pi/2$ we have $R_{\textrm{\tiny aux}} > R_{\mathcal{Q}}$ and $\hat{\gamma}_{A, \textrm{\tiny \,aux}}  \cap \mathcal{M}_3 =\emptyset$,
for $\pi/2 \leqslant \alpha \leqslant  \pi -\alpha_c$ it is possible that $\hat{\gamma}_{A, \textrm{\tiny \,aux}}  \cap \mathcal{M}_3 \neq \emptyset$,
while when $ \alpha \geqslant \pi -\alpha_c $ we have that some part of $\hat{\gamma}_{A, \textrm{\tiny \,aux}} $ always belongs to $\mathcal{M}_3$ because $R_{\textrm{\tiny aux}} < R_{\mathcal{Q}}$.

By employing the map \eqref{mapping}, analogous considerations can be done for the extremal surfaces anchored to a disk $A$ disjoint from a flat boundary,
considered in Sec.\;\ref{subsec:disk disjoint}.
The extremal surface is anchored to a pair of circles in $\mathbb{R}^2$ and one of them is $\partial A$.
For this configuration explicit examples are given in Fig.\;\ref{fig_shadow_alpha} and Fig.\;\ref{fig_shadow_dist}, 
where $\hat{\gamma}_{A, \textrm{\tiny \,aux}}$  are the shaded surfaces.

As for the auxiliary surfaces corresponding to the extremal surfaces anchored to the singular domains considered in Sec.\;\ref{sec corners}
we refer the reader to the exhaustive discussion reported in  \cite{Seminara:2017hhh}.
Here we just recall that for the half disk adjacent to the boundary (see Sec.\;\ref{sec app half disk})  
$\hat{\gamma}_A\cup \hat{\gamma}_{A, \textrm{\tiny \,aux}}$ is the hemisphere 
and for the infinite wedge adjacent to the boundary (see Sec.\;\ref{sec wedge}) 
$\hat{\gamma}_A\cup \hat{\gamma}_{A, \textrm{\tiny \,aux}}$ is the extremal surface 
anchored to an infinite wedge in $\mathbb{R}^2$ found in \cite{Drukker:1999zq}.

\bibliographystyle{nb}


\end{document}